\documentclass[aps,prb,a4paper,twocolumn]{revtex4}
\usepackage{graphics}% Include figure files

\begin{document}

%%%%%%%%%%%%%%%%%%%%%%%%%%%%%%%%%%%%%%%%%%%%%%%%%%%%%%%%%%%%
\title{Nature of Adsorption on TiC($111$)}

\author{Carlo Ruberto}
\email{ruberto@fy.chalmers.se}
\author{Bengt I. Lundqvist}

\affiliation{Department of Applied Physics, Chalmers University of Technology, 
SE-412 96 G\"{o}teborg, Sweden}

%%%%%%%%%%%%%%%%%%%%%%%%%%%%%%%%%%%%%%%%%%%%%%%%%%%%%%%%%%%%
\begin{abstract}

Extensive density-functional calculations are performed for chemisorption of 
atoms in the three first periods (H, B, C, N, O, F, Al, Si, P, S, and Cl) 
on the polar TiC($111$) surface.  Calculations are also performed for O on 
TiC($001$), for full O($1 \times 1$) monolayer on TiC($111$), as well as 
for bulk TiC and for the clean TiC($111$) and ($001$) surfaces.  
Detailed results concerning atomic structures, energetics, and 
electronic structures are presented.  For the bulk and the clean surfaces, 
previous results are confirmed.  In addition, new detailed results are given 
on the presence of C--C bonds in the bulk and at the surface, as well as on 
the presence of a Ti-based surface resonance (TiSR) at the Fermi level and 
of C-based surface resonances (CSR's) in the lower part of the surface upper 
valence band (UVB).  For the adsorption, adsorption energies $E_{\rm ads}$ and 
relaxed geometries are presented, showing great variations characterized by 
pyramid-shaped $E_{\rm ads}$ trends within each period.  An extraordinarily 
strong chemisorption is found for the O atom, $8.8$ eV/adatom.  
On the basis of the calculated electronic structures, a concerted-coupling 
model for the chemisorption is proposed, in which two different types of 
adatom--substrate interactions work together to provide the obtained strong 
chemisorption: (i) adatom--TiSR and (ii) adatom--CSR's.  This model is 
used to successfully describe the essential features of the calculated 
$E_{\rm ads}$ trends.  The fundamental nature of this model, based on 
the Newns-Anderson model, should make it apt for general application to 
transition-metal carbides and nitrides, and for predictive purposes in 
technological applications, like cutting-tool multilayer coatings and 
MAX phases.

\end{abstract}

%%%%%%%%%%%%%%%%%%%%%%%%%%%%%%%%%%%%%%%%%%%%%%%%%%%%%%%%%%%%

\maketitle

%%%%%%%%%%%%%%%%%%%%%%%%%%%%%%%%%%%%%%%%%%%%%%%%%%%%%%%%%%%%
\section{Introduction}

Titanium carbide (TiC) is an intriguing material.  Like other early transition-metal 
carbides (TMC) and nitrides (TMN), it is characterized by a combination of properties 
that are traditionally assigned to two completely different types of materials: ceramics 
and metals.  It possesses the physical properties (extreme hardness, 
high melting point, outstanding wear resistance, {\it etc.}) typical of ceramic 
materials, such as Al$_2$O$_3$, but exhibits at the same time electric and 
heat conductivities comparable to those of pure transition 
metals.\cite{Oyama}  Such a behavior originates from a unique combination of covalency, 
ionicity, and metallicity.\cite{Schwarz}  

TMC's and TMN's, and in particular TiC, are often used or suggested for use in applications 
where their interaction with the environment ({\it e.g.}, growth, reactivity, tribology) 
is a crucial factor.  Examples are:  coating materials on, {\it e.g.}, 
cemented-carbide cutting tools,\cite{Prengel} 
microelectromechanical systems (MEMS),\cite{MEMS} 
fusion-reactor walls,\cite{Fusion} 
biocompatible materials,\cite{Bio} and in space applications;\cite{Space}  
substrate or stabilizing interlayer material in recording heads,\cite{Heads} 
for deposited diamond films,\cite{Diamond} 
in electronics,\cite{Electronics} 
and for the growth of carbidic nanostructures.\cite{Gunster}  
They have been suggested as good catalyst materials.\cite{Catalysis}  
Also, they are one of the components in the MAX phases (such as Ti$_3$SiC$_2$ and 
Ti$_3$AlC$_2$), which are technologically interesting due to their unique combination 
of hardness and plasticity, properties that arise from the interaction between 
the layers of the M$_6$X (for example, Ti$_6$C) octahedra and of the A component 
(such as Si or Al) that make up their structures.\cite{MAX}  

In the interaction of a material with its environment, chemisorption is the first step.  
Thus, a proper description of this interaction calls for an understanding of the 
chemisorption properties and of the mechanisms behind them.  
This study focuses on TiC, since long considered a prototype for the early TMC's and 
TMN's, and on its ($111$) surface, which, although metastable, has shown to be the 
most reactive low-index surface of the early TMC's\cite{Jansen} and a frequently 
occurring surface in the above-mentioned applications.  

So far, studies on the adsorption properties on the early TMC's have focused 
on the nonpolar and energetically favored ($001$) surface.  
For the polar and reactive ($111$) surface, information is more limited.  
Studies of H and O monolayers on the ($111$) surface of TiC and of other TMC's 
and TMN's, as well as of the clean-surface electronic structures, have been 
published.\cite{Tsukada,Fujimori,Aono81,Oshima81_jlcm,Zaima,Tan,Bradshaw,Edamoto92_prb,Ahn,Edamoto90,Souda88,Oshima83,Edamoto92_ss,Edamoto92_prb_O}  
However, to our knowledge, little has been done on the systematics of 
the chemisorption and in particular on connecting these findings into a general model, 
in which adsorption of other species is taken into account.  
Indeed, a desirable goal is to obtain a descriptive and predictive model for the 
nature of the interaction between adsorbate and substrate.  

For these reasons, we perform an extensive theoretical investigation of atomic adsorption 
on the TiC($111$) surface, based on quantum-mechanical first-principles calculations 
within the density-functional theory (DFT).\cite{DFT}  In order to understand 
the nature of the chemisorption, we perform a trend study, in which adsorption 
of H, of the second-period elements B, C, N, O, and F, and of the third-period 
elements Al, Si, P, S, and Cl are considered.  For comparison, O adsorption on 
TiC($001$) is also considered.  In addition, in order to connect our results to 
the available experimental results, calculations on a full ($1 \times 1$) monolayer 
of O on TiC($111$) are performed and the results compared to the case of 
atomic adsorption.  Also, the atomic and electronic structures of bulk TiC and of the 
clean TiC($111$) and TiC($001$) surfaces are studied in detail.  

In this way, the nature of chemisorption is understood through detailed analyses of 
the changes in surface electronic structure upon adsorption and a model, describing 
the nature of the adatom--substrate interactions, is presented.  
By correlating the trends in calculated adsorption energies $E_{\rm ads}$ with 
the trends in calculated electronic structures for the different adatoms, 
indications about how $E_{\rm ads}$ depends on the adatom--TiC($111$) interaction 
are thus obtained.  

The organization of this paper is as follows.  
In Sec.\ II, the DFT computational method used is presented, including details of 
accuracy and convergence tests.  Various densities of states (DOS) and other key tools 
for the analysis of our calculated electronic structures are defined.  
In Sec.\ III, the results from our calculations are presented and analyzed.  
Section III.A presents the results for bulk TiC, with detailed emphasis on the 
electronic structure.  Earlier published results are confirmed.  However, 
more details are given, in particular on the existence of direct C$p$--C$p$ bonding 
states in the lower part of the upper valence band (UVB) of TiC.  
Section III.B describes the results for the clean TiC($111$) and ($001$) surfaces, 
including: surface relaxations, surface energetics, and electronic structures.  
Particular emphasis is given to the characterization of the surface electronic structures, 
with detailed descriptions of how they change compared to the bulk.  
Section III.C gives the results for the atomic adsorption on TiC($111$) and 
($001$), first describing in detail the calculated energetics, atomic structures 
(Sec.\ III.C.1), and electronic structures (Sec.\ III.C.2), and thereafter analyzing 
the obtained electronic-structure information (Sec.\ III.C.3).  The systematics 
of the calculated electronic structures are used to formulate a 
``concerted-coupling'' model, describing the nature of the interaction 
between the adsorbate atoms and the TiC($111$) surface.  
Section III.D presents the results for full ($1 \times 1$) monolayer coverage of O 
on TiC($111$), including: energetics, atomic and electronic structures, and 
an analysis of the applicability of our concerted-coupling model to the case 
of full monolayer adsorption.  
In Sec.\ IV, we relate our electronic-structure results to the calculated trends in 
adsorption energies and discuss how our concerted-coupling model can be 
used in a qualitative way to understand the calculated trends in adsorption energy.  
In Sec.\ V, the main conclusions from the paper are briefly summarized.

%%%%%%%%%%%%%%%%%%%%%%%%%%%%%%%%%%%%%%%%%%%%%%%%%%%%%%%%%%%%
\section{Computational method}

The calculations are performed with the DFT plane-wave pseudopotential 
code {\tt dacapo 1.30}.\cite{dacapo}  The generalized-gradient approximation (GGA) 
of Perdew-Wang 1991 (PW91)\cite{PW91} is used for exchange and correlation.  
The atomic nuclei and core electrons are described with Vanderbilt 
ultrasoft pseudopotentials \cite{Vanderbilt} and the Brillouin zone (BZ) is 
sampled with the Monkhorst-Pack scheme.\cite{Monkhorst}  Slab geometry and 
periodic boundary conditions are employed and the potential 
discontinuity in the vacuum region thus arising from the 
lack of slab mirror symmetry is corrected for.\cite{Bengtsson}  
Full atomic relaxations within the given lattice parameters are performed 
using a BFGS quasi-Newton method, as implemented by 
Bengtsson.\cite{Bengtsson_PhD}  
On each considered system, the relaxation is performed until the sum of the 
remaining forces on all relaxed atoms in the unit cell is less than $0.05$ eV/\AA , 
which corresponds to remaining forces on each individual atom of less 
than $0.01$ eV/\AA .  

The method and the Ti, C, O, and Al pseudopotentials 
have previously been shown to yield successful results for alumina bulk and surface 
structures,\cite{Yourdshahyan,Ruberto} as well as for TiC/Co interface 
studies.\cite{Dudiy}  The pseudopotential for B has been succesfully employed 
to describe the binding in boron nitride.\cite{Rydberg}  
The accuracy of the other pseudopotentials used in our calculations is 
tested by calculating the atomization energies of 
H$_2$, N$_2$, F$_2$, P$_2$, S$_2$, and Cl$_2$ and the lattice constant 
and binding energy of bulk Si and comparing these results to the experimental 
values and to the results from similar calculations.  

The electronic structures of the considered systems are analyzed by 
studying different types of local densities of states (LDOS), obtained from the 
calculated Kohn-Sham (KS) wavefunctions $\Psi_n$ and energy eigenvalues $E_n$:  
(i) atom- and/or orbital-projected densities of states [LDOS($E$)] 
are obtained by projecting the KS wavefunctions onto the individual atomic 
orbitals and plotted as functions of the energy relative to 
the Fermi level, $E - E_F$;  
(ii) the total density of states in real space, around a specific 
energy value, [DOS(${\bf r}$, $E$)] is obtained by plotting 
$|\Psi_n({\bf r})|^2$, integrated over the whole BZ, 
for the value of the band index $n$ that corresponds to the KS 
energy $E_n$ close to $E$.  
Also, band-structure plots are obtained by calculating the 
eigenvalues at ${\bf k}$ points located along the BZ 
symmetry lines.  These plots are analyzed in detail by 
plotting $|\Psi_{n {\bf k}}({\bf r})|^2$ in three-dimensional 
real space for the ${\bf k}$ points located along the BZ
symmetry lines.  

In addition, a measure of the localization of charge around 
individual atoms is obtained from the calculated DFT charge density by 
using the ``atoms-in-molecule''-method of Bader.\cite{Bader}  
The algorithm of Ref.\ \onlinecite{Malcolm} is used.  In these 
calculations, the charge density is sampled on a grid with grid-point 
separation of approximately $0.07$ \AA .  
Also, the total-electron distribution in real space is analyzed by 
plotting the differences between the electronic densities calculated for 
the considered systems and those calculated for the systems composed of the 
corresponding free atoms.  

For the calculations on bulk TiC (Sec.\ III.A), a plane-wave cutoff of $350$ eV and a 
$6 \times 6 \times 6$ ${\bf k}$-point sampling are used.  This has previously been shown 
to yield a good accuracy for bulk TiC.\cite{Dudiy}  

For the calculations on the clean TiC($111$) surface (Sec.\ III.B), a $450$ eV cutoff 
and a $6 \times 6 \times 1$ ${\bf k}$-point sampling are used.  The TiC structure is 
composed of Ti and C layers that alternate in the [$111$] direction (see Sec.\ III.B).  
In this paper, we define a ``TiC bilayer'' as the unit composed of one Ti and one 
C ($111$) atomic layer.  Our calculations on the clean TiC($111$) surface are performed 
on slabs composed of up to 15 such TiC bilayers.  All considered slabs are 
stoichiometric, that is, they are composed of equal amounts of Ti and C atoms.  
Therefore, one side of the ($111$) slabs is terminated by a C layer, while the other 
side is terminated by a Ti layer.  Because of this, only the ``cleavage energy'' 
$E_{\rm cleav}$, that is, the sum of the C-terminated and Ti-terminated surface 
energies, can be calculated:  
\[
E_{\rm cleav} = E_{\rm cell}(n) - n E_{\rm b},
\]
where $E_{\rm cell}(n)$ is the total energy of a slab with $n$ TiC bilayers, as 
obtained from our DFT calculations, and 
\[
E_{\rm b} = E_{\rm cell} (n) - E_{\rm cell} (n-1)
\]
is the bulk energy of one TiC bilayer.\cite{Gay,Boettger_Esurf,Fiorentini}  
The cleavage energy corresponds to the energetical cost of cleaving the infinite 
TiC crystal into two semi-infinite crystals.  
Convergence of $E_{\rm b}$ is obtained for $n \geq 4$, disregarding quantum-size 
oscillations,\cite{Fiorentini,Boettger_QSE} which make the calculated cleavage energy 
values vary by at most $\pm 0.3$ J/m$^2$.  
The thickness of the vacuum region is $8.75$ \AA , corresponding to the thickness of four 
TiC bilayers.  Increasing the cutoff energy from $350$ eV to $450$ eV and/or the 
vacuum thickness from four to six TiC bilayers changes the calculated $E_{\rm cleav}$ 
value for the unrelaxed slab by less than $0.1$ J/m$^2$.  

\begin{figure}
\scalebox{1}{\includegraphics{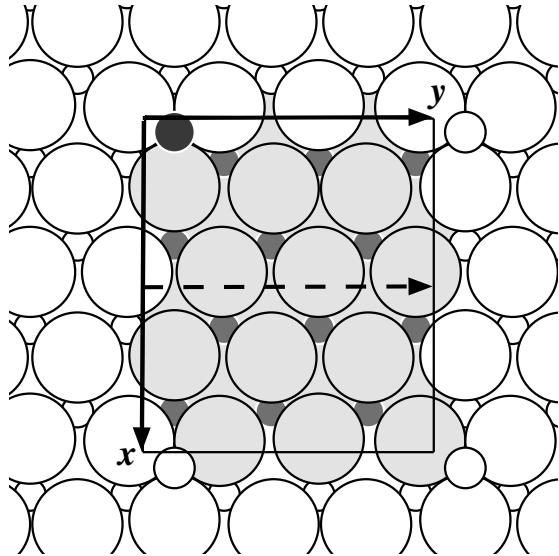}}
\caption{\label{fig:cell}Surface unit cell used for the adsorbate calculations on 
TiC($111$).  The ($4 \times 3$) unit cell used for the fcc adatoms is marked by the 
lines and the shaded atoms.  
Large lighter circles are Ti atoms, small darker circles are C atoms, 
and the small black circle is the adsorbed atom.  
The present picture shows the 
relaxed structure of an adsorbed O atom in the fcc site.  The ($2 \times 3$) 
unit cell, also used in the calculations (see text for details), is also shown 
(dashed arrow).}
\end{figure}

For the adatom calculations on TiC($111$) (Sec.\ III.C), a slab composed of four 
TiC bilayers is used.  In order to be able to compare our results with those from our 
study of the formation of thin films of $\alpha$- and $\kappa$-Al$_2$O$_3$ on 
TiC($111$),\cite{Ruberto_PhD} orthogonal surface lattice parameters, resembling 
the structure of the $\kappa$-Al$_2$O$_3$($001$)/($00{\overline 1}$) 
surface,\cite{Ruberto} are used in these 
calculations (see Fig.\ \ref{fig:cell}).  Each atomic Ti/C layer is thus composed of 
six atoms in a ($2 \times 3$) geometry.  The adatom and the top three TiC bilayers are 
allowed to relax in all directions.  A $4 \times 2 \times 1$ ${\bf k}$-point sampling, 
a $400$ eV cutoff energy, and a vacuum thickness of at least $9.6$ \AA\ are used.  
The adsorption energy of the adatom is calculated as
\[
E_{\rm ads} = E_{\rm slab+adatom} - E_{\rm slab} - E_{\rm atom},
\]
where $E_{\rm slab+adatom}$ is the calculated energy of the fully 
relaxed TiC slab with the adatom, $E_{\rm slab}$ is the 
energy of the relaxed clean TiC slab, and $E_{\rm atom}$ is the energy 
of the free adatom.  All values are calculated with the same supercell size, 
plane-wave cutoff energy, and ${\bf k}$-point sampling.  
The accuracy is tested by considering the system with strongest adsorption energy, 
an O adatom adsorbed in the fcc site, and repeating its relaxation with an 
$8 \times 4 \times 1$ {\bf k}-point sampling, a $700$ eV plane-wave cutoff, a $20.6$ \AA\ 
vacuum thickness, and a slab thickness of seven TiC bilayers, respectively.  None of 
these changes produces a change in calculated adsorption energy of more than $\pm 0.05$ 
eV/adatom.

%%%%%%%%% FIGURE %%%%%%%%%%%%%%%%%
%%%%%%%%% FIGURE %%%%%%%%%%%%%%%%%

In addition, in order to test any presence of adatom--adatom interactions, the 
relaxation of all adatoms in fcc site is repeated using supercells in which each atomic 
layer is composed of 12 atoms in a ($4 \times 3$) geometry (Fig.\ \ref{fig:cell}).  
A $2 \times 2 \times 1$ ${\bf k}$-point sampling is used.  Again, no change in 
adsorption energy larger than $0.05$ eV/adatom is obtained.  However, for the adatoms 
with large radii (atomic or ionic, depending on the degree of ionicity of the 
adsorbate--substrate bond, see Sec.\ III.C.2) (B, F, Al, Si, and Cl), 
slightly asymmetric adsorbate--substrate bond lengths are found after relaxation of 
the ($2 \times 3$) supercell.  Also, evidence for adatom--adatom interactions are 
found in the calculated DOS's for these adatoms.  Therefore, all results in 
this paper on the atomic adsorption in the fcc site on TiC($111$) are those 
obtained from the ($4 \times 3$) supercell.  

For the calculations on the clean and the O-adsorbed TiC($001$) surfaces 
(Secs.\ III.B and III.C), a slab composed of five atomic layers is used.  
Each TiC($001$) atomic layer consists of equal amounts of Ti and C atoms 
(see Sec.\ III.B).  
In order to allow sufficient distance between the adsorbed O adatoms, the surface cell 
dimensions are chosen such that each layer is composed of 16 atoms (eight Ti and eight C) 
in a ($4 \times 4$) geometry.  
A $450$ eV cutoff, a $4 \times 4 \times 1$ ${\bf k}$-point sampling, 
and a vacuum thickness of at least $10.9$ \AA\ are used.  Since the two surfaces of a 
TiC($001$) slab are equivalent, the calculated ($001$) cleavage energy corresponds 
to twice the ($001$) surface energy.  

Finally, the calculations on a full O($1 \times 1$) monolayer on TiC($111$) 
(Sec.\ III.D) are performed using both a slab with 15 TiC bilayers (and one atom per 
atomic layer) and a slab with 4 TiC bilayers [and six atoms per layer in a ($2 \times 3$) 
geometry].  The same parameters are used as for the above-described 15-bilayer and 
($2 \times 3$) 4-bilayer slabs, respectively.  
Between these two slab, the calculated interlayer relaxations of the O monolayer and of 
the top two TiC bilayers differ by less than $0.02$ \AA\ and the calculated 
adsorption energies differ by less than $0.1$ eV/adatom.

%%%%%%%%%%%%%%%%%%%%%%%%%%%%%%%%%%%%%%%%%%%%%%%%%%%%%%%%%%%%
\section{Results and analysis}

%-----------------------------------------------------------
\subsection{Bulk TiC}

%...........................................................
\subsubsection{Atomic Structure}

Bulk TiC adopts the NaCl structure.  
The lattice parameter obtained from our calculation ($4.332$ \AA ) 
is very close to the experimental value ($4.330$ \AA ).\cite{Dunand}  
Along the [$111$] direction, the structure is composed of alternating Ti and C 
layers, stacked in an $ABC$ sequence (see Sec.\ III.B), with calculated interlayer 
distance of $1.251$ \AA , Ti--C bond length of $2.166$ \AA , and Ti--Ti and C--C distances 
within each ($111$) layer of $3.063$ \AA .

%...........................................................
\subsubsection{Electronic Structure}

The nature of the bonding in bulk TiC and other TMC's and TMN's has been extensively 
studied over many years, both experimentally and 
theoretically.\cite{Dunand,Ihara,Johansson,Blaha83,Blaha85,Schwarz,Price,Eberhart}  
The common viewpoint that emerges from these studies is that the bonding in several 
TMC's and TMN's is dominated by strong, directional, covalent M--X bonds 
(M: metal atom, X: C or N), which account for the high hardness of these 
compounds.\cite{Jhi}  At the same time, a transfer of electrons from M to X 
takes place, implying a certain degree of ionic bonding.  
Further, the DOS does not vanish at the Fermi energy ($E_F$), giving the compounds 
a metallic character.  Thus, the bonding in TiC is mainly due to covalent Ti--C 
bonds, in which a certain degree of ionicity (electron transfer from Ti to C) 
can be detected, combined with a smaller amount of metallic Ti--Ti bonds.  
Most previous studies have excluded the presence of direct C--C interactions 
{\it a priori}.  However, more recently, a COOP (crystal orbital overlap populations) 
analysis has pointed out the presence of bonding and antibonding C--C interactions in 
the lower and upper parts of the TiC upper valence band (UVB), respectively.\cite{Matar}  

In the following, we present our results on the electronic structure of bulk TiC.  
The presence of iono-covalent Ti--C bonds and of metallic Ti--Ti bonds is 
confirmed and analyzed in detail.  In addition, our analysis clearly shows that 
bonding C--C states are present in the lower part of the UVB.  
Our subsequent analysis of the adsorption on the TiC($111$) surface (Sec.\ III.C.3) 
shows this to be important for the formulation of a model for 
adsorption on TiC.  Therefore, we pursue in this and the following subsection detailed 
analyses of both the previously characterized Ti--C and Ti--Ti states and the 
so far largely neglected C--C states.  

\begin{figure*}
\scalebox{2}{\includegraphics{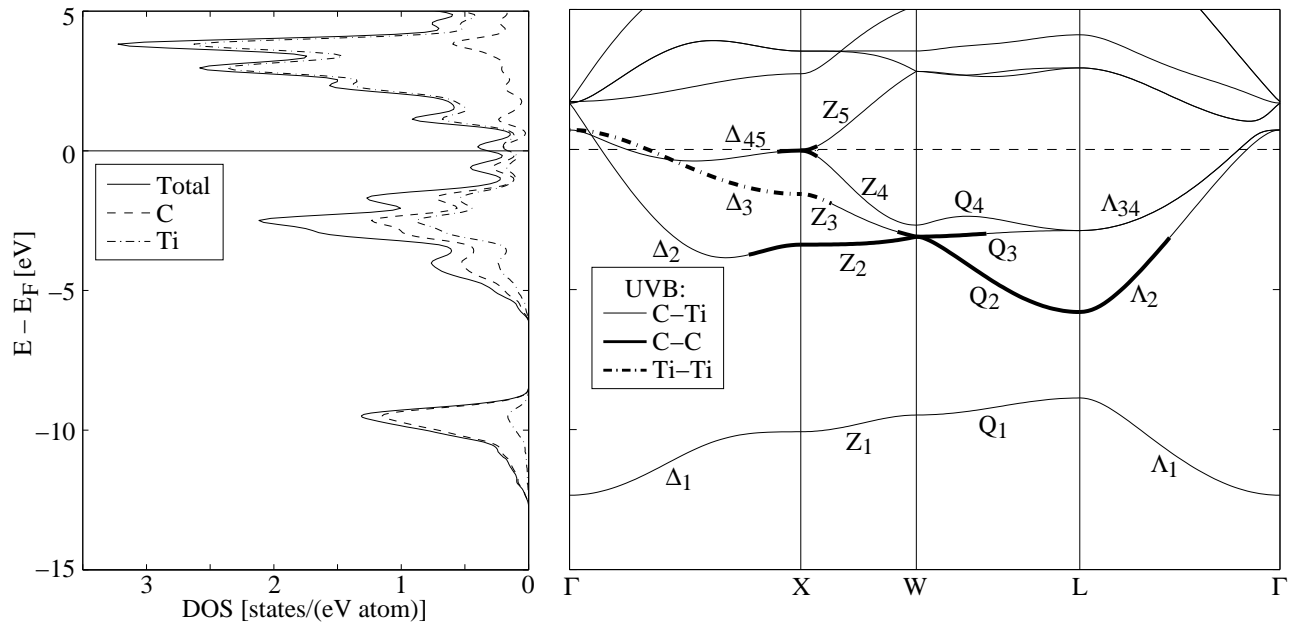}}
\caption{\label{fig:BAND_LDOS_bulk}Calculated band structure (right) and total and 
atom-projected DOS($E$) (left) for bulk TiC.  The high-symmetry points 
in the Brillouin zone are $\Gamma = (0,0,0)$, $X = \frac{2\pi}{a}(0,1,0)$, 
$W = \frac{2\pi}{a}(\frac{1}{2},1,0)$, and 
$L = \frac{2\pi}{a}(\frac{1}{2},\frac{1}{2},\frac{1}{2})$, 
where $a$ is the primitive-lattice parameter of bulk TiC.  
The band-structure plot shows also which parts of the upper valence 
band (UVB) correspond to C--Ti, C--C, and Ti--Ti bonding states, respectively, according 
to our state-resolved analysis of the DOS (described in the text and illustrated in 
Fig.\ \ref{fig:SPAC_bulk}).}
\end{figure*}

\begin{figure}
\scalebox{.45}{\includegraphics{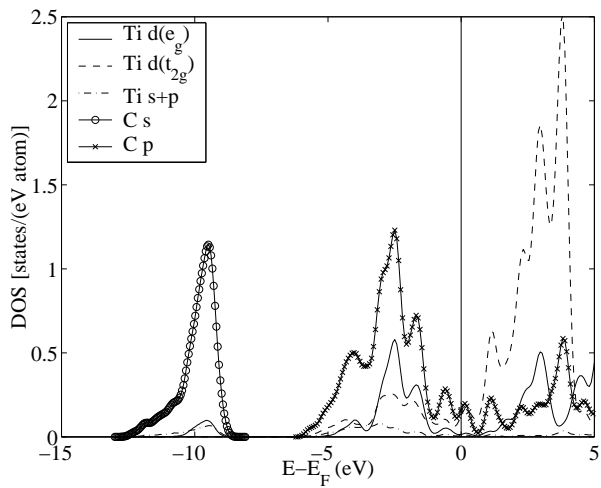}}
\caption{\label{fig:PDOS_bulk}Calculated DOS($E$) for bulk TiC, projected onto Ti $d$ 
valence orbitals with $e_g$ and $t_{2g}$ symmetry, respectively, onto Ti $s+p$ 
valence orbitals, and onto C $s$ and $p$ valence orbitals, respectively.}
\end{figure}

Our calculated band structure and DOS($E$) for bulk TiC are shown in Figs.\ 
\ref{fig:BAND_LDOS_bulk} and \ref{fig:PDOS_bulk}.  
The lower valence band (LVB), peaked at $9.5$ eV below $E_F$, 
is dominated by C $2s$ states, with only small contributions from Ti.  
The upper valence band (UVB) is mainly composed of C $2p$ and Ti $3d$ 
states and extends above $-6.1$ eV, with a major peak at $-2.5$ eV and smaller peaks 
at $-4.0$ and $-1.7$ eV (all energies here and in the following are given relative to 
$E_F$).  No C $sp$ hybridization is detected.  The main contribution to the UVB comes from 
the C $2p$ orbitals, indicating the partially ionic nature of the Ti--C 
bond.  However, the large overlap of Ti and C states, in particular around and 
above the main UVB peak, demonstrates a strong covalent character of the bond.  
The UVB and the conduction band (CB), on each side of $E_F$, form a continuum, with 
a wide, nonzero but deep, minimum in the DOS around $E_F$.  
The lowest value of the DOS is approximately $0.21$ 
states/(eV$\cdot$TiC-unit), in good agreement with previous theoretical 
and experimental results.\cite{Ihara,Price}  
This confirms the metallic character of the crystal.  
The CB has mainly Ti $3d$ character, with a smaller C $2p$ contribution, 
which again indicates the partially ionic character of the Ti--C bond.  
Still, an overlap of the Ti and C states can be seen, which is 
consistent with the view that the CB arises mainly from the antibonding 
states of the Ti--C covalent interaction.

%%%%%%%%% FIGURE %%%%%%%%%%%%%%%%%
%%%%%%%%% FIGURE %%%%%%%%%%%%%%%%%

%%%%%%%%% FIGURE %%%%%%%%%%%%%%%%%
%%%%%%%%% FIGURE %%%%%%%%%%%%%%%%%

Due to the octahedral symmetry around the Ti site in bulk TiC, the Ti $3d$ electrons 
redistribute into orbitals of $e_g$ ({\it i.e.}, $x^2 - y^2$ and 
$z^2$) and $t_{2g}$ ({\it i.e.}, $xy$, $yz$, and $zx$) 
symmetries.\cite{Dunand,Blaha83,Blaha85,Schwarz,Price}  
The $e_g$ orbitals point toward the nearest C-atom sites, while the $t_{2g}$ orbitals 
point toward the tetrahedral holes in the fcc Ti-atom network.\cite{Dunand,Schwarz}  
Our calculated LDOS, projected onto Ti $e_g$ and $t_{2g}$ orbitals 
(Fig.\ \ref{fig:PDOS_bulk}), shows a dominating $e_g$ character of the UVB.  
On the other hand, both the states around $E_F$ and in the CB are 
dominated by $t_{2g}$ orbitals.  

These results agree well with those obtained from the previous studies on the bonding 
nature of bulk 
TiC.\cite{Dunand,Ihara,Johansson,Blaha83,Blaha85,Schwarz,Price,Eberhart}  
According to them, the UVB is dominated by bonding Ti--C $\sigma$ states between 
Ti $3d(e_g)$ and C $2p$ orbitals ($pd\sigma$), with contributions from 
Ti--Ti $\sigma$ states between the Ti $3d(t_{2g})$ orbitals ($dd\sigma$) and 
from Ti--C $\pi$ states between Ti $3d(t_{2g})$ and C $2p$ orbitals 
($pd\pi$).  

However, our results show also that 
(i) the Ti DOS is very weak in the energetical region below the main UVB peak, 
(ii) no particular symmetry preference [$s$, $p$, $d(e_g)$, or $d(t_{2g})$] can be 
detected for the Ti states in this region, and 
(iii) C states are also present around $E_F$.  

\begin{figure*}
\scalebox{.7}{\includegraphics{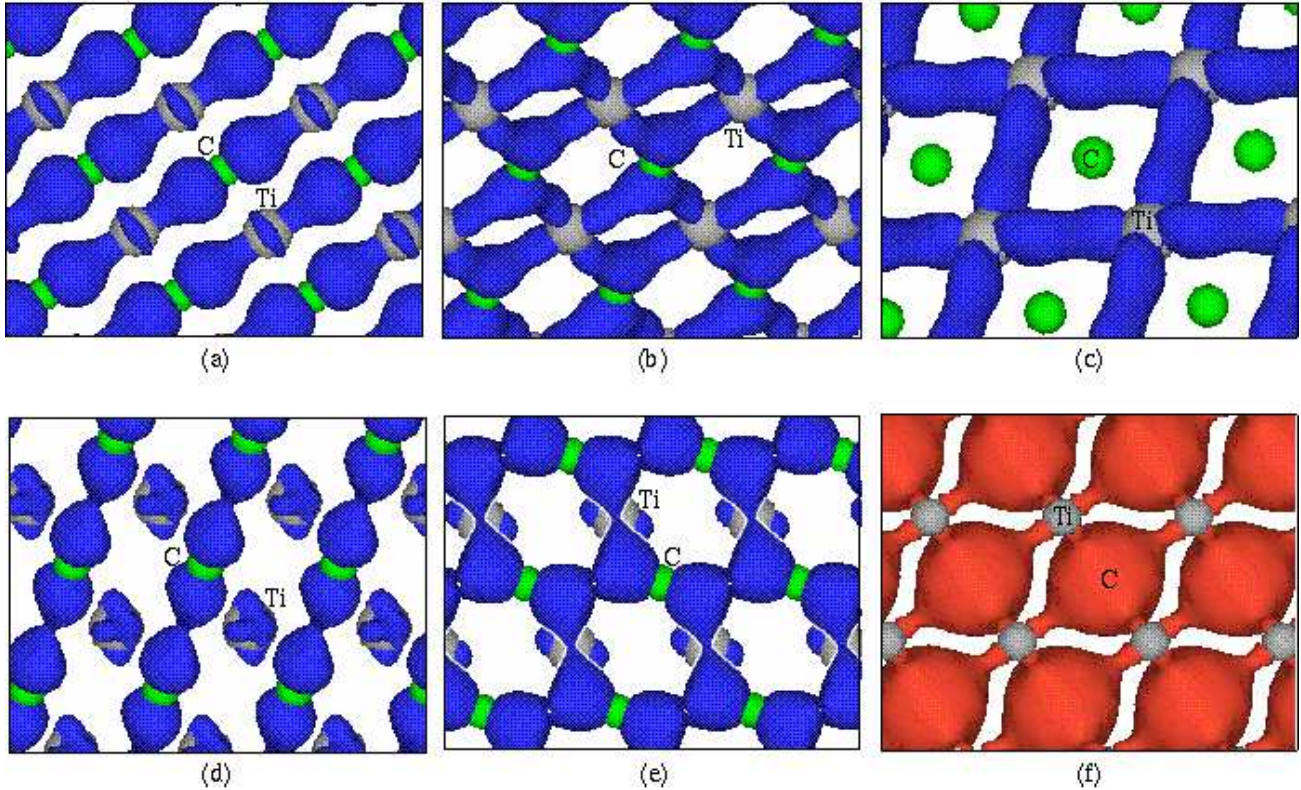}}
\caption{\label{fig:SPAC_bulk}(Color online).  
(a)--(e) Three-dimensional real-space contour plots of 
$|\Psi_{n{\bf k}}({\bf r})|^2$ in bulk TiC for the states $n{\bf k}$ at:  
(a) the point of lowest energy of the $\Delta_2$ band, showing Ti($e_g$)--C($p$) $\sigma$ 
bonds;   
(b) the $L_3$ point, showing Ti($t_{2g}$)--C($p$) $\pi$ bonds;   
(c) the $X_3$ point, showing Ti($t_{2g}$)--Ti($t_{2g}$) $\sigma$ bonds;   
(d) the $W_2$ point, showing C($p$)--C($p$) $\sigma$ bonds;   
(e) the $L_2$ point, showing C($p$)--C($p$) $\pi$ bonds.  
(f) Three-dimensional contour plot of the difference between the total electron density 
in bulk TiC and the electron density of free Ti and C atoms.  
All subfigures are viewed perpendicular to the $[111]$ direction, 
except (c), which shows a $\{001\}$ plane.  
Larger (grey/lighter) balls are Ti atoms, smaller (green/darker) balls are C atoms.  
In subfigure (f) the C atoms are hidden inside the large electron clouds.  
Note that in subfigure (e), the Ti electron clouds lie behind and are not connected to 
the C--C electron clouds.  
The contour plots correspond to density values of (a)--(e) $0.17$ electron states/\AA $^3$ 
and (f) $+0.02$ electron states/\AA $^3$.}
\end{figure*}

In order to further investigate these issues, we examine real-space, 
three-dimensional, plots 
of $|\Psi_{n {\bf k}}({\bf r})|^2$ (see Sec.\ II) for a number of states $n{\bf k}$ 
along all the valence bands of the TiC band structure.  
This analysis confirms the claims of previous studies, regarding the 
Ti--C and Ti--Ti bonds in bulk TiC.  
However, it also gives more detailed information on the types of bonding states present in 
the UVB and on where they are located in the band structure:  
(i) Ti--C bonds are mainly found along $\Delta_2$ ($pd\sigma$ bonds), 
$\Delta_{45}$ ($pd\pi$), $Z_4$ ($pd\sigma$), $Q_3$ and $Q_4$ ($pd\sigma$ and $pd\pi$), 
and $\Lambda_{34}$ ($pd\pi$) [Figs.\ \ref{fig:SPAC_bulk}(a) and (b) illustrate 
this by the $pd\sigma$ and $pd\pi$ bonds at the minimum energy point of $\Delta_2$ and 
at the $L_{34}$ point, respectively];  
(ii) Ti--Ti $dd\sigma$ bonds within the TiC\{$001$\} planes are 
found along the $\Delta_3$ band and around the $X_3$ point, as illustrated by 
the three-dimensional contour plot of $|\Psi_{n {\bf k}}({\bf r})|^2$ at the $X_3$ point 
[Fig.\ \ref{fig:SPAC_bulk}(c)].

%%%%%%%%% FIGURE %%%%%%%%%%%%%%%%%
%%%%%%%%% FIGURE %%%%%%%%%%%%%%%%%

In addition, our analysis shows the clear presence of C--C bonding states, 
mainly in the lower part of the UVB:  
the states around the $X_2$, $X_{45}$, $W_{23}$, and $L_2$ points, and along the $Z_2$, 
$Q_2$, and the lower part of the $\Lambda_2$ symmetry lines consist almost 
exclusively of C--C bonding states (Fig.\ \ref{fig:BAND_LDOS_bulk}).  
Of these states, only those along $Q_2$ 
and the lower part of $\Lambda_2$ contain some amount of Ti--C bonding states.  
These C--C bonding states are of both $pp\pi$ (at $X_2$, $X_{45}$, and along $Z_2$ and the 
lower part of $Q_2$ and of $\Lambda_2$) and $pp\sigma$ (at $W_{23}$ and along the higher 
part of $Q_2$) characters.  Figures \ref{fig:SPAC_bulk}(e) and (d) illustrate 
this, showing $|\Psi_{n {\bf k}}({\bf r})|^2$ at $L_2$ and at $W_{23}$, respectively.  
Thus, as already suggested by the low Ti LDOS in this region, the lower part of the TiC UVB 
is almost exclusively composed of bonding C--C states, with very small Ti--C 
contributions.  

Figure \ref{fig:SPAC_bulk}(f) shows the difference between the total electron density 
in bulk TiC and the electron density of free Ti and C atoms, that is, how the 
valence-electron distribution is changed by the formation of bonds in bulk 
TiC.  The plot confirms the partially ionic character of the bond, with a clear 
charge transfer from the Ti to the C atoms (this is also confirmed by an analysis of 
the regions of negative charge-density difference).  At the same time, it can be seen 
that the covalent interaction is dominated by the Ti--C $\sigma$ bonds.  

Thus, we can conclude that the main contribution to the cohesion of bulk TiC comes from 
the iono-covalent Ti--C $\sigma$ bonds (in agreement with previous studies), 
but that our state-resolved analysis of the electron density 
[Figs.\ \ref{fig:SPAC_bulk}(a)--(e)] shows that Ti--C $\pi$, Ti--Ti $\sigma$, 
C--C $\sigma$, and C--C $\pi$ bonding states also are present in the UVB of bulk TiC, 
in different energy regions (Fig.\ \ref{fig:BAND_LDOS_bulk}):  
C--C bonding states are present in the lower part of the 
UVB (between approximately $-5.8$ and $-3.1$ eV); 
Ti--C bonding states are present from $-3.9$ eV all the way up to $E_F$; Ti--Ti 
bonding states are present between approximately $-1.6$ and $+0.7$ eV.  
Some C--C bonding states are also found around the $X_{45}$ point, at $E_F$.  
As will be seen later, knowledge about the presence and energetical location 
of these different states is important for an understanding of the mechanism 
behind atomic adsorption on TiC.  

Finally, we estimate the ionic character of bulk TiC by calculating 
the charge around each atom with the ``atoms-in-molecule'' method of Bader 
(see Sec.\ II).  This yields a charge transfer of $1.5$ electrons from Ti to C, 
in excellent agreement with the results from Mulliken population analyses of 
other calculations.\cite{Tsukada,Fujimori}

%-----------------------------------------------------------
\subsection{Clean TiC(111) and TiC(001) surfaces}

\begin{figure*}
\scalebox{.47}{\includegraphics{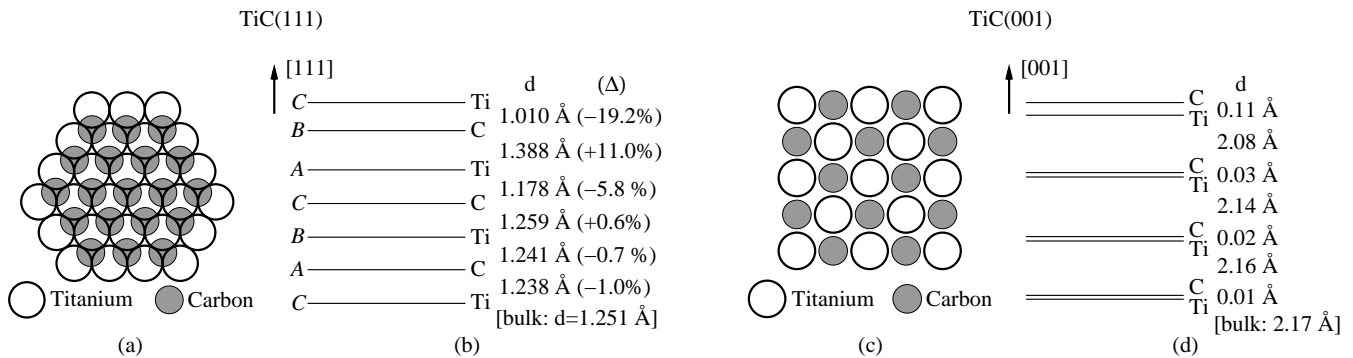}}
\caption{\label{fig:surf_struc}Surface structure of TiC($111$) and ($001$):  
(a) top view of the two atomic top layers of TiC($111$);  
(b) the top seven atomic surface layers of TiC($111$), seen perpendicularly 
to [$111$], showing the $ABC$ stacking of alternating Ti and C layers and 
the interlayer relaxations, in absolute values ($d$) and 
relative to the bulk distance ($\Delta$), as calculated with our 15-bilayer slab 
(see Sec.\ II);  
(c) top view of the surface atomic layer of TiC($001$);  
(d) the top four surface layers of TiC($001$), seen perpendicularly to [$001$], 
showing the rumpled relaxation of this surface, as obtained from our calculations.}
\end{figure*}

The TiC($111$) surface has been characterized, experimentally and theoretically, 
to be Ti($1 \times 1$) terminated 
[Figs.\ \ref{fig:surf_struc}(a-b)].\cite{Aono81,Oshima81_jlcm,Zaima,Tan}  
Due to the partial ionicity and NaCl structure of TiC, its ($111$) surface is 
polar, that is, it is energetically unstable due to the long-range electric 
field caused by the nonzero dipole moment, perpendicular to the surface, of the 
TiC bilayer.\cite{Tasker}  However, previous calculations have shown that 
this polarity is cancelled by a lowering of the ionic charge at the surface by 
approximately 50\%, due to the downward shift of the Ti antibonding CB LDOS at the 
surface.\cite{Tsukada,Fujimori}  
Experiments have indeed shown the presence of a Ti-centered surface resonance 
(TiSR) just below $E_F$.\cite{Oshima81_jlcm,Zaima,Bradshaw,Edamoto92_prb}\footnote{Note 
that this TiSR has been called surface state in the literature, but we 
prefer here to adopt the proper term surface resonance (SR), since its 
energy lies in the pseudogap between the bulk UVB and CB, that is, in a region 
of nonzero bulk DOS.}

%%%%%%%%% FIGURE %%%%%%%%%%%%%%%%%
%%%%%%%%% FIGURE %%%%%%%%%%%%%%%%%

In contrast, the TiC($001$) surface is nonpolar, being composed of equal 
amounts of positively charged Ti atoms and negatively charged C atoms [see Fig.\ 
\ref{fig:surf_struc}(c)].  Neither charge transfer nor surface resonance (SR) at 
$E_F$ have been reported for this surface.  

In the following, we present our results from DFT calculations on these two 
surfaces.  First, the relaxed atomic geometries and cleavage energies are presented 
and compared with previous results.  Then, we focus on the electronic structures of 
the two surfaces, in particular the ($111$) surface, describing in detail how they 
differ from the bulk electronic structure.  These results are in a general agreement 
with the few earlier 
studies.\cite{Tsukada,Fujimori,Aono81,Oshima81_jlcm,Zaima,Bradshaw,Edamoto92_prb}  
However, our analyses provide a much greater resolution of 
the surface electron-structure features and show a so far neglected wealth of 
different SR's, in particular on the ($111$) surface.  Knowledge about the presence 
and nature of these SR's is essential for the subsequent analysis of the electronic 
structures of the atomic adsorbates.

%...........................................................
\subsubsection{Geometry and Energetics}

For the ($111$) surface, our calculations on the 15-bilayer slab (see Sec.\ II) 
yield relaxations of the top six interlayer distances of 
$-19.2\%$, $+11.0\%$, $-5.8\%$, $+0.6\%$, $-0.7\%$, and $-1.0\%$ 
[Fig.\ \ref{fig:surf_struc}(b)].  
The large first-layer relaxation is in qualitative agreement 
with the one observed by Aono {\it et al.}\cite{Aono81} by impact-collision ion scattering 
spectroscopy (ICISS) ($-30\%$), 
and all values agree well with those from the tight-binding calculation of Tan 
{\it et al.}\cite{Tan}  The sum of the calculated top two interlayer 
distances ($2.40$ \AA ) agrees also very well with the value $2.3$ \AA\ 
from the STM investigation by Ahn {\it et al.}\cite{Ahn}  
The calculated relaxed bond length between the surface Ti and C atoms is 
$2.041$ \AA , which corresponds to a 6\% decrease from the bulk value of 
$2.166$ \AA .  

For the ($001$) surface, our calculations yield the rumpled relaxation described 
in the literature, with the C atoms relaxing outwards and the Ti atoms 
inwards.\cite{Price96,Tagawa}  This results in a perpendicular C--Ti distance of 
$0.11$ \AA\ in the surface layer, in excellent agreement with recent experimental 
results ($0.13 \pm 0.04$ \AA ).\cite{Tagawa}  Our relaxed perpendicular distances 
between the surface-layer Ti and subsurface-layer C ($2.08$ \AA ) and between 
C and Ti within the subsurface layer ($0.03$ \AA ) [see Fig.\ \ref{fig:surf_struc}(d)] 
are also in good agreement with the experimental results 
($2.11 \pm 0.04$ \AA\ and $0.01 \pm 0.04$ \AA , respectively).\cite{Tagawa}  

For the cleavage energies, {\it i.e.}, the energy required to cleave the infinite 
crystal along a chosen plane (as defined in Sec.\ II), we obtain
$E_{\rm cleav} = 11.74$ J/m$^2$ for the unrelaxed ($111$) surface and 
$E_{\rm cleav} = 11.43$ J/m$^2$ after relaxation of only the Ti-terminated side of the 
($111$) slab.  Since the ($111$) slab is stoichiometric, these values correspond to the 
sum of the surface energies of the Ti- and C-terminated TiC($111$) surfaces, respectively 
(see Sec.\ II).  However, the calculated energy gain of the relaxation, $0.31$ J/m$^2$, 
corresponds to the surface-energy gain of relaxing only the TiC($111$)-Ti surface.  
For the ($001$) surface, a value of $E_{\rm cleav} = 3.52$ J/m$^2$ is obtained after 
relaxation.  Since the ($001$) slab is mirror symmetric with respect to the ($001$) plane, 
$E_{\rm cleav}$ corresponds to twice the surface energy for this surface.  

These values of $E_{\rm cleav}$ agree well with the results from previous DFT calculations 
performed with the {\tt VASP} code:\cite{VASP} 
$E_{\rm cleav} = 3.46$ J/m$^2$ for the relaxed 
TiC($001$), $7.56$ J/m$^2$ for the relaxed TiC($011$), and $11.26$ J/m$^2$ for the 
relaxed TiC($111$).\cite{Dudiy04}  Thus, we see that the 
energetical order of the three low-index surfaces of TiC is ($001$) $<$ 
($110$) $<$ ($111$).  The lower stability of ($111$), with respect to ($001$), is 
expected, since the polar nature of the ($111$) surface should decrease its 
stability.\cite{Tasker}

%...........................................................
\subsubsection{Electronic Structure: TiC(111)}

\begin{figure}
\scalebox{.47}{\includegraphics{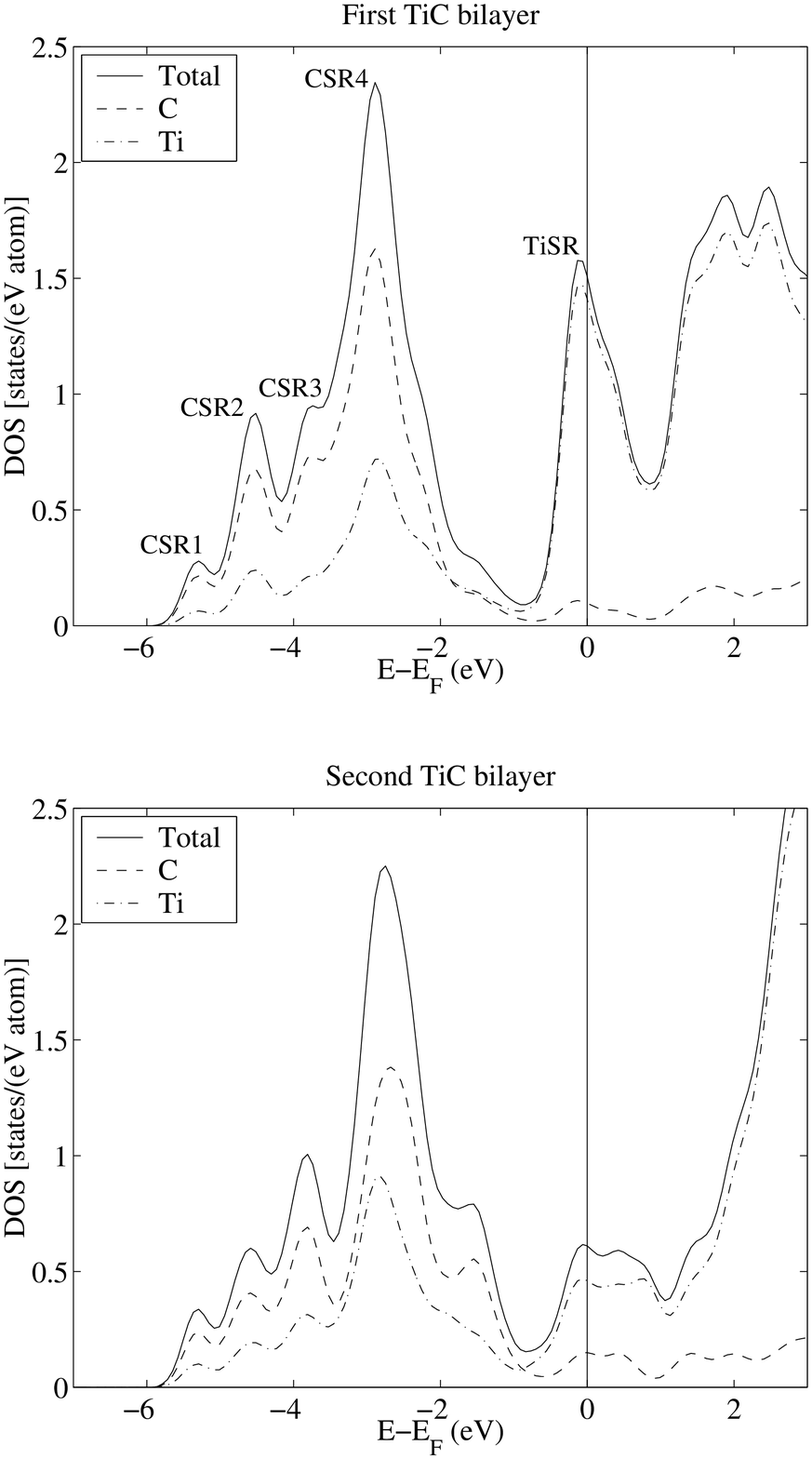}}
\caption{\label{fig:LDOS_111}Total and atom-projected DOS($E$) for the top two TiC 
bilayers of the clean TiC($111$) surface.  In the surface-bilayer DOS, the four UVB 
peaks and the peak at $E_F$ are identified as C-localized (CSR1--CSR4) and 
Ti-localized (TiSR) surface resonances, as described in the text.}
\end{figure}

\begin{figure*}
\scalebox{.47}{\includegraphics{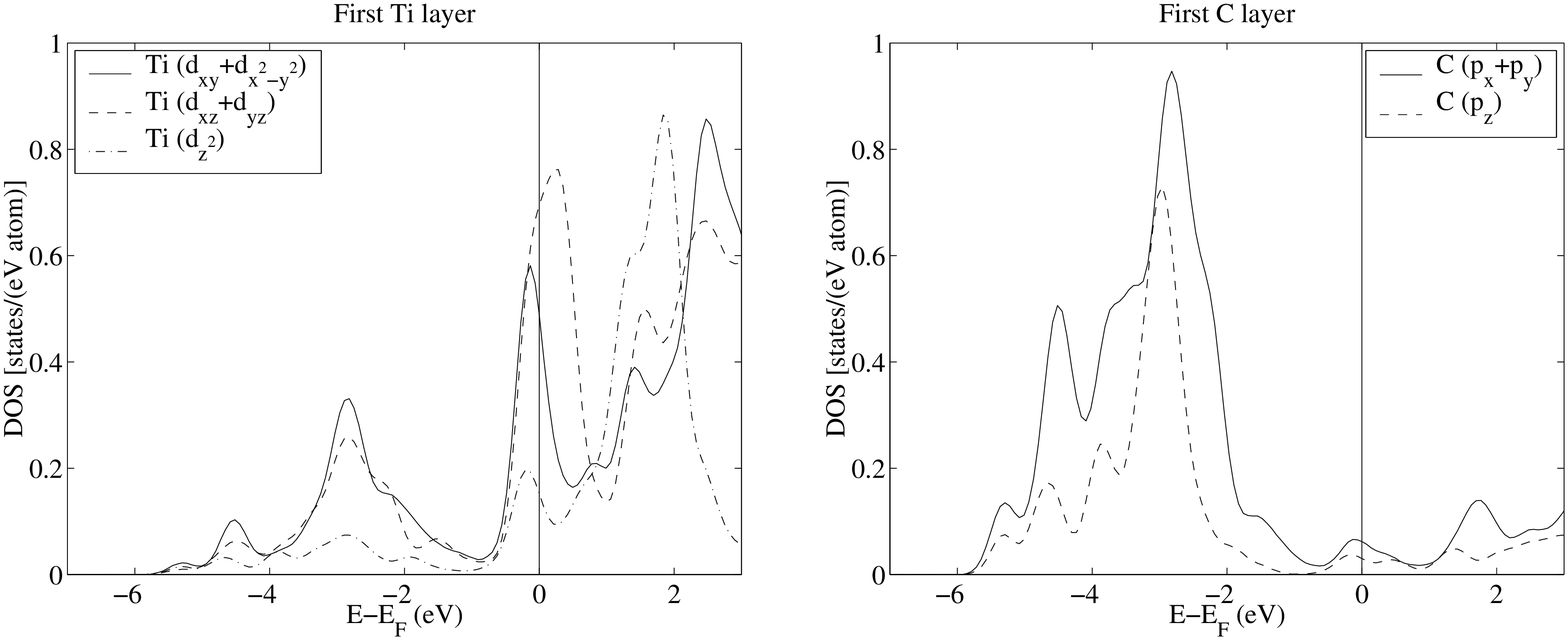}}
\caption{\label{fig:PDOS_111}Atom- and orbital-projected LDOS($E$)'s for the top TiC 
bilayer of the clean TiC($111$) surface.}
\end{figure*}

%.....
{\it (a) Fermi-level surface resonance.}  
The calculated DOS($E$) for the relaxed TiC($111$) surface is shown in 
Figs.\ \ref{fig:LDOS_111} and \ref{fig:PDOS_111}.  
The bulk DOS (Fig.\ \ref{fig:BAND_LDOS_bulk}) is recovered in the third 
TiC surface bilayer.  The top two bilayers are characterized by a strong 
SR, peaked right beneath $E_F$, of almost exclusively Ti $3d$ character (TiSR).  
The energetical position and Ti $d$ character of the TiSR are in good agreement with the 
results from previous experimental and theoretical 
studies.\cite{Zaima,Tsukada,Fujimori,Bradshaw,Edamoto92_prb}  
In addition, our orbital projection of the DOS (Fig.\ \ref{fig:PDOS_111}) shows that 
the TiSR can be decomposed into 
(i) a main peak at $-0.1$ eV that consists of mainly $d_{(xz,yz)}$ and 
$d_{(xy, x^2-y^2)}$ orbitals (the $z$ axis is here and in the following taken 
perpendicular to the surface) and 
(ii) a high-energy shoulder, above $E_F$, that consists almost exclusively of 
$d_{(xz,yz)}$ orbitals.  The highest density of $d_{(xz,yz)}$ orbitals is found at 
$+0.3$ eV.

%%%%%%%%% FIGURE %%%%%%%%%%%%%%%%%
%%%%%%%%% FIGURE %%%%%%%%%%%%%%%%%

%%%%%%%%% FIGURE %%%%%%%%%%%%%%%%%
%%%%%%%%% FIGURE %%%%%%%%%%%%%%%%%

\begin{figure*}
\scalebox{.88}{\includegraphics{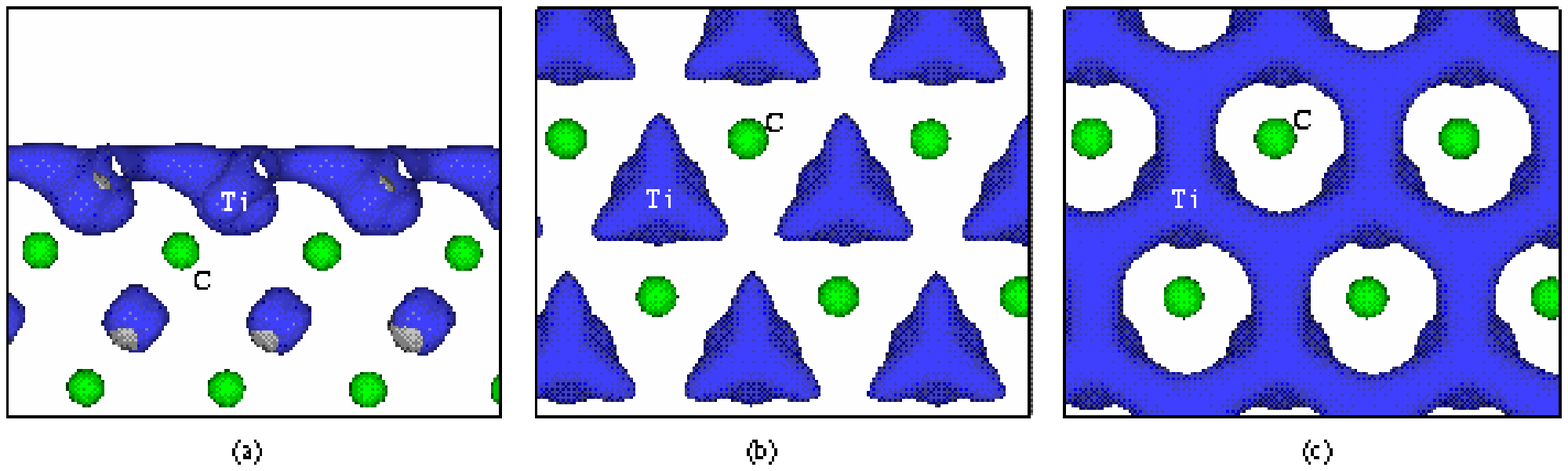}}
\caption{\label{fig:SPAC_SS}(Color online).  
Three-dimensional real-space contour plots of DOS(${\bf r}$, $E$), summed over all 
occupied electronic states above $-0.6$ eV, for the clean TiC($111$) surface, showing 
the Fermi-level surface resonance (TiSR).  The plot is viewed 
(a) perpendicularly to the surface and (b--c) from above the surface (showing only the top 
TiC bilayer).  Plot (b) shows a higher DOS value than plot (c).  
Green balls are C atoms.  The Ti atoms lie inside the electron clouds, as 
marked in the figures.}
\end{figure*}

The real-space shape of the TiSR (Fig.\ \ref{fig:SPAC_SS}) reveals 
this mixed $d_{(xz,yz)}+d_{(xy, x^2-y^2)}$ symmetry:  the electron distribution 
protrudes into the vacuum, while binding to neighboring surface Ti atoms.  
However, the TiSR electron density is not homogeneously distributed within the 
surface plane: the TiSR avoids the regions with C atoms directly underneath, that is, 
the hcp adsorption sites (see Sec.\ III.C.1).  Therefore, the surface Ti--Ti bonds 
are present only in the regions corresponding to the fcc adsorption sites.  
Also, similar plots show that, for lower DOS values, the TiSR density couples 
to states centered around Ti atoms lying deeper into the bulk, motivating 
our choice of calling these states SR's rather than surface states.

%%%%%%%%% FIGURE %%%%%%%%%%%%%%%%%
%%%%%%%%% FIGURE %%%%%%%%%%%%%%%%%

This spatial extent of the TiSR agrees with the one suggested by Aono 
{\it et al.}\cite{Aono81} and can be understood by considering the TiSR as 
containing dangling bonds that result from the breakage of the iono-covalent Ti--C bonds 
that cross the ($111$) cleavage plane.  
The breakage of these bonds causes the corresponding empty, antibonding, Ti-centered 
states of the bulk CB to collapse into more atomic-like orbitals, thus lowering their 
energy.  Therefore, the TiSR contains unsaturated Ti orbitals that point 
toward the sites that were previously occupied by the C atoms, that is, the surface 
fcc sites [Fig.\ \ref{fig:SPAC_SS}(b)].  At the same time, the TiSR can also be 
expected to contain broken Ti--Ti $dd\sigma$ bonds [Fig.\ \ref{fig:SPAC_bulk}(c)] 
that rearrange into new Ti--Ti hybridizations within the surface ($111$) plane.  

As noted above, previous calculations have led to the argument that the presence of a 
partially-filled TiSR causes an electron transfer to the surface Ti atoms and 
a resulting decrease in their ionicity of approximately $50\%$, compared to the 
bulk.\cite{Tsukada,Fujimori}  
Since TiC is partially ionic, the TiC($111$) surface is polar and should therefore 
be unstable, due to the presence of a macroscopic electric field caused by 
the nonzero perpendicular dipole moment of the TiC bilayer.\cite{Tasker}  
However, this field can be neutralized by inducing a surface charge that exactly 
cancels out the bulk macroscopic field.  For the ($111$) surface of a crystal with the 
NaCl structure, such as TiC($111$), this surface charge should be equal to $50\%$ of 
the bulk ionicity.\cite{Tsukada}  Based on the above-mentioned calculations, it has 
been argued that the extra electronic states needed to accomodate this charge at the 
Ti-terminated surface are provided by the TiSR.\cite{Tsukada,Fujimori}  

A Bader analysis (see Sec.\ II) of the atomic charges in our TiC($111$) slabs yields that 
(i) in the middle of the slabs, a charge transfer from 
Ti to C atoms of approximately $1.55$ electrons/atom takes place (in good agreement 
with our bulk results, see Sec.\ III.A.2); 
(ii) close to the Ti-terminated (C-terminated) surface of the slab, there is a total net 
charge of approximately $-0.8$ ($+0.8$) $|e|$/atom.  
Thus, approximately $0.8$ electrons/atom are transferred from the C-terminated side 
to the Ti-terminated side of the slab.  
This amount corresponds to $0.8 / 1.55 = 52\%$ of the ionicity in the middle of 
the slab, which is in excellent agreement with the previous results of a $50\%$ reduction 
in surface ionicity.  

However, a more detailed analysis of our Bader charges for the Ti-terminated surface 
shows that this surface charge is not located exclusively on the surface Ti atoms, 
but rather distributed among both Ti and C atoms of the top two TiC bilayers:  
the charge on the surface Ti atoms is $+1.1\, |e|$/atom, that is, only 
$\sim 0.45$ more electrons than on the bulk Ti atoms.  The remaining extra surface 
electrons are located on the C atoms of the first TiC bilayer ($0.25$ electrons/atom) and 
on the Ti and C atoms of the second TiC bilayer ($0.03$ electrons/atom on both Ti and C).  
Thus, the extra electrons do not only give rise to a TiSR but also an increased 
ionicity of the surface C atoms, compared to the bulk.  
This is characteristic of electron-structure changes also around the surface-bilayer C 
atoms, a point that is pursued further in the next subsection.  

\begin{figure}
\scalebox{.35}{\includegraphics{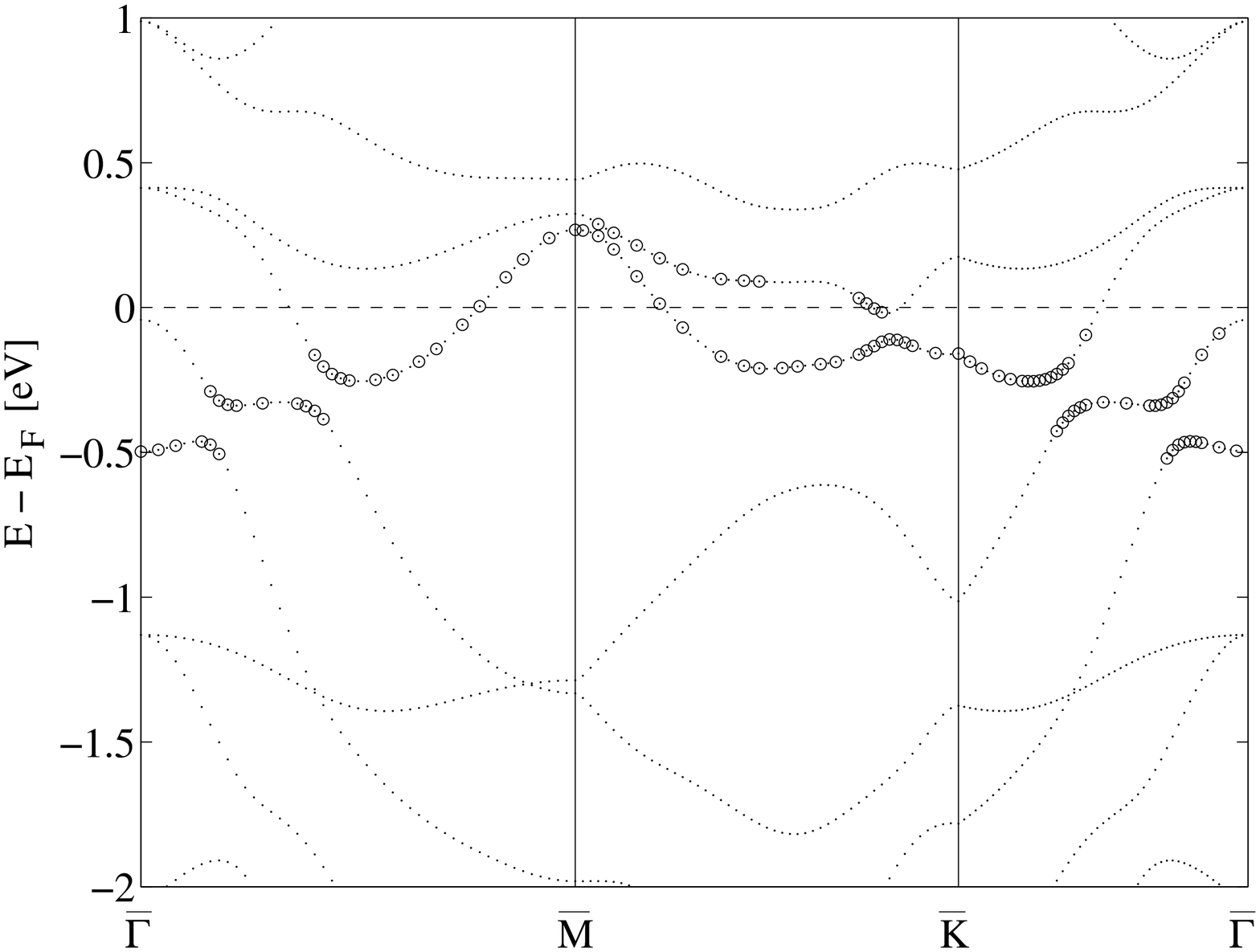}}
\caption{\label{fig:BAND_SS}Part of the calculated band structure around $E_F$ for 
a TiC(111) slab with four TiC bilayers.  Real-space plots of 
$|\Psi_{n{\bf k}}({\bf r})|^2$ for 
a number of the states belonging to the bands closest to $E_F$ have been inspected.  
Those states that show a strong surface localization in these plots are marked 
in the figure with circles.  The high-symmetry points in 
the two-dimensional Brillouin zone are $\overline{\Gamma} = (0,0)$, 
$\overline{M} = (\frac{1}{2},0)$, and 
$\overline{K} = (\frac{1}{3},\frac{1}{3})$, where the coordinates are given relative 
to the surface reciprocal unit cell.}
\end{figure}

Finally, it should be noted that the TiSR DOS peak is rather sharp, indicating a small 
dispersion and therefore a weak coupling between neighboring surface Ti atoms and 
with the bulk states at $E_F$.  
This is confirmed by our real-space plot of the TiSR (Fig.\ \ref{fig:SPAC_SS}), 
which shows that the electron density is very weak in the surface fcc sites, compared 
to the Ti sites, but also by an analysis of the band structure around $E_F$, 
as calculated for a slab with four TiC bilayers (Fig.\ \ref{fig:BAND_SS}).  
Analysis of contour plots of $|\Psi_{n{\bf k}}({\bf r})|^2$ for a number of 
${\bf k}$ points along this band structure allows an identification 
of the states that show a strong localization around the surface Ti atoms.  
These states, marked with circles in Fig.\ \ref{fig:BAND_SS}, show that the TiSR 
lies at $-0.5$ eV at $\overline{\Gamma}$ and disperses upwards in both the 
$\overline{\Gamma} \overline{M}$ and $\overline{\Gamma} \overline{K}$ 
directions, crossing $E_F$.  The dispersion is smaller than $1$ eV.  
Both the direction and the strength of the dispersion agree qualitatively with previous 
theoretical results for TiC($111$)\cite{Fujimori} and with experimental results for the 
largely similar NbC($111$) surface.\cite{Edamoto90}

%%%%%%%%% FIGURE %%%%%%%%%%%%%%%%%
%%%%%%%%% FIGURE %%%%%%%%%%%%%%%%%

%.....
{\it (b) UVB and LVB.}  
The UVB of the TiC($111$) surface bilayer extends between $-5.9$ and $-0.7$ eV and 
is largely similar to the bulk UVB.  However, a detailed analysis reveals a number 
of differences:  
(i) the position of the main peak lies at $-2.9$ eV, that is, $0.4$ eV lower than in the 
bulk UVB;  
(ii) the bulk-UVB high-energy peak (at $-1.7$ eV) is strongly quenched;  
(iii) the bulk-UVB low-energy peak (at $-4.0$ eV) is replaced by 
three peaks, at $-3.7$, $-4.5$, and $-5.3$ eV;  
(iv) the surface UVB DOS is more strongly localized around the C atoms than the 
bulk UVB DOS.  

Our analysis of the DOS(${\bf r}$, $E$) (see Sec.\ II) for all states within 
the surface UVB shows that the surface UVB can be divided into three regions, 
according to the bonding nature of its states: 
(i) a high-energy region, between $-1.7$ and $-0.7$ eV, consisting of C--Ti bonding 
states with higher electron concentration around the Ti atoms;  
(ii) a middle region, between $-3.3$ and $-1.8$ eV, consisting of C--Ti bonding 
states with higher electron concentration around the C atoms;  and 
(iii) a low-energy region, between $-5.9$ and $-3.3$ eV, consisting of 
primarily C--C bonding states, with some contribution from C--Ti interactions.  
The high-energy region corresponds to the quenched high-energy bulk peak, 
the middle region corresponds to the main peak, and the low-energy 
region corresponds to the three low-energy peaks.  

The differences between the bulk and surface UVB DOS's can be understood in a way 
analogous to our discussion on the origin of the TiSR.  
The high-energy region of the bulk UVB is composed of bonding Ti--C states, with 
higher localization on the C atoms.  Cleavage of the bulk structure to create the 
($111$) surface breaks these Ti--C bonds and removes the corresponding 
C atoms.  Therefore, these Ti--C bonding states collapse into more C-centered, 
atomic-like, states, which disappear as the corresponding C atoms are removed 
upon cleavage.  This shows up in our surface DOS as a quenching of the 
bulk-UVB peak at $-1.7$ eV.  

\begin{figure*}
\scalebox{.75}{\includegraphics{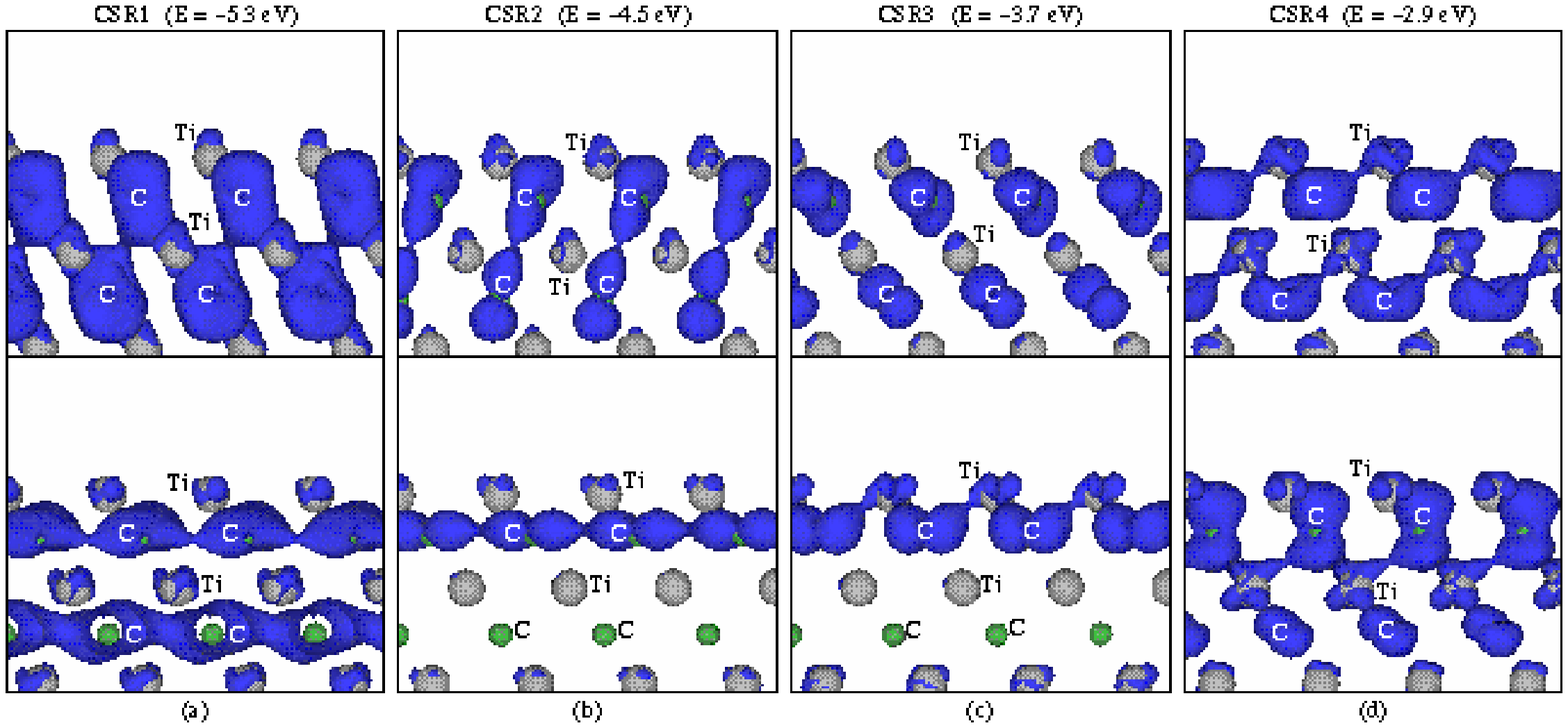}}
\caption{\label{fig:SPAC_UVB}(Color online).  
Three-dimensional real-space contour plots of DOS(${\bf r}$, $E$) for the clean 
TiC($111$) surface, at selected energies $E$ close to the energies of the four UVB 
peaks CSR1--CSR4 of the surfaces DOS($E$) (Fig.\ \ref{fig:LDOS_111}).  
For each UVB peak [(a)--(d)], two different electronic states, lying very closely in 
energy, are shown in order to illustrate the mixed $p_{xy} + p_z$ symmetry of each peak.  
The plots are viewed perpendicularly to the 
surface.  C (Ti) atoms are green/darker (grey/lighter) balls, or lie inside the electron 
clouds, as marked in the figures.  For different (same) CSR's, the plots correspond to 
different (same) values of DOS(${\bf r}$, $E$).}
\end{figure*}

Similarly, the formation of the surface causes the breakage of the C--C bonds 
present in the lower part of the bulk UVB (see Sec.\ III.A.2).  The DOS changes in 
the low-energy UVB region can therefore be interpreted as evidence for 
the formation of C-centered dangling bonds or SR's (CSR's).  Such an interpretation 
is corroborated by our calculated DOS(${\bf r}$, $E$) for the energetical regions 
around each of the UVB peaks in the surface DOS($E$) (illustrated by 
Fig.\ \ref{fig:SPAC_UVB}).  A higher localization and different orbital symmetry 
of these states around the surface-bilayer C atoms is clearly visible.  
Also, at low values of the DOS(${\bf r}$, $E$), these states connect to the 
second-bilayer C-atom states, giving these states the status of SR's.

%%%%%%%%% FIGURE %%%%%%%%%%%%%%%%%
%%%%%%%%% FIGURE %%%%%%%%%%%%%%%%%

Finally, the higher localization of the surface UVB DOS around the C atoms, compared to 
the bulk, is further evidence for the higher ionicity of the surface-bilayer C atoms, 
already discussed above in conjunction with our surface Bader analysis.  Our 
analysis above shows that this higher ionicity is caused by the presence of CSR's on 
these atoms.  

The LVB of the TiC($111$) surface is peaked at $-9.9$ eV and is almost exclusively
composed of C $2s$ orbitals, like the bulk LVB.  Again, we note a shift of $0.4$ eV
to lower energy, compared to the bulk LVB.

\begin{figure}
\scalebox{.75}{\includegraphics{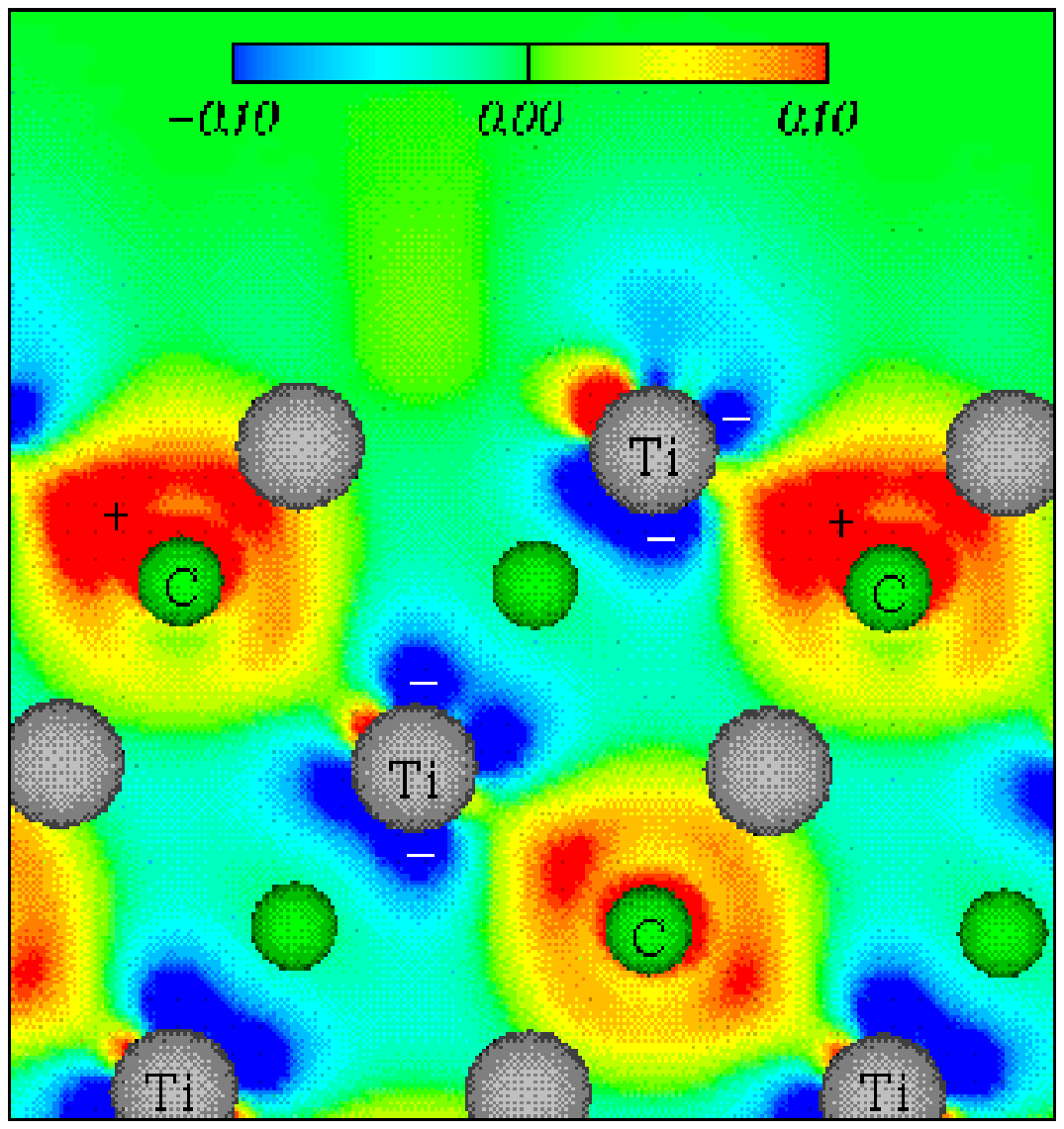}}
\caption{\label{fig:CHD_111}(Color online).  
Contour plot (through a plane containing the Ti and C atoms marked in the figure) 
of the difference between the total electron density at the TiC($111$) surface and 
the free-atomic densities.  Green/darker (grey/lighter) balls are C (Ti) atoms.  
Only the atoms lying in the drawn contour-plot plane are marked by names, 
the remaining atoms lying in front of the plane.  Red/dark (blue/black) regions 
correspond to positive (negative) electron-density differences, as 
exemplified in the figure by the $+$ and $-$ signs.}
\end{figure}

%.....
{\it (c) Total electron density.}  
The total effect of all these DOS changes on the surface electron distribution 
can be seen in our calculated total electron density (Fig.\ \ref{fig:CHD_111}).  
The SR's described above are clearly visible as increases in electron density, 
compared to the bulk, above both surface-bilayer Ti and C atoms.

%%%%%%%%% FIGURE %%%%%%%%%%%%%%%%%
%%%%%%%%% FIGURE %%%%%%%%%%%%%%%%%

%...........................................................
\subsubsection{Electronic Structure: TiC(001)}

\begin{figure}
\scalebox{.47}{\includegraphics{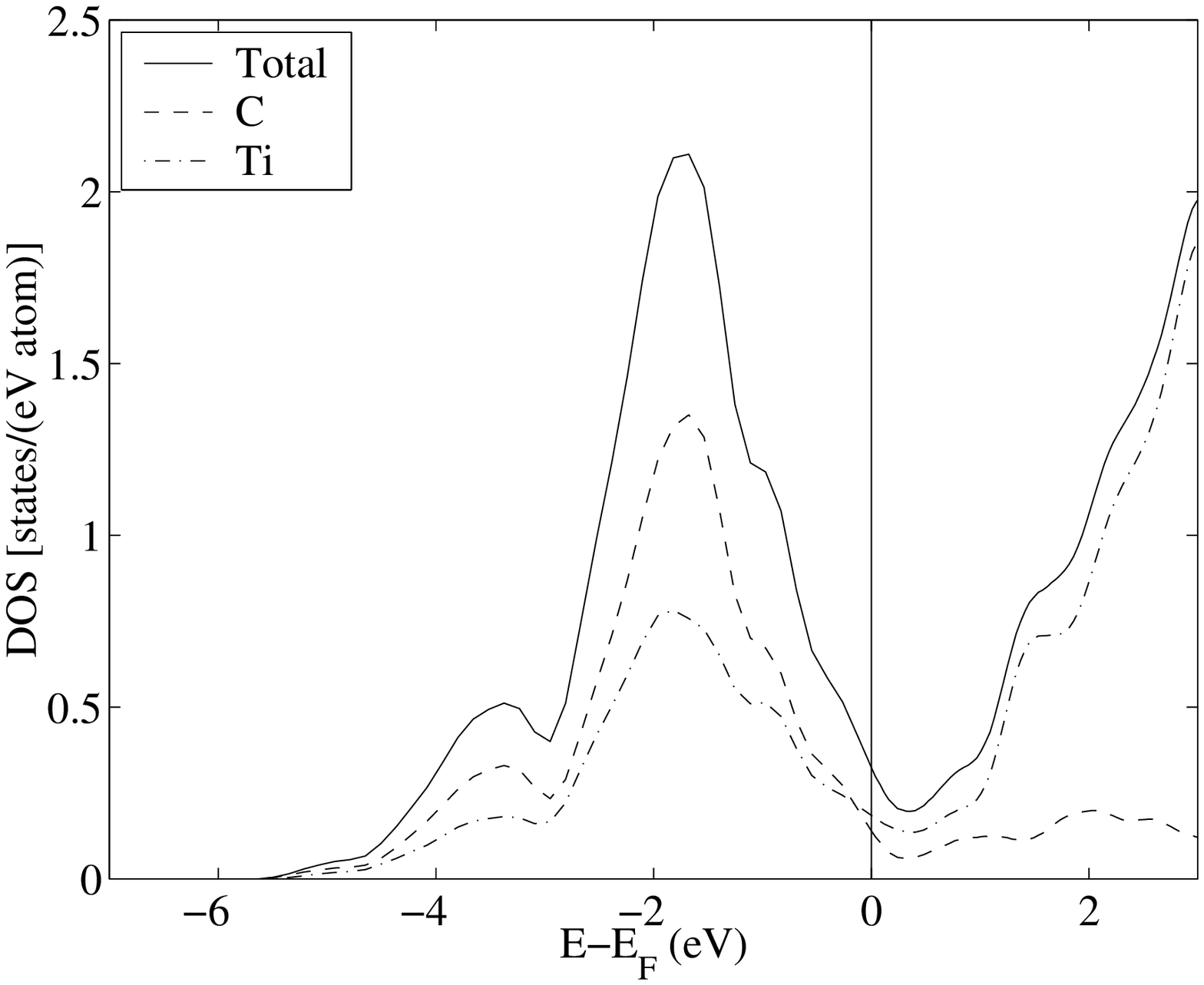}}
\caption{\label{fig:LDOS_001}Total and atom-projected DOS($E$) for the top atomic layer 
of the clean TiC($001$) surface.}
\end{figure}

For comparison, we have also calculated the DOS($E$) for the relaxed TiC($001$) surface 
(Fig.\ \ref{fig:LDOS_001}).  No clear SR is present at $E_F$ on this surface.  
The surface-layer DOS($E$) is very similar to the bulk DOS($E$), apart from a shift 
of the main LVB and UVB peaks by $0.8$ eV toward $E_F$.  In particular, the lower
part of the UVB is still characterized by only one peak, like the bulk UVB.  

Like on TiC($111$), however, changes occur 
in the high-energy part of the bulk UVB.  As discussed above, this part is 
characterized by the iono-covalent Ti--C bonds.  Like for TiC($111$), the creation 
of the ($001$) surface breaks these bonds and can therefore be expected to modify 
the electronic structure at these energies.  In particular, like on TiC($111$), 
the strong high-energy peak of the bulk UVB is quenched.  Also, orbital analysis 
shows energetical shifts of the Ti$d_{z^2}$ states lying in this region.  These are, 
in a similar way as for TiC($111$), indications of electron-structure changes due to 
the breakage of the bulk Ti--C bonds that cross the ($001$) cleavage plane.  

Thus, compared to the ($111$) surface, the nonpolar TiC($001$) surface shows 
no clear sign of CSR's in the lower part of the UVB nor of a TiSR around $E_F$.  On the 
other hand, SR's appear in the upper part of the UVB, due to the breakage of Ti--C 
bonds.

%%%%%%%%% FIGURE %%%%%%%%%%%%%%%%%
%%%%%%%%% FIGURE %%%%%%%%%%%%%%%%%

%-----------------------------------------------------------
\subsection{Atomic adsorption on TiC(111) and TiC(001)}

To understand the nature of chemisorption on TiC($111$), we perform a trend study, 
in which adsorption of the first-period element H, of the second-period elements 
B, C, N, O, F, and of the third-period elements Al, Si, P, S, Cl are considered.  
For comparison, the adsorption of O on TiC($001$) is also considered.  

For TiC($111$), all possible adsorption sites are considered: 
the two different three-foldly coordinated hollow sites, the two-foldly coordinated bridge 
site, and the top site.  The two hollow sites differ by their relation 
to the structure of the second atomic layer of the substrate: 
the site is either located directly above a second-layer C atom 
or it is not, corresponding to the hcp and the fcc sites, respectively.  
The only information present in the literature 
regarding the adsorption geometry on TiC($111$) concerns oxygen, which has been 
found to adsorb dissociatively in the fcc site,\cite{Souda88} and hydrogen, 
which has been found to adsorb in a three-fold hollow site, presumably the 
fcc site.\cite{Oshima83}  

For TiC($001$), two different sites are considered: on top of a C atom and on top of a 
Ti atom.  Previous studies have shown that oxygen adsorbs dissociatively on TiC($001$), 
and that the preferred adsorption site is on top of the C 
atom.\cite{Oshima81_ss,Souda91,Didzilius03,Zhang04}  

The structure of this subsection is as follows.  
First (Sec.\ III.C.1), the calculated energetics and relaxed geometries 
of all adsorbate systems are presented.  
Then (Sec.\ III.C.2), the calculated electronic structures of the systems with adatoms 
adsorbed in the stable fcc site of TiC($111$) are described and analyzed in detail.  
For comparison, the electronic structures of the O adatom adsorbed in the hcp and top 
sites on TiC($111$) and in the stable on-top-C site on TiC($001$) are also calculated and 
analyzed.  
Finally (Sec.\ III.C.3), a model for the adsorption mechanism on TiC($111$) is 
proposed, on the basis of these results.

%...........................................................
\subsubsection{Energetics and geometry}

For each adatom, total-relaxation calculations are performed, yielding the ground-state 
equilibrium geometry and adsorption energy $E_{\rm ads}$, corresponding to 
the energy gained by the formation of adatom--substrate bonds (see Sec.\ II).  
A higher value of $E_{\rm ads}$ implies a stronger adatom--substrate bond.  
Thus, analyses of the $E_{\rm ads}$ values give information about the preferred 
adsorption site for an adatom and a relative measure of the bond strengths of 
different adatoms on a surface.  

The calculated $E_{\rm ads}$ values for the adatoms in the different sites on TiC($111$) 
are listed in Table \ref{tab:Eads_111}.  For the fcc, hcp, and top sites, the 
values given are those obtained after full relaxation of the adsorbate and 
substrate atoms in all directions, with the exception of B, C, and N adatoms in top 
site, for which small forces ($\sim 0.05$ eV/\AA\ ) toward the neighboring fcc sites 
are still present after relaxation.  
No adatom binds in the bridge site, all of them relaxing to the neighboring fcc 
site.  Thus, the $E_{\rm ads}$ values given for the bridge site are those obtained after 
full relaxation of the adatoms in only the direction perpendicular to the substrate and 
of the substrate atoms in all directions.

%%%%%%%%% TABLE %%%%%%%%%%%%%%%%%%
\begin{table*}
\caption{\label{tab:Eads_111}Calculated atomic adsorption energies $E_{\rm ads}$ 
for the different adsorption sites on the TiC($111$) surface.  For all adatoms, 
the bridge site is unstable and relaxes to the neighboring fcc site.  
The values correspond to full relaxation of the adatom and surface-bilayer 
atomic coordinates, except for the bridge-site values, for which the adatom 
has only been relaxed in the direction perpendicular to the surface.  
Also given are the energy differences $\Delta E_{\rm ads}$ between the fcc/hcp and 
the bridge sites and between the fcc and the hcp sites.}
\begin{tabular}{cccccccccc}
\hline\hline
& & \multicolumn{4}{c}{$E_{\rm ads}$ (eV/atom)} & & 
\multicolumn{3}{c}{$\Delta E_{\rm ads}$ (eV/atom)} \\
\cline{3-6}
\cline{8-10}
atom & & fcc site & hcp site & top site & bridge site & & 
fcc--bridge & hcp--bridge & fcc--hcp \\
\cline{1-1}
\cline{3-6}
\cline{8-10}
H  & & $3.60$ & $3.34$ & $2.31$ & $3.27$ & & 
$0.33$ & $0.07$ & $0.26$ \\
\hline
B  & & $5.68$ & $5.55$ & $3.68$ & $4.92$ & & 
$0.76$ & $0.63$ & $0.13$ \\
C  & & $7.87$ & $7.15$ & $4.69$ & $6.73$ & & 
$1.14$ & $0.42$ & $0.72$ \\
N  & & $7.86$ & $6.87$ & $4.59$ & $6.77$ & & 
$1.09$ & $0.10$ & $0.99$ \\
O  & & $8.75$ & $7.93$ & $6.50$ & $7.95$ & & 
$0.80$ & $-0.02$ & $0.82$ \\
F  & & $6.92$ & $6.46$ & $6.01$ & $6.58$ & & 
$0.34$ & $-0.12$ & $0.46$ \\
\hline
Al & & $3.36$ & $3.31$ & $2.78$ & $3.24$ & & 
$0.12$ & $0.07$ & $0.05$ \\
Si & & $5.06$ & $4.89$ & $3.95$ & $4.74$ & & 
$0.32$ & $0.15$ & $0.17$ \\
P  & & $5.87$ & $5.61$ & $4.49$ & $5.00$ & & 
$0.87$ & $0.61$ & $0.26$ \\
S  & & $7.09$ & $6.78$ & $5.68$ & $6.75$ & & 
$0.34$ & $0.03$ & $0.31$ \\
Cl & & $5.48$ & $5.20$ & $4.51$ & $5.25$ & & 
$0.23$ & $-0.05$ & $0.28$ \\
\hline\hline
\end{tabular}
\end{table*}
%%%%%%%%% TABLE %%%%%%%%%%%%%%%%%%

All adatoms prefer to adsorb in the fcc site, followed by the metastable hcp site, 
where the adsorption energy in most cases is very close to that of the unstable 
bridge site.  The bond to the top site is significantly weaker than to the other sites 
and the top-site potential-energy surface (PES) has a rather shallow energy mininum, 
a small deviation from the optimum relaxed position being sufficient for destabilization 
and relaxation toward a neighboring fcc site.  
The overall energetical preference for the fcc site agrees with the experimental 
results for oxygen and hydrogen adsorption on TiC($111$).\cite{Souda88,Oshima83}  

The strongest fcc bond is obtained for oxygen, almost $9$ eV/atom, 
followed by C, N, S, and F, the energies for C and N being 
almost the same.  The weakest bonds are obtained for Al and H.  
The variation in $E_{\rm ads}$ values is very large, from $3.4$ eV/atom for Al to 
$8.8$ eV/atom for O.  

Also, the energy differences $\Delta E_{\rm ads}$ between the different adsorption sites 
(Table \ref{tab:Eads_111}) vary very strongly between the adatoms.  
In particular, the bridge site can be used as an approximation 
for the transition state between the fcc and hcp sites.  The energy barriers for 
diffusion between fcc and hcp sites can then be estimated from the differences 
between the $E_{\rm ads}$ values calculated for the fcc/hcp sites and the bridge site.  
The estimated barriers for diffusion from fcc to hcp site thus calculated vary 
from $0.12$ eV for Al to $1.14$ eV for C.  The smallest barriers are obtained for 
Al, Cl, Si, H, F, and S, in order of increasing barrier height.  

In addition, it is interesting to study the $E_{\rm ads}$ differences between 
fcc and hcp sites.  The smallest such differences are obtained for Al, B, and Si, 
in order of increasing difference.  On the other hand, the strongest preference for fcc 
site is obtained for N, O, and C.  

Thus, of all considered adatoms, Al and Si appear to possess the most planar 
PES's between the fcc and hcp sites, having \emph{both} small estimated diffusion barriers 
\emph{and} very similar $E_{\rm ads}$ values for the fcc and hcp sites.  
This indicates that the directionality of the adatom--substrate bond is weak for these 
two adatoms.  This is very interesting, considering the possibility that the high 
plasticity of MAX phases such as Ti$_3$SiC$_2$ and Ti$_3$AlC$_2$ could be caused by a 
good lateral mobility between the layers of Si/Al atoms and Ti$_6$C octahedra that make 
up their bulk structures.\cite{MAX}  

On the other hand, the C and N adatoms show a strong directionality of bonds 
to the substrate, having both high diffusion barriers and a high preference for the 
fcc site, with the C adatom having highest barriers.  This is consistent with the 
strong hardness and resistance against deformation of bulk TiC and TiN, TiC being 
the hardest one, described in the literature.\cite{Jhi}  

For O on TiC($001$) (Table \ref{tab:Eads_001}), our results confirm the preference for 
adsorption in the on-top C site that has been reported in the 
literature.\cite{Oshima81_ss,Souda91,Didzilius03,Zhang04}  
Also, our calculated energy difference between on-top C and on-top Ti 
site ($2.01$ eV/atom) compares well with the value from cluster DFT calculations 
($2.25$ eV/atom).\cite{Didzilius03}

%%%%%%%%% TABLE %%%%%%%%%%%%%%%%%%
\begin{table}
\caption{\label{tab:Eads_001}Calculated, fully relaxed, atomic adsorption energies 
$E_{\rm ads}$ for O in top-C and top-Ti sites on the TiC($001$) surface.}
\begin{tabular}{ccc}
\hline\hline
& \multicolumn{2}{c}{$E_{\rm ads}$ (eV/atom)} \\
\cline{2-3}
& top-C site & top-Ti site \\
\cline{2-3}
O/TiC($001$) & $4.96$ & $2.95$ \\
\hline\hline
\end{tabular}
\end{table}
%%%%%%%%% TABLE %%%%%%%%%%%%%%%%%%

\begin{figure}
\scalebox{.47}{\includegraphics{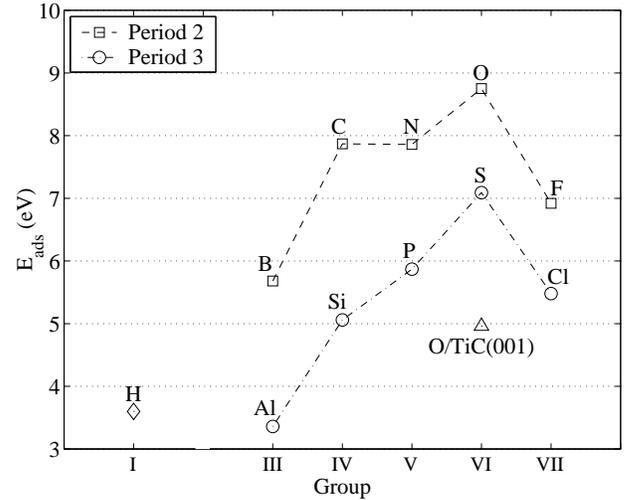}}
\caption{\label{fig:E_ads}Calculated, fully relaxed, atomic adsorption energies 
$E_{\rm ads}$ for the fcc site on the TiC(111) surface, showing the adsorption-strength 
trends.  Also shown is the adsorption energy for O in the most stable TiC($001$) site 
(on-top C).}
\end{figure}

For the TiC($111$) fcc site, the adsorption energies for second- and third-period elements 
show similar trends (Fig.\ \ref{fig:E_ads}), the only difference being the very similar 
values for C and N.  In both periods, the binding is strongest for the group-VI 
elements (O and S) and decreases monotonically (again, with the exception of C and N) when 
moving away from group VI.  Also, it decreases from second- to third-period elements.  
For O in the preferred site (on-top C) on TiC($001$), the adsorption is $43\%$ weaker 
than on TiC($111$) (Fig.\ \ref{fig:E_ads}).

%%%%%%%%% FIGURE %%%%%%%%%%%%%%%%%
%%%%%%%%% FIGURE %%%%%%%%%%%%%%%%%

The calculated relaxed atomic distances between the adsorbates in the fcc site and their 
nearest neighbors on the TiC(111) substrate (Table \ref{tab:ads_geom}) show similar trends 
within the two adatom periods:  
both the adsorbate--substrate distances and the substrate Ti--Ti distances are 
smallest for group-V and/or group-VI elements and increase monotonically when moving away 
from them.  This is again an indication of the particularly strong chemisorption of 
group-V and VI elements.  For most adatoms, the Ti--Ti distances are smaller than the 
clean-surface values, indicating a strong adatom--Ti attraction.  
Overall, the adatom--substrate distances are larger for period 3, due to the extra filled 
electron shell in the third-period adatoms.

%%%%%%%%% TABLE %%%%%%%%%%%%%%%%%%
\begin{table}
\caption{\label{tab:ads_geom}Calculated geometries of the relaxed adsorbate-TiC(111) 
systems for all considered adatoms in fcc site:  $d_{\rm ad-Ti}$ and $d_{\rm ad-C}$ are 
the distances between the adatoms and their nearest-neighbor Ti and C atoms, 
respectively;  $Z_{\rm ad-TiC}$ is the perpendicular distance between the 
adatoms and the TiC(111) surface; $d_{\rm Ti-Ti}$ and $d_{\rm C-C}$ are the 
Ti-Ti and C-C distances within the surface TiC(111) bilayer, close to the 
adatom; and $d_{\rm Ti-C}$ is the distance between the Ti and C atoms closest to 
the adatom.  The bond distances in the first TiC bilayer on the clean TiC($111$) surface 
are included for comparison.  All values are in \AA .}
\begin{tabular}{ccccccc}
\hline\hline
atom & $d_{\rm ad-Ti}$ & $d_{\rm ad-C}$ & $Z_{\rm ad-TiC}$ & 
$d_{\rm Ti-Ti}$ & $d_{\rm C-C}$ & $d_{\rm Ti-C}$ \\
\hline
H  & $2.01$ & $2.73$ & $1.03$ & 
$2.98$ & $3.08$ & $2.04$ \\
\hline
B  & $2.16$ & $3.01$ & $1.32$ & 
$2.95$ & $3.08$ & $2.06$ \\
C  & $1.99$ & $2.86$ & $1.10$ & 
$2.88$ & $3.09$ & $2.07$ \\
N  & $1.94$ & $2.80$ & $1.02$ & 
$2.85$ & $3.08$ & $2.06$ \\
O  & $1.98$ & $2.81$ & $1.07$ & 
$2.88$ & $3.08$ & $2.05$ \\
F  & $2.16$ & $2.93$ & $1.28$ & 
$3.00$ & $3.07$ & $2.04$ \\
\hline
Al & $2.67$ & $3.54$ & $2.02$ & 
$3.03$ & $3.07$ & $2.05$ \\
Si & $2.52$ & $3.39$ & $1.82$ & 
$3.00$ & $3.07$ & $2.05$ \\
P  & $2.44$ & $3.32$ & $1.72$ & 
$2.99$ & $3.07$ & $2.06$ \\
S  & $2.44$ & $3.30$ & $1.72$ & 
$3.00$ & $3.07$ & $2.05$ \\
Cl & $2.54$ & $3.35$ & $1.82$ & 
$3.06$ & $3.07$ & $2.04$ \\
\hline
Clean surface & --- & --- & --- & $3.06$ & $3.06$ & $2.04$ \\
\hline\hline
\end{tabular}
\end{table}
%%%%%%%%% TABLE %%%%%%%%%%%%%%%%%%

%...........................................................
\subsubsection{Electronic Structure}

In order to understand the nature of the chemisorption bond and trends, 
we analyze the electronic structure of 
all the considered adsorbates in the fcc site on TiC($111$), 
of the O adatom in hcp and top site on TiC($111$), and 
of the O adatom in the on-top-C site on TiC($001$).  

For each considered system, two different types of DOS's are calculated 
(see also Sec.\ II):  
(i) the difference in total, atom-, and orbital-projected DOS($E$) between 
the relaxed adsorbate+slab system and the corresponding slab with no adsorbate but with 
Ti and C atoms in the same positions [$\Delta$DOS($E$)];  
(ii) the distribution in real space of the contribution to the total DOS from 
specific energy values [DOS(${\bf r}$, $E$)].  
Also, the distribution of the total electron density around the studied adatoms, 
compared to the free-atom density, is shown.  A measure of the localization of 
charge around each atom is obtained with the Bader method (Sec.\ II).  

The $\Delta$DOS($E$) plots (also called adsorbate-induced DOS's)\cite{BIL_Tosi} 
show how the adsorption modifies the 
electronic structure in the surface region, for instance, by the appearance of 
chemical bonds between the adsorbate and the substrate.  
A negative $\Delta$DOS peak at a specific energy $E_0$ corresponds to a 
quenching of the electronic states having energy $E_0$, while 
a positive $\Delta$DOS peak at $E_1$ corresponds to the appearance of new states 
at $E_1$.  Thus, for instance, a negative peak at energy 
$E_0$, surrounded by two positive peaks at energies $E_1 < E_0$ and 
$E_2 > E_0$, can be interpreted as evidence for the hybridization of the states at 
$E_0$ with adatom states, resulting in the formation of bonding and 
antibonding states at $E_1$ and $E_2$, respectively.  

Projections of the $\Delta$DOS($E$) onto individual atoms and atomic orbitals 
in the surface-bilayer region give more detailed information on the nature of 
these bonds.  Such information is complemented by visualizations in three-dimensional 
real space of the DOS(${\bf r}$, $E$) for all electronic states within the 
TiC($111$) UVB and TiSR regions, thus providing vivid state-resolved illustrations 
of the bonds.  
Finally, the Bader and total charge-density analyses give information on the amount 
of charge transfer from substrate to adatom and on the charge polarization around each 
adatom, thus providing indications on the degree of covalency {\it vs.} ionicity of the 
adatom--substrate bonds.  

Starting with the H adatom, which demonstrates all the key features, this subsection
describes all our electron-structure results for the studied adatoms in deep detail 
and compares them with available experimental information.  
In the following subsection (Sec.\ III.C.3), this information is analyzed and used 
as basis for the formulation of a model describing the adsorbate--substrate interaction.

\begin{figure}
\scalebox{.47}{\includegraphics{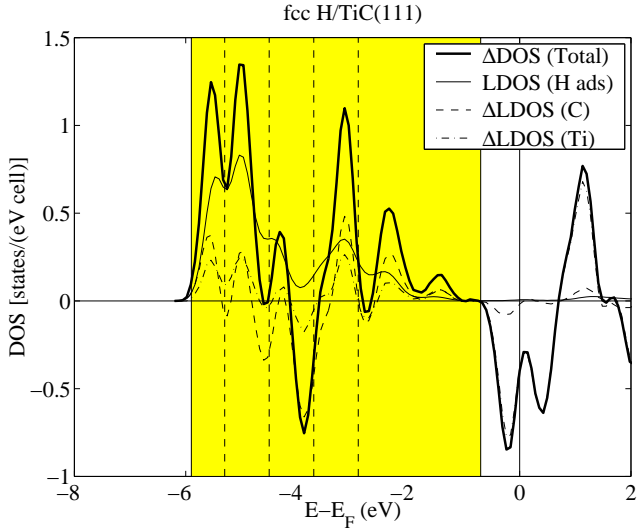}}
\caption{\label{fig:DeltaLDOS_Hads}Calculated density of states (DOS) for an H atom 
adsorbed in the fcc site on the TiC($111$) surface.  
The thin line shows the DOS projected on the H adatom.  The other three lines show the 
difference in DOS (``$\Delta$DOS'') between the relaxed slab \emph{with} adsorbed fcc H and 
a slab \emph{without} H adatom but with Ti and C coordinates fixed to the same values 
as those of the relaxed slab+adsorbate system.  The dashed (dot-dashed) line shows the 
$\Delta$DOS projected on all the C (Ti) atoms of the first TiC bilayer of the TiC($111$) 
surface.  The thick line shows the total $\Delta$DOS for the adsorbate+first-TiC-bilayer 
system.  The shaded area shows the energetical location of the UVB of the clean 
TiC($111$) surface.  The dashed vertical lines mark the energetical locations of 
the four UVB peaks of clean TiC($111$).}
\end{figure}

%.....
{\it (a) H adatom in fcc site on TiC(111).}  
The calculated $\Delta$DOS($E$) for an H adatom adsorbed in the fcc site on TiC($111$) 
(Fig.\ \ref{fig:DeltaLDOS_Hads}) is characterized by 
(i) two strong negative peaks around $E_F$, of almost exclusively Ti character 
(one peak, of mainly $d_{(xy, x^2-y^2)}$ symmetry, lies at $-0.2$ eV, and the other one, 
of almost exclusively $d_{(xz,yz)}$ symmetry, lies at $+0.4$ eV);  
(ii) a strong positive peak at $+1.1$ eV, of almost exclusively Ti $d$ character 
(with mixed $d$ symmetry);  
(iii) a strong negative peak at $-3.9$ eV, of mainly C character;  
(iv) a broad positive region between $-6.1$ and $-4.1$ eV 
(with two main peaks at $-5.5$ and $-5.0$ eV and a smaller one at $-4.3$ eV), 
composed mainly of H $s$ states, with minor contributions from the substrate C and Ti atoms 
for the two main peaks; and  
(v) a positive region between $-3.6$ and $-0.7$ eV (with peaks at $-3.1$, $-2.3$, and 
$-1.4$ eV), composed mainly of C-centered states.

%%%%%%%%% FIGURE %%%%%%%%%%%%%%%%%
%%%%%%%%% FIGURE %%%%%%%%%%%%%%%%%

\begin{figure*}
\scalebox{.85}{\includegraphics{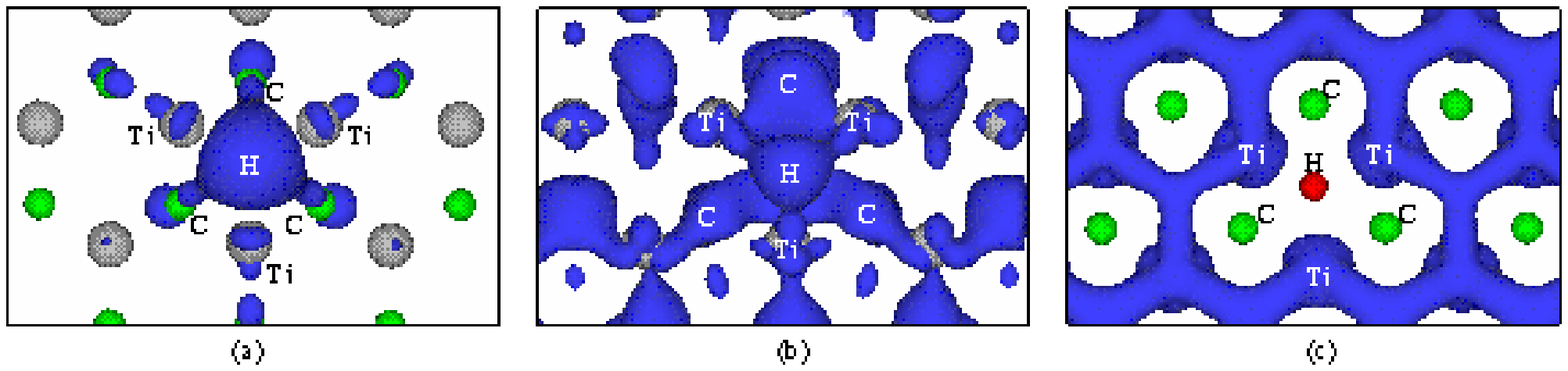}}
\caption{\label{fig:SPAC_Hads}(Color online).  
Three-dimensional contour plots of the calculated DOS(${\bf r}$, $E$), at different 
energies $E$, for an H atom adsorbed in the fcc site on the TiC($111$) surface.  
All three plots are viewed from above the 
surface, showing only the adatom and the top TiC bilayer, and correspond to 
(a) $E = -5.5$ eV, showing the strong H--C bonding character of the lowest $\Delta$DOS peak 
(Fig.\ \ref{fig:DeltaLDOS_Hads}); 
(b) $E = -3.1$ eV, showing the H--Ti bonding character of the $\Delta$DOS peak at 
$-3.1$ eV; 
and (c) the sum of all occupied states above $E = -0.3$ eV, showing the strong depletion 
of TiSR in the region around the H adatom.  
As marked in each plot, the smaller (green/darker) balls are C atoms, the larger 
(grey/lighter) balls are Ti atoms, and the red/darkest ball is the H adatom [in (a)--(b), 
the H adatom lies inside the electron cloud].  
In (b), the electron cloud around the H atom connects only to the neighboring Ti atoms, 
while the C-centered electron clouds connect only to the neighboring C and Ti atoms and 
lie below the H--Ti bonding clouds.}
\end{figure*}

The three-dimensional, real-space, plots of DOS(${\bf r}$, $E$) show that 
(i) the peaks at $-5.5$ and $-5.0$ eV correspond mainly to strong H--C bonding states 
[Fig.\ \ref{fig:SPAC_Hads}(a)], with weaker contributions from H--Ti bonding states;  
(ii) weaker H--Ti and H--C bonding states are present around the peak at $-4.3$ eV;  
(iii) bonding H--Ti states are present around the peaks at $-3.1$ 
[Fig.\ \ref{fig:SPAC_Hads}(b)] and $-2.3$ eV.  
In addition, the DOS(${\bf r}$, $E$) summed over all occupied states above $-0.3$ eV 
[Fig.\ \ref{fig:SPAC_Hads}(c)] clearly shows the depletion of TiSR around the adsorbed 
H atom.

%%%%%%%%% FIGURE %%%%%%%%%%%%%%%%%
%%%%%%%%% FIGURE %%%%%%%%%%%%%%%%%

Our prediction of strong H $s$-derived peaks at $-5.5$ and $-5.0$ eV can be compared 
with the experimental finding of an H $1s$ state between approximately $-5.5$ and 
$-6.5$ eV for full ($1 \times 1$) monolayer coverage of H on 
TiC($111$).\cite{Edamoto92_ss}  Also, our prediction of a sharp DOS decrease around 
$E_F$ compares well with published experimental results.\cite{Bradshaw}  

\begin{figure}
\scalebox{.75}{\includegraphics{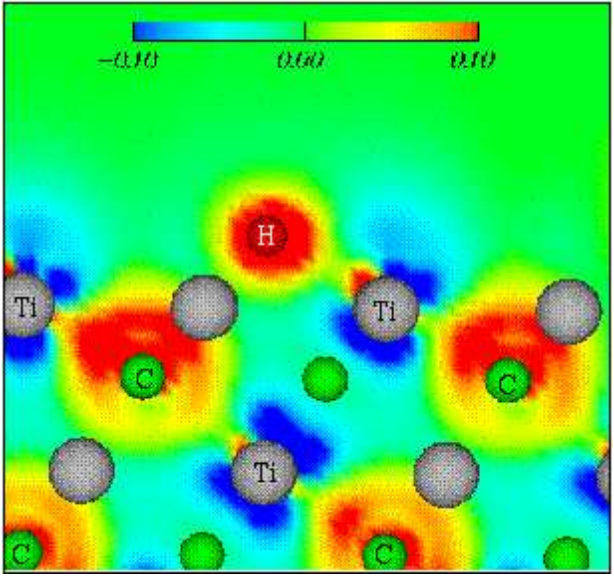}}
\caption{\label{fig:CHD_Hads}(Color online).  
Same as Fig.\ \ref{fig:CHD_111} but for the system with an H adatom adsorbed 
in fcc site on the TiC($111$) surface.  The (darkest) red ball is the adatom.}
\end{figure}

The total electron density around the H adatom (Fig.\ \ref{fig:CHD_Hads}) shows 
an excess of electrons (compared to a free H atom) around the adatom.  
Our Bader analysis of the charge localization yields an ``ionicity'' ({\it i.e.}, charge 
difference compared to the neutral atom) of $-0.64 |e|$ (where $|e|$ is the 
electronic charge) around the H adatom and a decrease of charge, compared to the 
clean TiC($111$) surface, around its nearest-neighbor (NN) Ti and C atoms 
(Table \ref{tab:Bader_ads}).  
Thus, electrons have transferred from both the Ti and C atoms in the 
surface bilayer to the adatom.  The extra electrons around the adatom 
show an almost spherical distribution, typical of the $s$ orbital, 
although a polarization toward the NN Ti atom can be seen.

%%%%%%%%% FIGURE %%%%%%%%%%%%%%%%%
%%%%%%%%% FIGURE %%%%%%%%%%%%%%%%%

%%%%%%%%% TABLE %%%%%%%%%%%%%%%%%%
\begin{table}
\caption{\label{tab:Bader_ads}Charge localization around all the considered adatoms and 
their nearest-neighbor (NN) Ti and C atoms, as calculated from our Bader analyses.  
The values given in the table are the ``ionicities'' of the different atoms, 
that is, the atomic charge values compared to the neutral atoms, given in units of 
electron charge $|e|$.  For comparison, the corresponding values for the Ti and 
C atoms of the surface bilayer of the clean TiC($111$) surface are also given.}
\begin{tabular}{cccc}
\hline\hline
adatom & \multicolumn{3}{c}{ionicity} \\
\cline{2-4}
& adatom & NN Ti & NN C \\
\hline
H  & $-0.64$ & $+1.24$ & $-1.75$ \\
\hline
B  & $-1.09$ & $+1.20$ & $-1.65$ \\
C  & $-1.35$ & $+1.32$ & $-1.65$ \\
N  & $-1.35$ & $+1.36$ & $-1.68$ \\
O  & $-1.17$ & $+1.34$ & $-1.68$ \\
F  & $-0.80$ & $+1.28$ & $-1.72$ \\
\hline
Al & $-0.47$ & $+1.08$ & $-1.71$ \\
Si & $-0.93$ & $+1.19$ & $-1.69$ \\
P  & $-1.12$ & $+1.25$ & $-1.68$ \\
S  & $-1.06$ & $+1.29$ & $-1.70$ \\
Cl & $-0.75$ & $+1.26$ & $-1.73$ \\
\hline
Clean surface & --- & $+1.09$ & $-1.79$ \\
\hline\hline
\end{tabular}
\end{table}
%%%%%%%%% TABLE %%%%%%%%%%%%%%%%%%

\begin{figure*}
\scalebox{.95}{\includegraphics{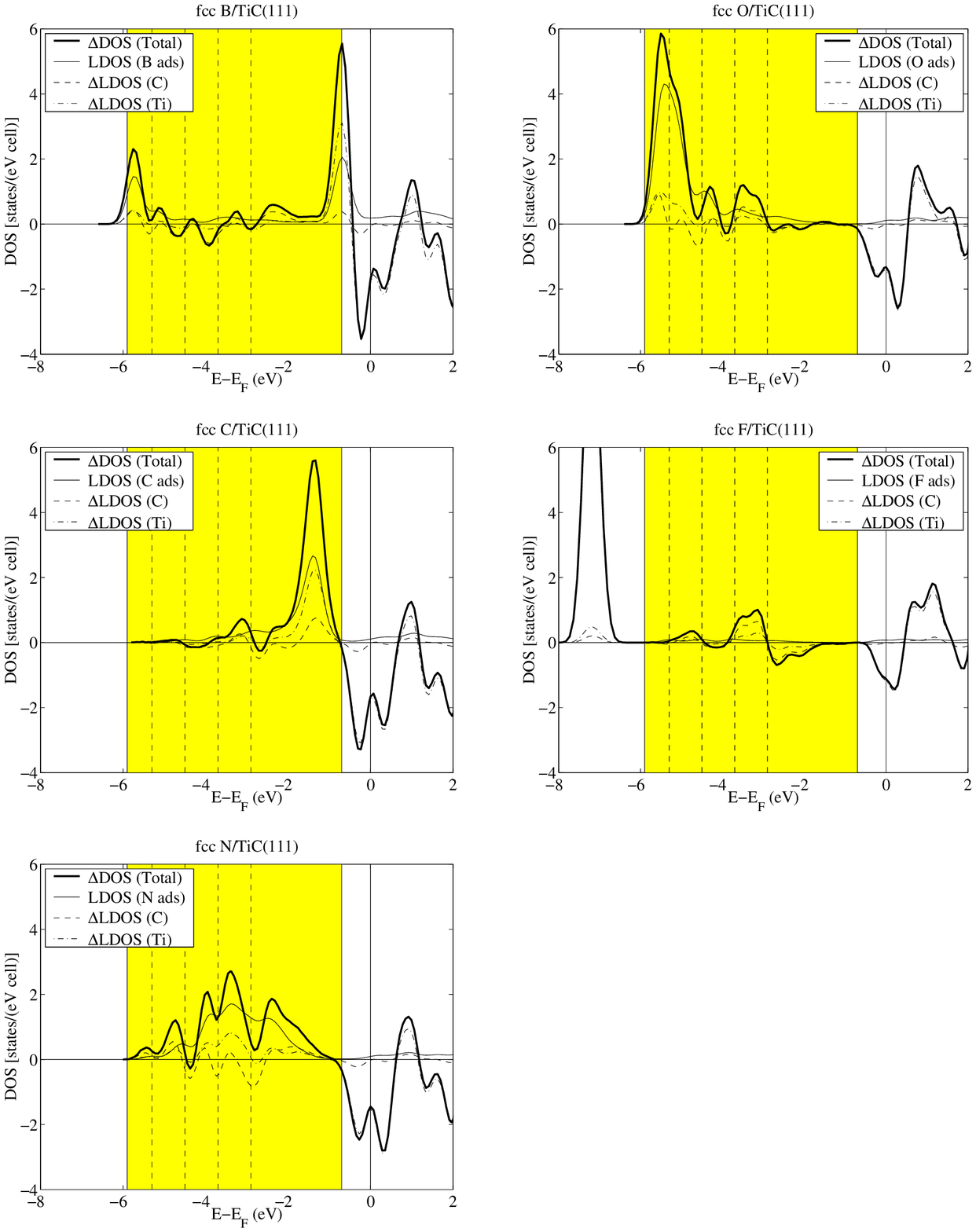}}
\caption{\label{fig:DeltaLDOS_PER2}Same as Fig.\ \ref{fig:DeltaLDOS_Hads} but for the 
second-period elements B, C, N, O, and F.}
\end{figure*}

\begin{figure*}
\scalebox{.7}{\includegraphics{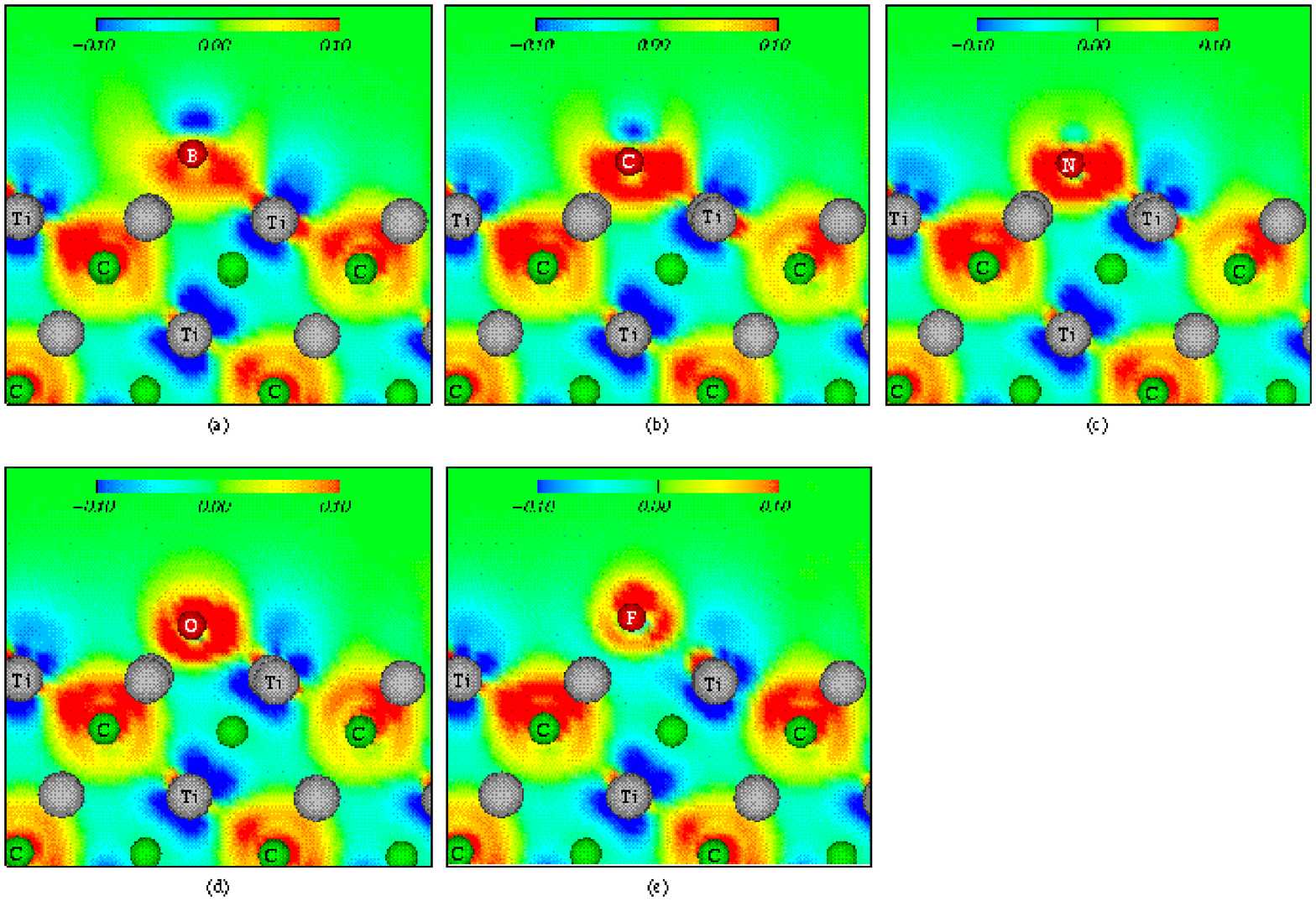}}
\caption{\label{fig:CHD_PER2}(Color online).  
Same as Fig.\ \ref{fig:CHD_Hads} but for the systems with the second-period adatoms 
(a) B, (b) C, (c) N, (d) O, and (e) F adsorbed in fcc site on the TiC($111$) surface.}
\end{figure*}

%.....
{\it (b) Second-period adatoms (B, C, N, O, and F) in fcc site on TiC(111).}  
The calculated $\Delta$DOS($E$) and total electron densities for the second-period 
adatoms adsorbed in the fcc site on TiC($111$) are shown in Figs.\ 
\ref{fig:DeltaLDOS_PER2} and \ref{fig:CHD_PER2}.  Also, the calculated ionicities 
({\it i.e.}, charge differences compared to the neutral atoms) for the adatoms and their 
Ti and C NN's, obtained from Bader analyses, are given in Table \ref{tab:Bader_ads}.

%%%%%%%%% FIGURE %%%%%%%%%%%%%%%%%
%%%%%%%%% FIGURE %%%%%%%%%%%%%%%%%

%%%%%%%%% FIGURE %%%%%%%%%%%%%%%%%
%%%%%%%%% FIGURE %%%%%%%%%%%%%%%%%

\underline{B adatom}:  
For B, the $\Delta$DOS($E$) (Fig.\ \ref{fig:DeltaLDOS_PER2}) is characterized by  
(i) a strong negative double-peak structure at $-0.2$ and $+0.3$ eV, of almost exclusively 
Ti character (with mainly mixed $d_{(xy, x^2-y^2)}$ and $d_{(xz,yz)}$ symmetry for the 
peak at $-0.2$ eV, and almost exclusively $d_{(xz,yz)}$ symmetry for the peak at 
$+0.3$ eV);  
(ii) a positive peak, of mainly Ti$d_{(xz,yz)}$ character, at $+1.0$ eV;  
(iii) much smaller negative peaks, of mainly C character, at $-4.7$, $-3.9$, and 
$-2.9$ eV;  
(iv) a main positive peak at $-0.7$ eV, mainly composed of Ti and B $p$ states;  
(v) a smaller positive peak at $-5.8$ eV, of mainly B $s$ character;  
(vi) much smaller positive peaks at $-5.1$ eV (of mainly B $s$ character) 
and at $-3.2$ eV (of mainly C and B $s$ character);  
(vii) a positive region between $-2.7$ eV and the main 
positive peak, including a peak at $-2.4$ eV, of mainly C character.  

The DOS(${\bf r}$, $E$) plots show that 
(i) the peak at $-0.7$ eV consists of very strong B--Ti bonding states; 
(ii) the peak at $-5.8$ eV consists of very strong B--C bonding states and of weaker 
B--Ti bonding states;  
(iii) the peak at $-5.1$ eV consists of weak B--Ti and B--C bonding states;  
(iv) only B--Ti bonding states are detected around $-3.2$ eV;  
(v) the only type of bonding states detected between $-2.7$ eV and the main 
positive peak are weak B--Ti states; 
(vi) similarly to the case of H [Fig.\ \ref{fig:SPAC_Hads}(c)], 
the large negative peak just below $E_F$ corresponds to a strong depletion 
of the TiSR around the B adatom.  

The total electron-density plot [Fig.\ \ref{fig:CHD_PER2}(a)], 
together with the calculated Bader charges (Table \ref{tab:Bader_ads}), 
show, similarly to the H adatom, a significant charge transfer from the 
surface-bilayer Ti and C atoms to the B adatom, yielding a B adatom ionicity 
of $-1.09|e|$.  In contrast to the H adatom, 
however, the distribution of the excess electrons around the B adatom shows a 
$p_{xy}$-like symmetry, extending clearly toward the NN Ti atoms.  Also, 
an electron depletion can be observed directly above the B adatom, again 
indicating a strong polarization of the adatom electronic distribution toward 
the substrate.  

\underline{C adatom}:  
For C, the $\Delta$DOS($E$) (Fig.\ \ref{fig:DeltaLDOS_PER2}) shows  
(i) again a large negative double-peak structure at $-0.2$ and $+0.3$ eV, of 
predominantly Ti character (with mixed $d_{(xy, x^2-y^2)}$ and $d_{(xz,yz)}$ symmetry 
for the peak at $-0.2$ eV, and only $d_{(xz,yz)}$ symmetry for the peak at $+0.3$ eV), 
similar to the one for B but with a slightly wider low-energy peak;  
(ii) a positive peak, of predominantly Ti$d_{(xz,yz)}$ character, at $+1.0$ eV;  
(iii) a main positive peak, of mainly adatom-C $p$ and Ti character, at $-1.3$ eV;  
(iv) a smaller positive peak, of mixed adatom and substrate character, at $-3.1$ eV.  

The DOS(${\bf r}$, $E$) plots show that 
(i) the main positive peak corresponds to very strong adatom-C--substrate-Ti bonding 
states;  
(ii) the smaller positive peak is composed of weak adatom-C--substrate-Ti and 
adatom-C--substrate-C bonding states;  
(iii) the large negative peak just below $E_F$ is characterized by a depletion 
of the TiSR around the C adatom.  

The total electron-density plot [Fig.\ \ref{fig:CHD_PER2}(b)] 
and Bader charges (Table \ref{tab:Bader_ads}) are similar to those for the 
B adatom, with a significant electron transfer from the surface-bilayer Ti and C atoms 
to the C adatom, these electrons assuming a clear $p_{xy}$-like symmetry around the 
adatom and extending toward the NN Ti atoms.  Also, an electron depletion is 
again found directly above the adatom, indicating a strong electronic polarization 
toward the substrate.  Compared to the B adatom, however, the electron transfer is 
larger, the calculated C adatom ionicity being $-1.35|e|$.  Also, the polarization 
toward the NN Ti atoms is slightly smaller.  

\underline{N adatom}:  
For N, the $\Delta$DOS($E$) (Fig.\ \ref{fig:DeltaLDOS_PER2}) displays 
(i) again a negative double-peak structure, of Ti character, at $-0.3$ and $+0.3$ eV, 
with the same orbital symmetries as for B and C, but with the low-energy peak 
slightly weaker than that for B and C;  
(ii) a positive peak, of mainly Ti$d_{(xz,yz)}$ character, at $+1.3$ eV;  
(iii) in contrast to the other period-two adatoms, a broad positive region that extends 
between $-5.9$ and $-0.9$ eV, composed of five peaks (a main peak, at $-3.4$ eV, 
which shows mainly N $p$ and Ti character, two minor peaks at $-5.4$ and $-4.7$ eV, 
which show mainly C character, and two intermediate peaks, at $-3.9$ and $-2.4$ eV, 
which show mainly N $p$ character).  

The DOS(${\bf r}$, $E$) plots show that 
(i) the states below $\sim -3.7$ eV consist of mainly N--C bonding states, together with 
weaker N--Ti bonding states;  
(ii) the region around $-3.4$ eV consists of approximately equally strong N--C and N--Ti 
bonding states;  
(ii) the region above $-2.9$ eV consists exclusively of N--Ti bonding states;  
(iii) the region of the negative peak just below $E_F$ is characterized by a depletion 
of the TiSR around the N adatom and by the presence of a very small N-centered dangling 
bond.  

The total electron-density plot for N [Fig.\ \ref{fig:CHD_PER2}(c)]
is very similar to that for C.  However, 
the adatom electrons assume a slightly more spherically symmetric distribution, 
that is, the polarization toward the substrate, and in particular toward the NN Ti atoms, 
although still present, is smaller than in the case of the C adatom.  
The total electron 
transfer from the surface-bilayer Ti and C atoms to the N adatom is identical to that 
for the C adatom, yielding the same adatom ionicity ($-1.35|e|$).  

\underline{O adatom}:  
For O, the $\Delta$DOS($E$) (Fig.\ \ref{fig:DeltaLDOS_PER2}) shows 
(i) a negative, Ti-dominated, double-peak structure at $-0.2$ and $+0.3$ eV, 
with the same orbital symmetries as B, C, and N, but with the low-energy peak weaker 
than for B, C, and N;  
(ii) a positive double-peak structure, of Ti character, above $E_F$ 
(one peak, at $+0.8$ eV, shows mainly $d_{(xz,yz)}$ symmetry, while the other peak, at 
$+1.6$ eV, shows almost exclusively $d_{z^2}$ symmetry);  
(iii) a main positive peak, of mainly O $p$ character, at $-5.5$ eV;  
(iv) smaller positive peaks at $-4.3$ (of almost exclusively O $p$ character) 
and at $-3.5$ eV (of mainly C and O$p_z$ character).  

The main positive peak, at $-5.5$ eV, is slightly broadened, with 
a shoulder on its high-energy side.  While the main peak displays equally large 
Ti and C character, the higher-energy shoulder shows almost no C character.  
A projection of the peak onto O $p_{xy}$ and $p_z$ orbitals reveals that these 
are displaced from one another: the O $p_{xy}$ peak lies at $-5.4$ eV, while the 
O $p_z$ peak lies at $-5.1$ eV.  The main part of the peak is thus composed of 
O $p_{xy}$ orbitals, while the shoulder at higher energy is caused by the O $p_z$ 
orbitals.  

This prediction of strong O $p$ states around $-5.5$ eV agrees well with the 
experimental finding of O $2p$ states between approximately $-5.0$ and $-6.5$ eV 
for full ($1 \times 1$) monolayer coverage of O on TiC($111$).\cite{Edamoto92_prb_O}  
Also, the same experimental study indicates that the O $2p_z$ state has a higher energy  
than the O $2p_{xy}$ states, in good agreement with our results.  In addition, 
our predicted sharp decrease in DOS around $E_F$ agrees also very well with published 
experimental results.\cite{Zaima}  

\begin{figure}
\scalebox{.8}{\includegraphics{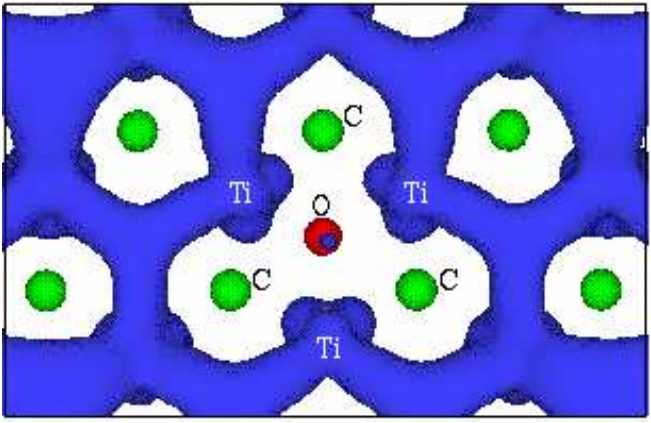}}
\caption{\label{fig:SPAC_SS_Oads}(Color online).  
Same as Fig.\ \ref{fig:SPAC_Hads}(c) but for an O adatom adsorbed in fcc site on 
TiC($111$).  A small DOS above the O adatom, of clear dangling-bond character, 
can be seen.}
\end{figure}

The DOS(${\bf r}$, $E$) plots for adsorbed O show that 
(i) the main, broadened, positive peak consists of very strong O--Ti and O--C bonding 
states;  
(ii) only O--Ti bonding states are present around the smaller positive peaks;  
(iii) the region of the negative peak just below $E_F$ is characterized by a depletion 
of the TiSR around the O adatom and by a small O-centered dangling bond (see Fig.\ 
\ref{fig:SPAC_SS_Oads}), slightly more pronounced than the one for N.  

The total electron-density plot [Fig.\ \ref{fig:CHD_PER2}(d)] 
shows a more symmetric distribution of the O adatom 
electrons than those of the previous second-period adatoms.  Still, a polarization 
toward the NN Ti atoms can be seen.  
Again, the Bader charges (Table \ref{tab:Bader_ads}) 
show a significant electron transfer from the surface-bilayer 
Ti and C atoms, yielding an O adatom ionicity of $-1.17|e|$.

%%%%%%%%% FIGURE %%%%%%%%%%%%%%%%%
%%%%%%%%% FIGURE %%%%%%%%%%%%%%%%%

\underline{F adatom}:  
For F, the $\Delta$DOS($E$) (Fig.\ \ref{fig:DeltaLDOS_PER2}) shows 
(i) that the negative double peak around $E_F$ is significantly 
smaller than for the previous period-two adatoms but still composed of a 
Ti $d_{(xy, x^2-y^2)}+d_{(xz,yz)}$ peak at $-0.1$ eV and of a Ti $d_{(xz,yz)}$ peak at 
$+0.3$ eV;  
(ii) a positive double-peak structure above $E_F$, at $+0.7$ and $+1.1$ eV, with 
mainly Ti $d_{(xz,yz)}$ symmetry;  
(iii) a huge positive peak, of almost exclusively F $p$ character, at $-7.2$ eV;  
(iv) much smaller positive peaks, of mainly C character, at $-4.8$, at $-3.5$, and 
at $-3.1$ eV;  
(v) small negative peaks, of C character, at $-2.7$ and at $-2.1$ eV.  

The DOS(${\bf r}$, $E$) plots show that 
(i) the main peak consists of mainly F-centered states, with only slight bonding to 
the neighboring Ti and C atoms; 
(ii) no F--Ti or F--C bonding states are found around the smaller positive peaks;  
(iii) the negative-peak region just below $E_F$ is again characterized by a depletion 
of the TiSR around the F adatom and by an F-centered dangling bond, stronger than 
for the previous second-period adatoms.  

The total electron-density plot for F [Fig.\ \ref{fig:CHD_PER2}(e)] 
is qualitatively different from those 
for the previous second-period adatoms.  No polarization of the F adatom 
electron density toward the substrate is found, rather, there is a polarization 
toward the vacuum side of the adatom.  Still, some polarization toward the 
NN Ti atom can be detected, but this is much weaker than for the previous 
second-period adatoms, the electron-density difference being almost zero in-between 
the F--Ti bond.  
Again, the Bader charges (Table \ref{tab:Bader_ads}) 
show a significant electron transfer from 
the surface-bilayer Ti and C atoms to the F adatom, yielding an almost completely 
ionized F adatom, its ionicity being $-0.80|e|$.

\begin{figure*}
\scalebox{.95}{\includegraphics{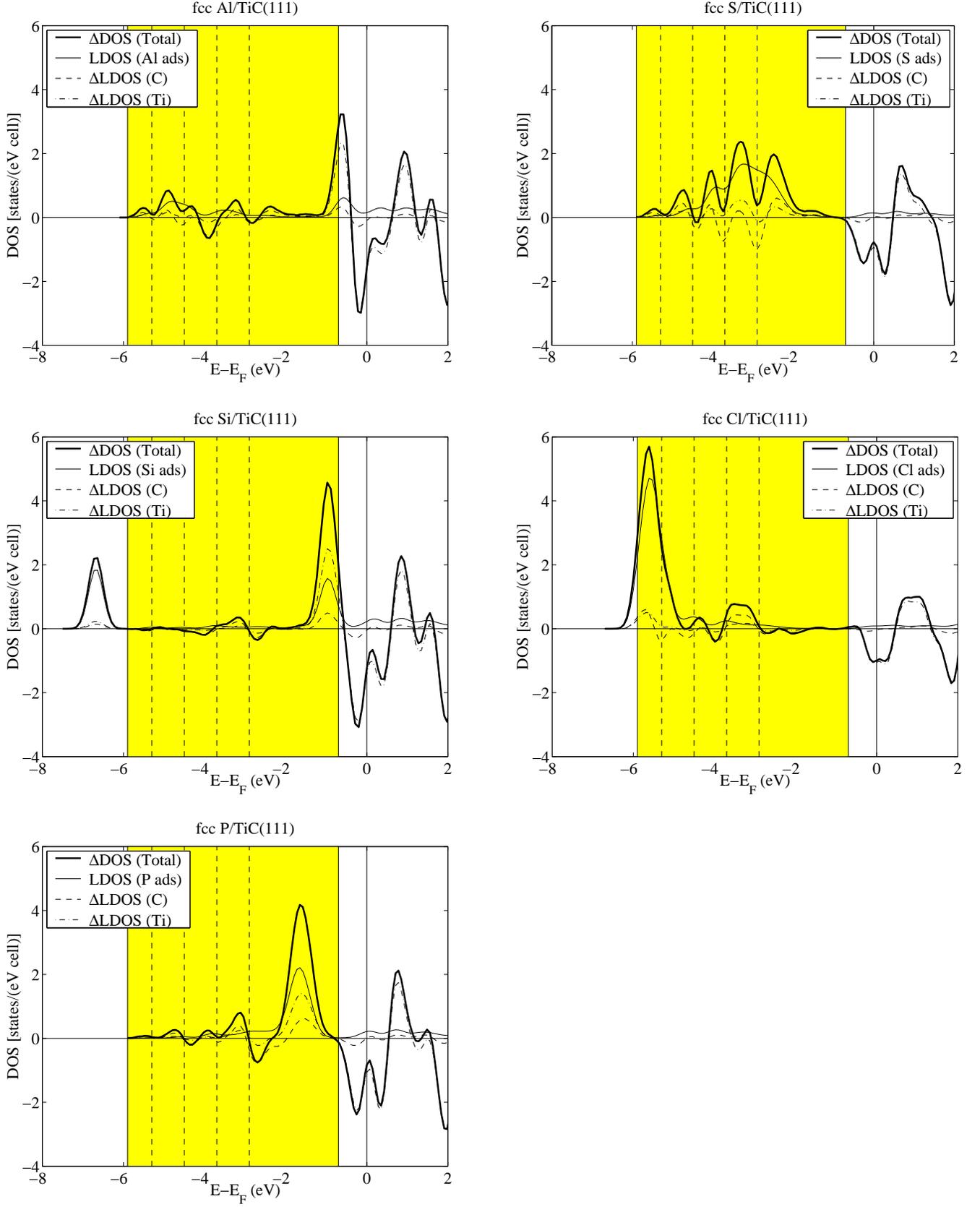}}
\caption{\label{fig:DeltaLDOS_PER3}Same as Fig.\ \ref{fig:DeltaLDOS_Hads} but for the 
third-period elements Al, Si, P, S, and Cl.}
\end{figure*}

\begin{figure*}
\scalebox{.7}{\includegraphics{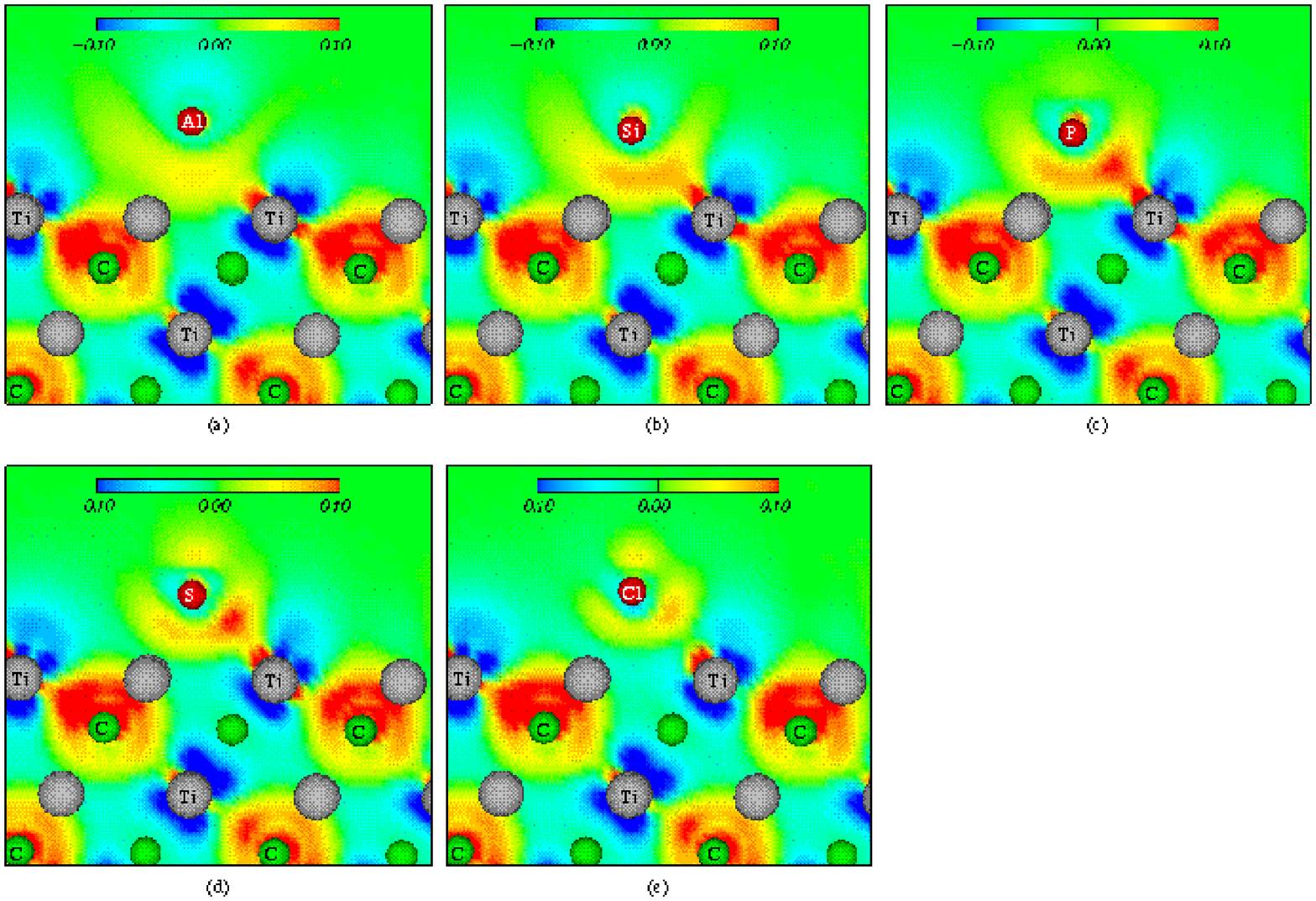}}
\caption{\label{fig:CHD_PER3}(Color online).  
Same as Fig.\ \ref{fig:CHD_Hads} but for the systems with the third-period adatoms 
(a) Al, (b) Si, (c) P, (d) S, and (e) Cl adsorbed in fcc site on the TiC($111$) surface.}
\end{figure*}

%.....
{\it (c) Third-period adatoms (Al, Si, P, S, and Cl) in fcc site on TiC(111).}  
The calculated $\Delta$DOS($E$) and total electron densities 
for the third-period adatoms adsorbed in the fcc site on TiC($111$) 
(Figs.\ \ref{fig:DeltaLDOS_PER3} and \ref{fig:CHD_PER3}) 
show strong similarities to those calculated for the second-period elements.

%%%%%%%%% FIGURE %%%%%%%%%%%%%%%%%
%%%%%%%%% FIGURE %%%%%%%%%%%%%%%%%

%%%%%%%%% FIGURE %%%%%%%%%%%%%%%%%
%%%%%%%%% FIGURE %%%%%%%%%%%%%%%%%

\underline{Al adatom}:  
For Al, the $\Delta$DOS($E$) (Fig.\ \ref{fig:DeltaLDOS_PER3}) shows 
(i) a large negative double-peak structure at $-0.1$ eV (of mainly 
Ti$d_{(xy, x^2-y^2)}$+$d_{(xz,yz)}$ symmetry) and at $+0.4$ eV (of exclusively 
Ti$d_{(xz,yz)}$ symmetry);  
(ii) a positive double peak at $+0.9$ and $+1.6$ eV, both of predominantly Ti 
character, with mixed $d$ symmetry;  
(iii) a main positive peak, of mainly Ti character, at $-0.6$ eV;  
(iv) smaller negative peaks at $-3.9$ eV, of mainly C character, and at $-2.8$ eV, 
of C and Ti character;  
(v) small positive peaks, of mainly Al $s$ character, at $-5.5$, $-4.9$, $-4.3$, 
and $-3.2$ eV;  
(vi) a region of positive $\Delta$DOS between $-2.7$ eV and the main positive peak, 
including a C-dominated peak at $-2.4$ eV.  
Orbital projection of the Al states along the whole UVB region reveals that they 
display $p$ symmetry above $-1.2$ eV and $s$ symmetry below $-1.2$ eV.  

The DOS(${\bf r}$, $E$) plots show that 
(i) the main peak consists of very strong Al--Ti bonding states;  
(ii) only Al--C bonding states are found in the region below $-4.8$ eV;  
(iii) mainly Al--Ti bonding states (with weaker Al--C contributions) are found between 
$-4.8$ and $-4.2$ eV;  
(iv) only weak Al--Ti bonding states are present around the peak at $-3.2$ eV;  
(v) almost no Al-centered states are found between $-2.7$ eV and the main positive peak;  
(vi) the negative peak just below $E_F$ is characterized by a depletion of the TiSR 
around the Al adatom.  

The electron-density plots (Fig.\ \ref{fig:CHD_PER3})
show clearly the overall larger radii of the third-period 
adatoms, compared to the second-period adatoms.  Despite this, similarities 
are present.  Similarly to B, the plot for Al [Fig.\ \ref{fig:CHD_PER3}(a)] 
shows a strong polarization 
of the adatom charge toward the substrate and in particular toward the NN Ti atoms.  
Also, the Bader charges (Table \ref{tab:Bader_ads}) 
show again an electron transfer from the 
surface-bilayer Ti and C atoms to the Al adatom.  However, this is significantly 
weaker than for the second-period adatoms, yielding an Al adatom ionicity of 
$-0.47|e|$.  

\underline{Si adatom}:  
For Si, the $\Delta$DOS($E$) (Fig.\ \ref{fig:DeltaLDOS_PER3}) shows 
(i) a large, Ti-dominated, negative double-peak structure at $-0.2$ and $+0.4$ eV, 
with same orbital symmetries as for Al but with a slightly wider low-energy peak;  
(ii) a positive, Ti-dominated, double peak at $+0.8$ eV (of mainly $d_{(xz,yz)}$ symmetry) 
and at $+1.6$ eV (of mixed $d$ symmetry);  
(iii) a sharp positive peak, of mainly Ti and Si $p$ character, at $-1.0$ eV;  
(iv) a smaller positive peak, of almost exclusively Si $s$ character, at $-6.6$ eV;  
(v) much smaller negative peaks at $-4.0$ and $-2.7$ eV, of mainly C character; 
(vi) a small positive peak at $-3.2$ eV, of mainly C character.  

The DOS(${\bf r}$, $E$) plots show that 
(i) very strong Si--Ti bonding states are present around $-1.0$ eV;  
(ii) the states at $-6.6$ eV are almost exclusively Si centered, although some weak 
Si--Ti bonding overlap can be detected;  
(iii) almost no Si-centered states are found around the smaller negative and positive 
peaks;  
(iv) the negative-peak region just below $E_F$ corresponds again to a depletion of 
the TiSR around the Si adatom.  

The total electron-density plot for Si [Fig.\ \ref{fig:CHD_PER3}(b)] 
is similar to that for Al, with a strong 
polarization toward the substrate and in particular toward the NN Ti atoms.  
Compared to Al, the electron transfer from the surface-bilayer 
Ti and C atoms to the Si adatom is larger, yielding a Si adatom ionicity of 
$-0.93|e|$.  This is clearly visible in the electron-density plot, showing a 
higher electron density in the region below the Si adatom than below the Al 
adatom.  

\underline{P adatom}:  
For P, the $\Delta$DOS($E$) (Fig.\ \ref{fig:DeltaLDOS_PER3}) shows 
(i) again a negative, Ti-dominated, double-peak structure at $-0.3$ and $+0.3$ eV, 
with same orbital symmetries as for Al and Si but with weaker low-energy peak than 
for Al and Si;  
(ii) a positive, Ti-dominated, double peak at $+0.8$ eV (of mainly $d_{(xz,yz)}$ 
symmetry) and at $+1.5$ eV (of mixed $d$ symmetry);  
(iii) a main positive peak, slightly broader than for Si and of mainly P $p$ and Ti 
character, at $-1.6$ eV;  
(iv) smaller positive peaks, of mainly C character, at $-3.1$, $-3.9$, and $-4.7$ eV;  
(v) small negative peaks, of mainly C character, at $-2.7$ and $-4.4$ eV.  

The DOS(${\bf r}$, $E$) plots show
(i) very strong P--Ti bonding states around the main peak;  
(ii) weak P--C and P--Ti bonding states around the peaks at $-3.1$ and $-3.9$ eV;  
(iii) weak P--Ti bonding overlap around $-4.7$ eV;  
(iv) a depletion of the TiSR around P and a weak P-centered dangling bond in 
the negative-peak region just below $E_F$.  

The total electron-density plot for P [Fig.\ \ref{fig:CHD_PER3}(c)] 
is similar to those for the previous 
third-period adatoms, with a strong polarization toward the substrate and in 
particular toward the NN Ti atoms.  However, the charge density around the 
P adatom is much higher than around Al and Si.  This is reflected by the 
Bader charges (Table \ref{tab:Bader_ads}), 
yielding a stronger electron transfer from the surface-bilayer 
atoms to the P adatom than to Al and Si, the P adatom ionicity being $-1.12|e|$.  

\underline{S adatom}:  
For S, the $\Delta$DOS($E$) (Fig.\ \ref{fig:DeltaLDOS_PER3}) resembles that 
for N and shows 
(i) the usual negative, Ti-dominated, double-peak structure at $-0.3$ and $+0.3$ eV, 
with the same orbital symmetries as for Al, Si, and P, but with weaker 
peaks than for Al, Si, and P;  
(ii) a Ti-dominated double peak above $E_F$, smaller than for previous third-period adatoms 
and composed of a main peak at $+0.7$ eV (of mainly $d_{(xz,yz)}$ 
symmetry) and of a secondary peak around $+1.1$ eV (of mixed $d$ symmetry);  
(iii) a broad positive region between $-5.9$ and $-0.9$ eV, which, similarly to N, 
is composed of five peaks (at $-5.4$ and $-4.7$ eV, which show mainly C character, and 
at $-4.0$, $-3.3$, and $-2.5$ eV, which show mainly S $p$ character).  

The DOS(${\bf r}$, $E$) plots show that 
(i) the region below $-3.8$ eV is dominated by S--C bonding states, with weaker 
S--Ti contributions;  
(ii) the region around $-3.3$ eV shows equally strong amounts of S--C and S--Ti bonding 
states;  
(iii) the region above $-3.2$ eV displays only S--Ti bonding states;  
(iv) the negative-peak region just below $E_F$ is characterized by a depletion of the 
TiSR around S and by a weak S-centered dangling bond, slightly stronger than in 
the case of P.  

The total electron-density plot for S [Fig.\ \ref{fig:CHD_PER3}(d)] 
shows similarities to those for 
the previous third-period adatoms.  However, the electron density around the S adatom, 
although still strongly polarized toward the NN Ti atoms, is smaller than for the 
previous third-period adatoms.  This is reflected by the smaller electron 
transfer from the surface-bilayer Ti and C atoms to the S adatom, compared to 
the previous third-period adatoms, as obtained from the Bader charges 
(Table \ref{tab:Bader_ads}), which yield a S adatom ionicity of $-1.06|e|$.  

\underline{Cl adatom}:  
For Cl, finally, the $\Delta$DOS($E$) (Fig.\ \ref{fig:DeltaLDOS_PER3}) shows 
(i) again a negative double peak around $E_F$ (at $-0.1$ and $+0.2$ eV), 
much smaller than for the other third-period adatoms but with the same Ti $d$ symmetries;  
(ii) a positive double peak above $E_F$ (at $+0.8$ and $+1.1$ eV, both with mainly 
Ti $d_{(xz,yz)}$ symmetry);  
(iii) a main positive peak at $-5.6$ eV, with an almost exclusively Cl $p$ 
character;  
(iv) smaller positive peaks at $-4.4$ eV (of predominantly Cl $p$ character), and 
at $-3.5$ and $-3.3$ eV (of mainly C character);  
(v) a small negative peak, of mainly C character, at $-4.0$ eV.  

The DOS(${\bf r}$, $E$) plots show that 
(i) the main peak consists of mainly Cl-centered states, which nevertheless form strong 
bonding coupling to neighboring C and Ti atoms;  
(ii) the states around $-4.4$ eV show weak Cl--Ti bonding coupling;  
(iii) all other states show little or no Cl contribution;  
(iv) the negative-peak region just below $E_F$ is again characterized by a 
depletion of the TiSR around the Cl adatom and by a weak Cl-centered dangling bond, 
stronger than for the previous third-period adatoms.  

The total electron-density for Cl [Fig.\ \ref{fig:CHD_PER3}(e)] 
shows a significantly smaller electron density 
around the adatom, compared to the other third-period adatoms.  Almost zero 
electron-density difference is found in between the Cl adatom and its NN Ti atoms, 
indicating an ionic bond.  Indeed, the Bader analysis (Table \ref{tab:Bader_ads}) 
yields an almost completely 
ionized Cl adatom, its ionicity being $-0.75|e|$.

%.....
{\it (d) O adatom in hcp and top sites on TiC(111).}  
In order to understand whether the nature of the adsorption is affected by 
the adsorption site, we also calculate and examine the electronic 
structures of the most stable adatom, O, in hcp and top sites and 
compare them to the results for O in fcc site.  

\begin{figure*}
\scalebox{.95}{\includegraphics{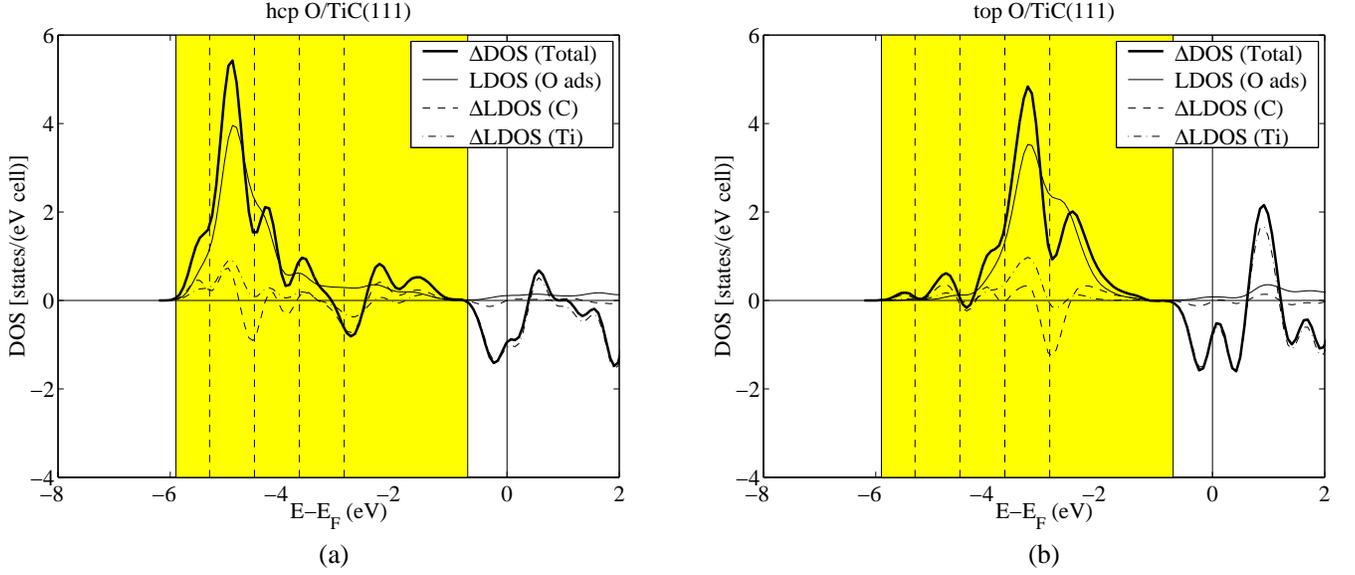}}
\caption{\label{fig:DeltaLDOS_Ohcptop}Same as Fig.\ \ref{fig:DeltaLDOS_Hads} but for an 
O atom adsorbed in (a) the hcp site and in (b) the top site on the TiC($111$) surface.}
\end{figure*}

\underline{O adatom in hcp site}:  
The $\Delta$DOS($E$) for an O adatom adsorbed in the hcp site on the TiC($111$) surface 
[Fig.\ \ref{fig:DeltaLDOS_Ohcptop}(a)] shows that 
(i) the negative, Ti-dominated, double peak around $E_F$ (at $-0.2$ and $+0.1$ eV, of 
$d_{(xz,yz)}+d_{(xy, x^2-y^2)}$ and $d_{(xz,yz)}$ symmetry, respectively) is 
considerably smaller than in the $\Delta$DOS for the fcc O adatom;
(ii) the positive, Ti-dominated, positive peak at $+0.6$ eV (of $d_{(xy, x^2-y^2)}$ 
symmetry) is also considerably smaller than for fcc O;  
(iii) an additional negative peak, of mainly C $p_z$ character, is present at $-2.8$ eV;
(iv) the main positive peak, mainly composed of O $p_{xy}$ states, with 
a smaller contribution from O $p_z$ states, lies at $-4.9$ eV;  
(v) smaller positive peaks are present at $-4.3$ and $-3.7$ (of mainly O$p_z$ character) 
and at $-2.3$ and $-1.6$ eV (of mainly mixed O $p_z$ and C $p_z$ character);  
(vi) there is a shoulder on the low-energy side of the main peak, caused 
by a positive peak of C $p$ symmetry located at $-5.5$ eV.

%%%%%%%%% FIGURE %%%%%%%%%%%%%%%%%
%%%%%%%%% FIGURE %%%%%%%%%%%%%%%%%

\begin{figure*}
\scalebox{.75}{\includegraphics{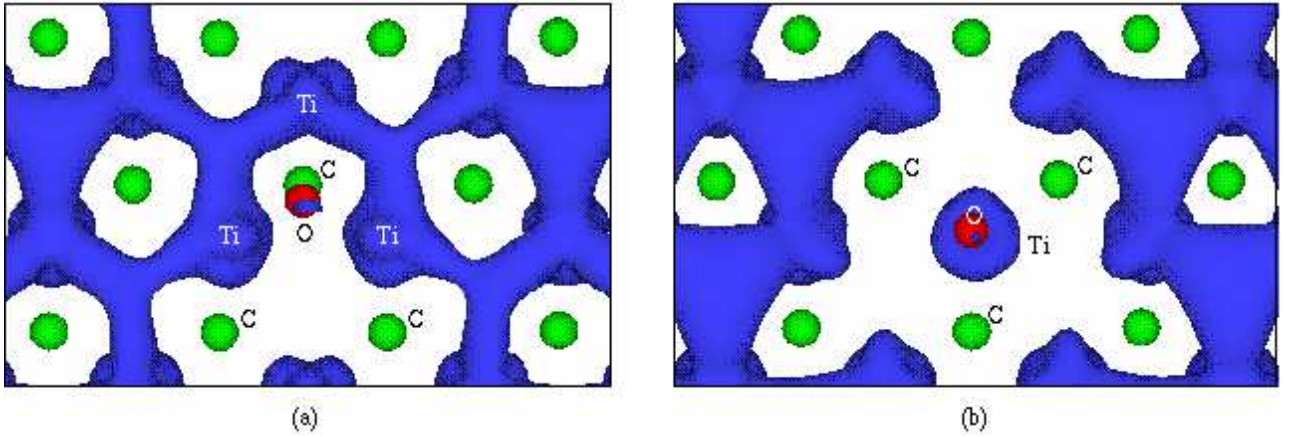}}
\caption{\label{fig:SPAC_SS_Ohcptop}Same as Fig.\ \ref{fig:SPAC_Hads}(c) but for an O 
atom adsorbed in (a) the hcp site and in (b) the top site on the TiC($111$) surface.  
The contour plots correspond to the same value of the DOS(${\bf r}$, $E$) as the 
contour plot in Fig.\ \ref{fig:SPAC_SS_Oads}.}
\end{figure*}

The DOS(${\bf r}$, $E$) plots show that 
(i) the main peak is composed of strong O--C and O--Ti bonding states;  
(ii) the low-energy shoulder of the main peak is dominated by very strong 
O--C bonding states; 
(iii) the peak at $-4.3$ eV contains both O--Ti and O--C bonding states;  
(iv) only O--Ti bonding states are found around the remaining positive $\Delta$DOS peaks;  
(v) the region of the negative peak just below $E_F$ consists of a depletion of the 
TiSR around the O adatom and of a small O-centered dangling bonds [Fig.\ 
\ref{fig:SPAC_SS_Ohcptop}(a)].

%%%%%%%%% FIGURE %%%%%%%%%%%%%%%%%
%%%%%%%%% FIGURE %%%%%%%%%%%%%%%%%

\underline{O adatom in top site}:  
The $\Delta$DOS($E$) for an O adatom adsorbed in the top site on TiC($111$) 
[Fig.\ \ref{fig:DeltaLDOS_Ohcptop}(b)] shows that 
(i) the negative, Ti-dominated, double peak around $E_F$ (at $-0.2$ and $+0.4$ eV, 
of essentially the same symmetries as for hcp O, although a slightly larger 
$d_{z^2}$ contribution is present for the peak at $-0.2$ eV), is again considerably 
smaller than for the fcc O adatom;  
(ii) the positive, Ti-dominated, peak at $+0.9$ eV (of mainly $d_{(xz,yz)}$ symmetry) 
has a similar height as the corresponding peak for fcc O;  
(iii) the main positive peak (of mainly O $p_{xy}$ character) lies at a considerably higher 
energy than for fcc and hcp O, at $-3.3$ eV;  
(iv) other positive peaks are present at $-5.5$ and at $-4.7$ (of mainly C character), 
and at $-2.5$ eV (of mainly O $p_z$ character);  
(v) there is a shoulder on the low-energy side of the main peak, caused by a C-centered 
peak located at $-4.0$ eV.  

The DOS(${\bf r}$, $E$) plots show almost exclusively O--Ti bonding states.  Still, a 
clear O--C coupling is present at the low-energy shoulder of the main peak, at $-4.0$ eV, 
and a weaker O--C coupling is present at the main peak, at $-3.3$ eV.  
The sum of the DOS(${\bf r}$, $E$) for all occupied states above $-0.3$ eV
[Fig.\ \ref{fig:SPAC_SS_Ohcptop}(b)] shows that the top O adatom causes a depletion of the 
TiSR in the three neighboring fcc sites.  As opposed to fcc and hcp O, however, the TiSR is 
non-vanishing directly below the O adatom, due to the presence of the nearest-neighbor Ti 
atom.  

%.....
{\it (e) O adatom on TiC(001).}  
For comparison, the electronic structure for O on TiC($001$) is also examined.  

\begin{figure}
\scalebox{.47}{\includegraphics{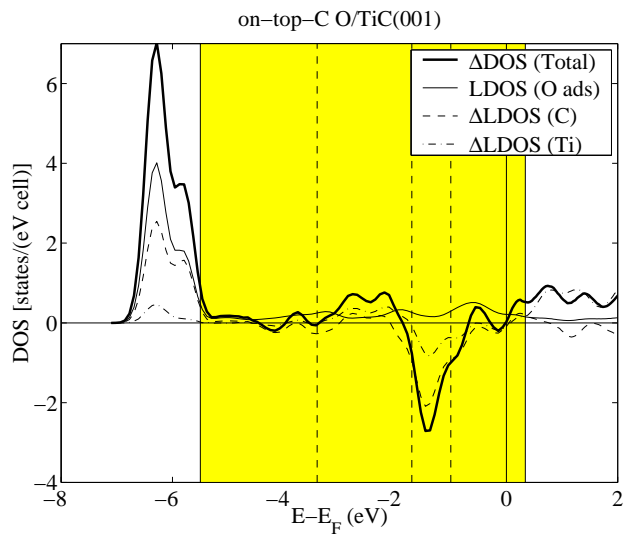}}
\caption{\label{fig:DeltaLDOS_OadsC_001}Same as Fig.\ \ref{fig:DeltaLDOS_Hads} but for 
an O atom adsorbed in the on-top-C site on the TiC($001$) surface.}
\end{figure}

The $\Delta$DOS($E$) for an O adatom adsorbed in the on-top-C site of the TiC($001$) 
surface (Fig.\ \ref{fig:DeltaLDOS_OadsC_001}) is characterized by 
(i) a strong negative peak, of predominantly C character, at $-1.5$ eV;  
(ii) a positive double-peak structure, of mainly O and C character, at $-6.3$ and 
$-5.9$ eV (orbital projection reveals that the more bound peak, at $-6.3$ eV, consists 
of O and C states with exclusively $p_{xy}$ symmetry, while the peak at $-5.9$ eV shows 
only $p_z$ symmetry);  
(iii) smaller positive peaks, of mixed adatom and substrate character, at $-3.7$, 
$-2.8$, and $-2.2$;  
(iv) an O-dominated positive peak at $-0.6$ eV;  
(v) small positive peaks, of mainly Ti character, above $E_F$ (at $+0.2$, $+0.7$, and 
$+1.4$ eV).

%%%%%%%%% FIGURE %%%%%%%%%%%%%%%%%
%%%%%%%%% FIGURE %%%%%%%%%%%%%%%%%

%...........................................................
\subsubsection{Analysis of the Electronic Structure: The Concerted-Coupling Model}

In this subsection, the above-described electron-structure results for the atomic 
adsorption on TiC($111$) are analyzed and interpreted in terms of the 
Newns-Anderson (NA) chemisorption model.  This yields a model for the nature 
of the chemisorption on TiC($111$) that we call the ``concerted-coupling 
model'', since it describes the adatom--substrate interaction as arising from 
the concerted action of two contributions:  
(i) a strong, in the NA sense, coupling between the adatom state(s) and the 
substrate TiSR, and (ii) a coupling, whose strength varies between the different adatoms, 
between the adatom state(s) and the substrate CSR's.  

First [Sec.\ III.C.3(a)], the essential points of the NA model are briefly summarized, 
providing a theoretical framework for our subsequent analysis of the electron-structure 
results for fcc H on TiC($111$) [Sec.\ III.C.3(b)].  A model for 
the nature of the H--TiC($111$) interaction is thus proposed.  
Then [Secs.\ III.C.3(c)--(f)], this model is succesfully applied to 
the description of our electron-structure results for the other systems described 
in Sec.\ III.C.2.  Thus, we conclude that this model is suited for a 
general description of the nature of chemisorption on TiC($111$).  In Sec.\ III.D, the 
model is also shown to yield a good description of the chemisorption of a full 
($1 \times 1$) monolayer of O on TiC($111$).  In Sec.\ IV, the energetical consequences 
of the model are discussed in a qualitative way.  
By correlating the trends in electronic structures with the 
trends in calculated adsorption energies, we point out which features of 
the concerted-coupling model are important for the chemisorption strength on 
TiC($111$).

%...
{\it (a) The Newns-Anderson (NA) model for chemisorption.}  
In the NA model,\cite{Newns} chemisorption on metal surfaces is described with a 
Hamiltonian originally introduced for the description of magnetic 
impurities.\cite{Anderson}  The solutions to the NA Hamiltionian provide 
a theoretical framework for the description of the basic mechanisms behind 
chemisorption on metal surfaces.  It has been used, for example, to successfully 
describe chemisorption trends on transition-metal surfaces (the $d$-band 
model).\cite{Hammer95,Stokbro97,Hammer00}  

Here, the basic concepts arising from the NA model are presented.  These are then 
used in our analysis of the electron-structure results for chemisorption on 
TiC($111$).  

Traditionally, two limiting cases are considered in the solution of the 
NA Hamiltonian: ``weak'' and ``strong'' chemisorption.\cite{Newns,Desj_Spanj}  

In the weak limit, the coupling matrix element between adatom and substrate 
states is small compared to the width of the substrate band.  A single 
solution is thus obtained, 
$E \approx \varepsilon_a^* + \Lambda(\varepsilon_a^*)$, where $\varepsilon_a^*$ is the 
effective atomic level of the adatom (that is, the free atomic level shifted due to the
electrostatic interaction of the adatom with the metal surface), and
$\Lambda(\varepsilon_a^*)$ is small compared to the substrate band width.  
If $E$ falls within the energetical
region of the substrate band, the resulting adsorbate level is broadened.  Otherwise,
the resulting DOS consists of a single state outside the substrate energy band,
together with a typically small and continuous contribution extending over the whole 
substrate band.

In the strong limit, the adatom--substrate coupling matrix element is large compared to 
the substrate band width.  This yields three solutions to the NA Hamiltonian:  
two single states, one on each side of the substrate band, together with a weak continuous 
part extending over the whole substrate band.  These two single states 
can be regarded as the ``bonding'' and ``antibonding'' states, respectively, of a 
``surface molecule'' formed by the adatom and its neighboring substrate atoms.  

By applying the NA model in a tight-binding approximation, Ref.\ 
\onlinecite{Desj_Spanj} shows that these bonding and antibonding 
states appear gradually with increasing size of the ratio $\beta^{\prime} / \beta$, 
where $\beta^{\prime}$ is the adatom--substrate hopping integral and $\beta$ is the 
substrate hopping integral.  
At small $\beta^{\prime} / \beta$ values, the coupling is weak and the adsorbate level 
is broadened.  As $\beta^{\prime} / \beta$ increases, this level broadens further 
and splits gradually into two separate levels.  

In addition to the weak and strong chemisorption, it is therefore reasonable to 
speak also of an ``intermediate'' chemisorption, in which the adsorbate level is strongly 
broadened and split into bonding and antibonding subpeaks.  The stronger the 
chemisorption is, the more are these subpeaks separated from one another.

\begin{figure*}
\scalebox{.9}{\includegraphics{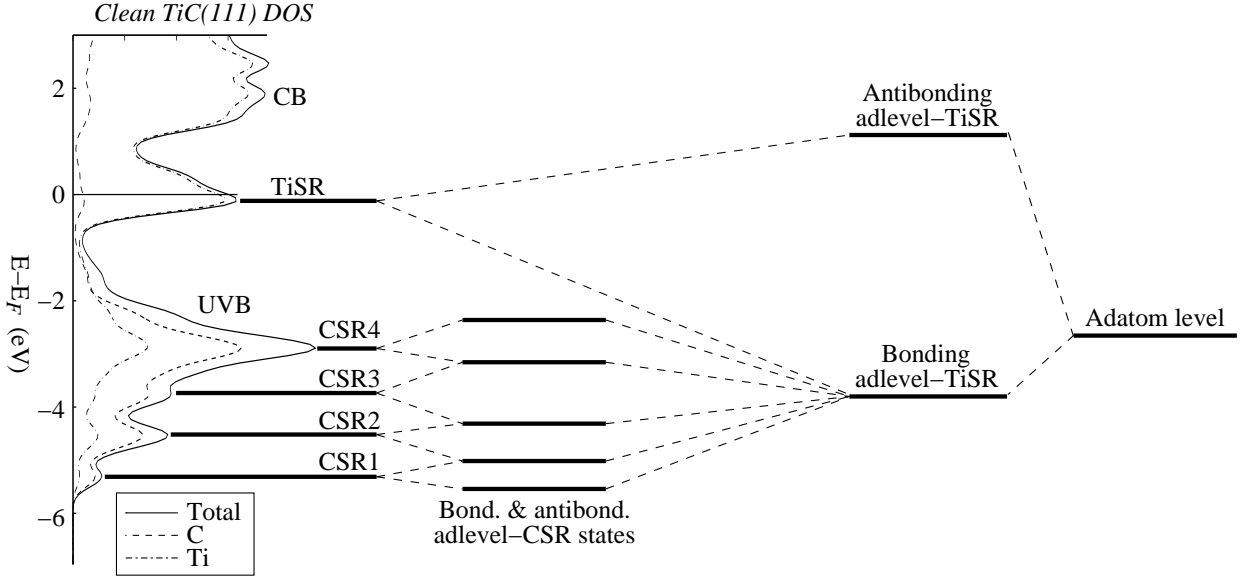}}
\caption{\label{fig:concerted_coupling}Schematic diagram of the concerted coupling 
between the adatom state and the TiC($111$) Ti- and C-centered surface resonances 
(TiSR and CSR1--CSR4).}
\end{figure*}

%...
{\it (b) H adatom in fcc site on TiC(111).}  
The $\Delta$DOS($E$) for fcc H on TiC($111$) (Fig.\ \ref{fig:DeltaLDOS_Hads}) 
shows a strong negative peak in the energetical region of the TiSR.  
This indicates that the adsorption of H causes a strong quenching of the TiSR 
around the adatom, a fact confirmed by our real-space 
visualization of the states in this energetical region [Fig.\ \ref{fig:SPAC_Hads}(c)].  
Thus, the TiSR participates in the formation of adatom--substrate bonds.  Since the 
TiSR is rather sharp, having weak dispersion and being well localized around the 
surface Ti atoms [Sec.\ III.B.2(a)], the coupling between the adatom state and the TiSR 
can be expected to be strong, in the NA sense, and should therefore give rise 
to well separated bonding and antibonding surface-molecule states 
(Fig.\ \ref{fig:concerted_coupling}).  The bonding state 
should lie below the atomic level of the adatom and the antibonding state should 
lie above the TiSR energy.

%%%%%%%%% FIGURE %%%%%%%%%%%%%%%%%
%%%%%%%%% FIGURE %%%%%%%%%%%%%%%%%

Indeed, the $\Delta$DOS($E$) shows the appearance of a positive peak above the TiSR 
and of a broad positive region below it.  The peak above the TiSR can thus 
be interpreted as the antibonding adatom--TiSR state, while 
the positive region below the TiSR can be interpreted as bonding adatom--TiSR states.  

At the same time, the adatom level can also be expected to hybridize with the substrate 
UVB states, in a way similar to the weak coupling to the $sp$ band on transition-metal 
surfaces.\cite{Hammer95,Hammer00}  
On TiC($111$), however, such an interaction should be stronger than on pure transition 
metals, due to the more localized nature of the TiC($111$) UVB, compared to the 
transition-metal $sp$ band.  
Indeed, our $\Delta$DOS($E$) shows a strong negative peak at the energetical location 
of one of the TiC($111$) UVB peaks, indicating a strong coupling between this peak 
and the adatom state.  As shown in Sec.\ III.B.2(b), this UVB peak (called CSR3 
in Fig.\ \ref{fig:LDOS_111}) corresponds to surface-localized C states, or CSR's.  
Indeed, the negative $\Delta$DOS peak is mainly localized around the surface-layer 
C atoms.  

Furthermore, a closer inspection of the $\Delta$DOS reveals that negative, C-centered, 
peaks are also present at the energetical locations of the two lowest TiC($111$) 
UVB peaks, CSR1 and CSR2, and that the energies of the positive peaks lie in-between 
the energies of the UVB peaks.  Thus, this indicates that the adatom level 
couples rather strongly with the TiC($111$) CSR's, giving 
rise to bonding and antibonding states that combine to form the observed 
positive $\Delta$DOS peaks (Fig.\ \ref{fig:concerted_coupling}).  

According to this interpretation, the low-energy $\Delta$DOS peaks should be 
characterized by bonding adatom--C states, while the high-energy peaks should lack 
such states.  Our DOS(${\bf r}$, $E$) plots confirm this:  
at lower energies, bonding adatom--C states dominate over adatom--Ti states 
[Fig.\ \ref{fig:SPAC_Hads}(a)]; 
as the energy increases, this adatom--C character diminishes in favor of the 
bonding adatom--Ti states; above the main negative $\Delta$DOS($E$) peak, only 
adatom--Ti states are found [Fig.\ \ref{fig:SPAC_Hads}(b)].  

Our electron-structure results for fcc H on TiC($111$) indicate therefore that the 
chemisorption arises from the concerted action of two different types of 
adatom--substrate couplings:  
(i) a strong coupling between the adatom state and the TiSR and 
(ii) a somewhat weaker coupling between the adatom state and the CSR's that are 
present in the low-energy part of the TiC($111$) UVB.  
We call this a \emph{concerted coupling}.  
This mechanism has similarities to the $d$-band model for chemisorption on 
transition metals.\cite{Hammer95,Stokbro97,Hammer00}  
Here, however, the weak coupling to the $sp$ band is replaced 
by a stronger coupling to the more localized TiC($111$) CSR's, increasing the 
importance of this coupling for the chemisorption strength.  In particular, the 
following analyses show that the strength of this coupling varies significantly 
between the different adatoms and can therefore be expected to play a role in the 
chemisorption-strength trends.  

In order to analyze these variations and trends, 
we now turn to the analysis of the electronic structures for the second- and 
third-period adatoms and of whether the concerted-coupling model 
is valid also for these adatoms, which are characterized by frontier orbitals 
of $p$ symmetry, in contrast to the simpler $s$ symmetry of the H orbital.

%...
{\it (c) Second-period adatoms in fcc site on TiC(111).}
Like for H, the $\Delta$DOS's for the second-period adatoms in fcc 
site on TiC($111$) (Fig.\ \ref{fig:DeltaLDOS_PER2}) show a strong negative peak in the 
energetical region of the TiSR, together with positive peak(s) just above the TiSR and 
regions of sharp or broadened positive peaks below it.  Thus, also for the second-period 
adatoms, there is a strong coupling between the adatom state and the TiSR.  
The area of the negative $\Delta$DOS peak increases slightly as B $\rightarrow$ C 
and decreases as C $\rightarrow$ N $\rightarrow$ O $\rightarrow$ F, 
indicating corresponding increase and decreases, respectively, of the adatom--TiSR 
coupling strength.  

The energetical region above the TiSR looks generally similar for all 
second-period adatoms, with a Ti-centered positive single- or double-peak 
around $\sim +1.0$ eV.  Like for H, this can be interpreted as the antibonding 
solution of the adatom--TiSR coupling.  On the other hand, the positive region below 
the TiSR, which can correspondingly be interpreted as composed of 
bonding adatom--TiSR states, differs between the different adatoms.  For all adatoms, 
it contains a positive region characterized by states of mixed substrate-Ti $d$, 
substrate-C $p$, and adatom $p$ symmetries.  This region is successively shifted to 
lower energies as the adatom nuclear charge $Z$ increases.  
This shift agrees with the interpretation of this region as the bonding 
solution of the adatom--TiSR coupling.  Such a bonding solution must lie below the 
energy of the free adatom state.  As $Z$ increases, the valence electrons 
of the free adatom are more strongly bound to its nucleus, that is, their 
energies are lowered.  As the free-adatom state decreases in energy, so must also 
the bonding solution of the adatom--TiSR coupling.  

As $Z$ increases, the bonding adatom--TiSR state overlaps therefore with 
different parts of the substrate electronic structure, at successively decreasing 
energies.  In particular, our calculated $\Delta$DOS's show that:  
for B, it lies at the upper edge of the substrate UVB region and is very sharp; 
for C, it lies in the high-energy region of the UVB and is slightly 
broadened; 
for N, it lies in the middle of the UVB region and is strongly broadened 
and split into subpeaks; 
for O, it lies in the low-energy region of the UVB and is broadened and 
somewhat split into subpeaks; and 
for F, it lies well below the UVB region and is very sharp.  
In addition, our DOS(${\bf r}$, $E$) plots show that, for all 
adatoms, only adatom--Ti bonding states are present in the high-energy region of 
the UVB, while both adatom--Ti and adatom--C bonding states are present in the 
middle- and low-energy regions of the UVB, and that the adatom--C contribution 
increases successively as the energy decreases.  

Thus, the strength and nature of the coupling between the adatom and the 
TiC($111$) UVB states vary significantly between the different adatoms.  
For B, the sharpness of the positive $\Delta$DOS peak 
indicates that it only corresponds to bonding adatom--TiSR states [as confirmed by the 
DOS(${\bf r}$, $E$) plots] and that no coupling takes place with the UVB states.  

For C, the positive $\Delta$DOS 
peak is slightly broadened, indicating a weak coupling of the adatom state to the states 
in the upper region of the UVB.  Since these states are Ti-dominated 
[see Sec.\ III.B.2(b)], no strong adatom--C bonds are formed.  However, a small 
subpeak is formed just below the main UVB peak, containing very weak adatom--C bonding 
states.  This indicates a very weak hybridization with the C-centered states of the 
main UVB peak.  

For N, the strong broadening and partial splitting of the positive 
$\Delta$DOS peak indicate a rather strong adatom--UVB coupling.  
Like for the fcc H adatom, positive peaks are formed in between the peaks of 
the substrate-UVB DOS.  At the same time, negative C-localized peaks are present at 
the energies of the substrate UVB peaks, indicating that the CSR's in the substrate 
UVB hybridize with the adatom state.  The formation of subpeaks in the 
positive $\Delta$DOS can therefore be interpreted as the formation of bonds 
between the adatom and the C atoms in the surface TiC bilayer.  Similarly to 
the case of the H adatom, these subpeaks correspond to combinations of bonding 
and antibonding adatom--C states.  This is confirmed by the DOS(${\bf r}$, $E$) plots, 
which show a successively stronger dominance of bonding adatom--C 
states with decreasing energy, as well as the absence of any bonding adatom--C states 
at energies higher than the main substrate-UVB peak (CSR4).  

At the same time, the differences between the N and H $\Delta$DOS's can be attributed 
to the different symmetries of the adatom valence states ($s$ for H and 
$p$ for N).  For example, the much stronger coupling to the TiSR for N (as well as 
for the other second- and third-period adatoms) than for H (illustrated by the much 
stronger negative $\Delta$DOS peaks around $E_F$ for N) can be understood from the 
better spatial overlap of the TiSR $d_{xz, yz} + d_{xy, x^2-y^2}$ orbitals with the 
$p$ (mainly, the $p_{xy}$) orbitals of the second- and third-period adatoms than with 
the $s$ orbital of the H adatom.  

For O, the splitting of the positive $\Delta$DOS peak indicates again an 
adatom--UVB coupling.  Like for N, positive subpeaks are present in between 
the substrate-UVB peaks, while negative, C-localized, peaks are present at 
the energies of the UVB peaks.  As described in Sec.\ III.C.2(b), the main, 
broad, O $p$ peak, which straddles the lowest-energy UVB peak, is actually composed 
of an O $p_{xy}$ peak and of an O $p_z$ peak, each lying on opposite sides of the 
lowest-energy UVB peak.  Therefore, 
also for O, the $\Delta$DOS can be interpreted as evidence for rather strong 
hybridizations between the adatom state and the CSR's.  In particular, the main 
positive $\Delta$DOS peaks lie here in the UVB region that is dominated by the CSR's, 
that is, below the main peak (CSR4) of the substrate UVB.  As our DOS(${\bf r}$, $E$) plots 
show, it is in this region that the strongest bonding adatom--C states are found.  

It is also interesting to notice that the most strongly bonding O--C states 
have O $p_{xy}$ symmetry.  This agrees with the experimental findings for full 
($1 \times 1$) monolayer coverage of O on TiC($111$) 
[see Sec.\ III.C.2(b)].\cite{Edamoto92_prb_O}  
Apparently, the stronger bonding of the O $p_{xy}$ orbitals, compared to the O $p_z$ 
orbitals, is due to the geometry of the fcc site, which favors lateral ($xy$) interaction 
with both the Ti and the C atoms of the TiC($111$) surface bilayer.  

Finally, for F, the sharpness of the positive $\Delta$DOS peak shows that no 
adatom--UVB coupling occurs, due to the fact that the adatom level lies well 
outside the energetical region of the UVB.  

Thus, like for the fcc H adatom, two different types of couplings can take place also 
for the second-period adatoms:  one between the adatom state and the substrate TiSR 
and one between the adatom state and the substrate CSR's.  However, their strengths 
vary significantly with varying $Z$ within the adatom period.  In particular, the 
strength of the coupling to the CSR's varies from no coupling (B and F) to rather 
strong coupling (O and N).  

Finally, we note that, for B, the B $s$ state lies in the low-energy 
region of the substrate UVB.  This causes a slight broadening of the B $s$ level, 
as well as the formation of a weak subpeak slightly above the main B $s$ peak.  
The DOS(${\bf r}$, $E$) plots show that, like for the adatom $p$ states, this is 
due to a hybridization between the adatom $s$ state and the CSR's of the surface 
TiC bilayer.

%...
{\it (d) Third-period adatoms in fcc site on TiC(111).}
The calculated electronic structures for the third-period adatoms in fcc site on 
TiC($111$) show trends that are very similar to those for the second-period adatoms.  
Overall, however, two main differences can be detected:  
(i) the negative $\Delta$DOS peak is smaller than for the second-period 
adsorbates and (ii) the positive $\Delta$DOS peak below the TiSR lies at higher energies 
than for the second-period adsorbates.  Both differences are due to core 
orthogonalization, (i) resulting from the larger radii of the third-period adatoms, 
which cause smaller overlaps between adatom and TiSR wavefunctions and thus weaker 
adatom--TiSR couplings and (ii) from the higher energy, within a group, of the valence 
electron state of a free third-period atom.  

Still, however, the two types of couplings described above are present.  
Like for the second-period adatoms, the adatom--TiSR coupling strength appears to 
slightly increase as Al $\rightarrow$ Si and to decrease as 
Si $\rightarrow$ P $\rightarrow$ S $\rightarrow$ Cl.  
The coupling between the adatom $p$ state and the CSR's follows also a similar trend 
as for the second-period adatoms, however, with differences due to the higher 
free-adatom energy levels.  The coupling to the UVB is now very weak 
for both Al and Si, both with positive $\Delta$DOS peaks around the top of the 
substrate-UVB region.  The $\Delta$DOS for P resembles that for C, the slightly lower 
energy of the P peak, however, causing a slightly stronger adatom--UVB coupling, 
shown by the slightly stronger peak broadening and the formation of a low-energy 
subpeak with slightly stronger adatom--C bonding character.  In the same way, 
the $\Delta$DOS for S resembles that for N, with strong subpeaks formed by the 
hybridization between the adatom state and the CSR's.  

For Cl, the main positive $\Delta$DOS peak lies in 
the low-energy region of the substrate UVB, at a slightly lower energy than 
that of the O main peak.  Thus, the Cl state might be expected to behave similarly to 
the O state and couple strongly to the CSR's.  However, our calculated 
$\Delta$DOS's show a smaller broadening and subpeak formation for Cl than for O.  
Further, the DOS(${\bf r}$, $E$) plots show that the Cl peak is composed of states 
that are strongly localized around the Cl adatom, with a very weak coupling to 
substrate atoms.  The adatom--UVB coupling is thus much weaker for Cl than for O.  
The reason is clearly seen in our Bader calculations of the atomic charges 
(Table \ref{tab:Bader_ads}).  For all adatoms, electrons are transferred from 
the surface-bilayer Ti and C atoms to the adatom.  For the group-VII elements 
(F and Cl) this transfer is large enough to almost completely fill the outer 
valence-electron shell of the adatom, thus almost completely ionizing these adatoms.  
This is confirmed by our analysis 
of the total electronic density around each adatom (see Sec.\ III.C.2 and 
Figs.\ \ref{fig:CHD_PER2} and \ref{fig:CHD_PER3}), which shows an increase in 
electron density along the adatom--Ti bond for all adatoms except for F and Cl.  
This is indicative of covalent adatom--Ti bonds for all adatoms except for F and Cl, 
which show a prevalently ionic type of bond.  The almost filled valence-electron 
shell of Cl is therefore almost inert toward further hybridization with the CSR's, 
resulting in a weaker adatom--UVB coupling than in the case of O.  

Thus, also for the third-period adatoms, chemisorption is characterized by 
the concerted action of the two different types of coupling: adatom--TiSR and 
adatom--CSR's.  Again, within the period, the coupling strengths vary significantly 
with $Z$.  In addition, they differ between the two periods, due to the extra 
electron shell in period three.  

Finally, it can be noted that for Al, the adatom $s$ state lies inside the 
energetical region of the substrate UVB and is therefore broadened and 
split into subpeaks of adatom--C and adatom--Ti bonding character.

%...
{\it (e) O adatom in hcp and top sites on TiC(111).}
In order to test whether the concerted-coupling model can also be used to understand 
our results for adsorption in other sites than fcc, we now analyze the electronic 
structures for the O adatom in hcp and top sites.  

For both hcp and top O adatom, the $\Delta$DOS's show negative peaks in 
the energetical region of the TiSR, together with positive peaks just above and 
broader positive regions below.  Thus, the coupling to the TiSR is 
active in these adsorption sites as well.  However, for both hcp and top O, 
the resulting bonding states ({\it i.e.}, the positive $\Delta$DOS region below the 
TiSR) lie at higher energies than for fcc O, implying a weaker adatom--TiSR coupling.  

For the hcp adatom, the reason for this can be found in the symmetry of the TiSR.  
As shown in Sec.\ III.B.2(a), due to its dangling-bond character, the TiSR extends 
toward the fcc site and avoids the hcp site.  Thus, the overlap between the adatom 
states and the TiSR is smaller in the hcp site than in the fcc site, resulting in a 
weaker hopping integral $\beta^{\prime}$ between the adatom and its NN Ti atoms.  

This is confirmed by our DOS(${\bf r}$, $E$) plots for the energetical region of the 
TiSR, which show that also for the hcp O adsorbate [Fig.\ \ref{fig:SPAC_SS_Ohcptop}(a)], 
the TiSR depletion occurs around the fcc site close to the adatom.  
Also, the relaxation displaces the hcp adatom slightly toward the 
neighboring fcc site, indicating that the adatom--TiSR hybridization is 
stronger in the fcc site.  

As described in Sec.\ III.C.3(a), the degree of broadening of the adsorbate level 
and of its splitting into bonding and antibonding states depends on the strength 
ratio $\beta^{\prime} / \beta$, where $\beta$ is the substrate hopping 
integral.\cite{Desj_Spanj}  A smaller value of $\beta^{\prime} / \beta$ implies 
a smaller splitting.   Thus, the weaker adatom--TiSR overlap in the hcp site, compared 
to the fcc site, causes a smaller downward shift of the resulting bonding-state 
energy from the free adatom-state energy.  Indeed, the main positive peak below 
the TiSR in the hcp $\Delta$DOS lies $0.6$ eV higher than in the fcc 
$\Delta$DOS.  Because of this, the coupling with the UVB states is also changed, 
the hcp O state coupling more strongly to the CSR2 peak of the substrate UVB, 
compared to the fcc O state.  

In the top site, the adatom--TiSR overlap is larger than in the hcp site, 
implying a higher value for $\beta^{\prime}$.  However, the atomic coordination is 
lower in the top site.  The smaller splitting of the bonding and antibonding adatom--TiSR 
states can therefore be understood from the smaller value of the second moment of the 
adatom LDOS, $\mu^{(2)} = N \beta^{\prime}$, where $N$ is the coordination of the 
adatom.  

The difference between the top and fcc $\Delta$DOS's is much more pronounced 
than that between hcp and fcc, the main positive peak of the top $\Delta$DOS lying 
$2.2$ eV higher than in the fcc $\Delta$DOS.  
Again, the interaction with the substrate UVB is considerably changed because of this, 
the strongest coupling now being to the main UVB peak (CSR4).  
Correspondingly, the adatom--UVB coupling is more localized around the surface-bilayer 
Ti atoms than the C atoms.  This is also due to the site geometry, which makes overlap 
with C-localized states negligible.  Still, though, some clear O--C coupling 
is detected in the lower part of the main positive $\Delta$DOS peak, showing that 
the adatom--CSR coupling is weakly active also for top O.  

We conclude that the concerted-coupling model is capable of explaining the 
electronic-structure features for adsorption in hcp and top sites as well.  
Like for the fcc site, the chemisorption in hcp and top sites can be 
described as composed of two interacting components: an adatom--TiSR coupling and 
an adatom--CSR coupling.

%.....
{\it (f) O adatom on TiC(001).}  
Since the nonpolar TiC($001$) surface lacks the TiSR characteristic of 
the polar ($111$) surface [Sec.\ III.B.3], a comparison between the electronic 
structures of O adsorbed on the two surfaces can provide some information on the 
importance of the TiSR for the adsorption.  

Our calculated $\Delta$DOS for O in the most favorable TiC($001$) site (on-top C) 
shows the presence of only one negative peak (Fig.\ \ref{fig:DeltaLDOS_OadsC_001}), 
which lies just above the main peak of the substrate UVB and is mainly localized 
around the substrate C atom.  Thus, only one type of coupling is present here, between 
the adatom state and C-localized states in the upper part of the substrate UVB.  
As discussed above [Sec.\ III.B.3], this is the only UVB region that shows signs, in our 
results, of SR's.  The $\Delta$DOS shows that this coupling forms bonding peaks just 
below the UVB region, of mixed O and C character, and an antibonding region around 
and above $E_F$.  

Thus, the weaker adsorption on TiC($001$), compared to TiC($111$), can be 
understood to arise from a lack of coupling to Ti-centered SR's on ($001$).

%-----------------------------------------------------------
\subsection{Oxygen ($1 \times 1$) monolayer on TiC($111$)}

So far, experimental studies of adsorption on TiC($111$) have primarily 
been concerned with full oxygen ($1 \times 1$) monolayer 
coverage.\cite{Souda88,Edamoto92_prb_O}  Therefore, 
we conclude our investigation with calculations on the O monolayer on TiC($111$).  
As for the case of atomic adsorption, first, the adsorption energetics and geometry are 
investigated.  Then, the electronic structure is analyzed in order to understand the 
bonding nature.  In particular, we consider to what extent our concerted-coupling model 
for atomic chemisorption can be used in the situation of full monolayer coverage.  

%...........................................................
\subsubsection{Energetics and geometry}

\begin{figure}
\scalebox{.46}{\includegraphics{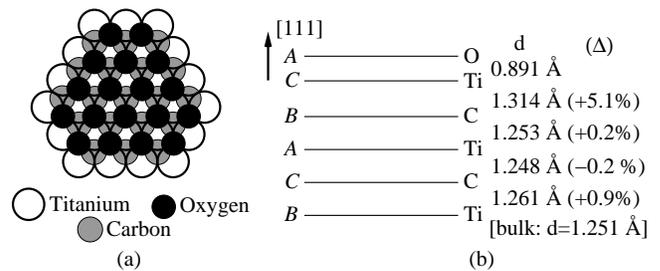}}
\caption{\label{fig:TiC_O_surf}Structure of TiC($111$)-O($1 \times 1$):  
(a) top view of the three top surface layers;  (b) the top six 
surface layers, seen perpendicularly to [$111$], showing the stacking 
sequence and the interlayer relaxations, in absolute values 
($d$) and relative to the TiC bulk distance ($\Delta$), calculated with our 
15-bilayer slab (see Sec.\ II).}
\end{figure}

On TiC($111$), full monolayer coverage of O($1 \times 1$) has been 
observed experimentally.\cite{Souda88}  The unreconstructed O($1 \times 1$) structure is 
reported to be stable up to approximately $1000^{\circ}$C,  where it starts 
to reconstruct to ($\sqrt{3} \times \sqrt{3}$)-R$30^{\circ}$.  The O($1 \times 1$) 
atoms are found to occupy the hollow sites with no C directly underneath, that is, 
the fcc sites [see Fig.\ \ref{fig:TiC_O_surf}(a)].\cite{Souda88}

%%%%%%%%% FIGURE %%%%%%%%%%%%%%%%%
%%%%%%%%% FIGURE %%%%%%%%%%%%%%%%%

Our calculated adsorption energies for O($1 \times 1$) in both hollow sites 
(fcc and hcp) and in top site (Table \ref{tab:O1x1_en}) agree with the experimental 
results of the fcc site being the stable one.

%%%%%%%%% TABLE %%%%%%%%%%%%%%%%%%
\begin{table}
\caption{\label{tab:O1x1_en}Adsorption energies $E_{\rm ads}$ for a full monolayer of 
O ($1 \times 1$) on TiC(111), as calculated with our 15-bilayer slab (see Sec.\ II).  
Adsorption in fcc, hcp, and top sites is considered.}
\begin{tabular}{ccc}
\hline\hline
\multicolumn{3}{c}{$E_{\rm ads}$ [eV/atom]} \\
fcc site & hcp site & top site \\
\hline
$8.12$ & $7.31$ & $5.48$ \\
\hline\hline
\end{tabular}
\end{table}
%%%%%%%%% TABLE %%%%%%%%%%%%%%%%%%

According to our calculations, the TiC($111$) surface relaxation is significantly 
affected by the fcc O overlayer.  The four top interlayer relaxations of TiC($111$)
(as calculated with our 15-bilayer slab, see Sec.\ II) are now: $+5.1\%$, $+0.2\%$, 
$-0.2\%$, and $+0.9\%$ [Fig.\ \ref{fig:TiC_O_surf}(b)], as compared to the 
values $-19.2\%$, $+11.0\%$, $-5.8\%$, and $+0.6\%$ for the clean TiC($111$) 
surface [Fig.\ \ref{fig:surf_struc}(b)].  
The distance between the O layer and the TiC($111$) surface is $0.891$ \AA .  
The O--Ti bond is thus quite strong, pulling up the 
top Ti layer away from the second C layer and toward the chemisorbed O layer.  
Also, we note that the relaxation is limited to the first surface layer (the interlayer 
distances change by less than $0.02$ \AA\ from the second TiC bilayer on).  

The ICISS study of Souda {\it et al.}\cite{Souda88} reports about the presence of two 
different O sites, with different Ti--O interlayer distances ($1.0$ \AA\ and $0.8$ \AA , 
respectively).  No sign of this is present in our results, our calculated interlayer 
distance lying in the middle of these two values.  We stress here that the 
convergence of our calculations has been carefully tested and that full relaxation 
using both a ($1 \times 1$) and a ($2 \times 3$) surface unit cell has been 
performed [see Sec.\ II for details].  In addition, the calculation with the ($2 \times 3$) 
cell has been performed both with and without symmetry constraints 
on the calculated KS wavefunctions.  

Our relaxed O--Ti and O--C bond lengths are $1.98$ \AA\ and $2.83$ \AA , 
respectively.

%...........................................................
\subsubsection{Electronic Structure}

Like for the case of atomic adsorption, we now turn to the analysis 
of the electronic structure for full ($1 \times 1$) monolayer of O on TiC($111$), in 
order to understand the nature of the bond.  

\begin{figure}
\scalebox{.47}{\includegraphics{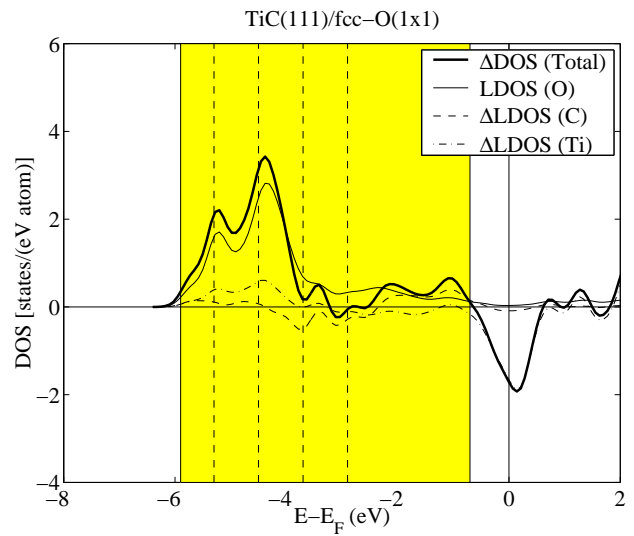}}
\caption{\label{fig:DeltaLDOS_fccO_4lay}Same as Fig.\ \ref{fig:DeltaLDOS_Hads} but for 
a full ($1 \times 1$) monolayer of O atoms adsorbed in the fcc sites of the 
TiC($111$) surface.}
\end{figure}

In particular, the calculated $\Delta$DOS for fcc O($1 \times 1$) 
(Fig.\ \ref{fig:DeltaLDOS_fccO_4lay}) is compared with that for the fcc O adatom 
(Fig.\ \ref{fig:DeltaLDOS_PER2}).  For fcc O($1 \times 1$), we note that 
(i) the negative peak around $E_F$ is broadened and consists of a single peak, 
at $+0.1$ eV, of almost exclusively Ti $d_{(xz,yz)}$ and $d_{(xy, x^2-y^2)}$ character;  
(ii) the positive peak in the low-energy part of the UVB is broadened and split into 
two subpeaks, at $-5.2$ and $-4.4$ eV, of almost exclusively O character 
(the smaller peak, at $-5.2$ eV, has almost exclusively O $p_{xy}$ symmetry);  
(iii) smaller positive peaks are present at $-3.4$ eV (of mainly O $p_z$ symmetry), 
at $-2.1$ eV (of mainly O $p_z$ and C $p_z$ symmetry), and at $-1.1$ eV (of mainly 
C $p_z$ symmetry).  

The DOS(${\bf r}$, $E$) plots for the fcc O ($1 \times 1$) monolayer show that 
(i) the peak at $-5.2$ eV is composed of O--O, O--Ti, and O--C bonding states; 
(ii) the peak at $-4.4$ eV is composed of O--Ti and O--C bonding states; 
(iii) the remaining positive peaks contain only O--Ti bonding states.  

These results agree with experimental studies that show that the 
O $p_{xy}$ orbitals lie at lower energy than the O $p_z$ 
orbitals.\cite{Edamoto92_prb_O}

%%%%%%%%% FIGURE %%%%%%%%%%%%%%%%%
%%%%%%%%% FIGURE %%%%%%%%%%%%%%%%%

%...........................................................
\subsubsection{Analysis of the Electronic Structure}

Our calculated $\Delta$DOS (Fig.\ \ref{fig:DeltaLDOS_fccO_4lay}) shows a strong 
depletion of states in the energetical region of the TiSR.  This indicates that the 
adatom--TiSR coupling is active also in the case of full monolayer ($1 \times 1$) 
adsorption of O on TiC($111$).  In addition, the DOS(${\bf r}$, $E$) plots show 
the presence of adatom--C bonding states in the low-energy region of the UVB, implying 
that the adatom--CSR coupling is also still present.  
Thus, our concerted-coupling model appears to be valid also for the 
case of full monolayer coverage of O on TiC($111$).  

However, the calculated $\Delta$DOS for O($1 \times 1$) adsorption shows a 
considerably stronger broadening of the adatom $p$ peak than for atomic O 
(Fig.\ \ref{fig:DeltaLDOS_PER2}), as well as a splitting of the 
adatom $p$ peak into two subpeaks.  The DOS(${\bf r}$, $E$) plots show that 
O--O bonding states are present in the lower subpeak and absent in the 
higher subpeak.  Thus, the broadening and splitting of the $p$ peak arise from the 
coupling between neighboring O atoms within the O monolayer, which gives rise to the 
formation of bonding and antibonding O--O states.  The bonding O--O states dominate 
in the lower part of the O $p$ peak, while antibonding O--O states dominate in the 
upper part of the O $p$ peak.  

It is interesting to notice that the O $p_{xy}$ states are more important for the 
most bonding states, while the O $p_z$ states are more important at higher energies.  
Obviously, the O $p_{xy}$ states overlap more strongly with the neighboring 
O atoms in the monolayer, thus providing stronger binding.  However, the O $p_{xy}$ 
states are more bonding also in the case of atomic adsorption (as discussed in 
Secs.\ III.C.2 and III.C.3) and are therefore more favorable also for the coupling 
with the TiC($111$) surface.  This can be understood by 
considering that the O adatom lies in the fcc site, that is, in the middle of the 
two triangles formed by the NN Ti and C atoms.  Thus, the interaction 
with the NN Ti and C atoms is predominantly lateral, as exemplified 
by Figs.\ \ref{fig:SPAC_Hads}(a)--(b), and as discussed also in Sec.\ III.C.3(c).

%%%%%%%%%%%%%%%%%%%%%%%%%%%%%%%%%%%%%%%%%%%%%%%%%%%%%%%%%%%%
\section{Discussion}

In the previous sections, a concerted-coupling model is presented to describe the 
mechanisms behind chemisorption on the polar TiC($111$) surface.  
The model is based on detailed analyses, within the theoretical framework of the 
Newns-Anderson (NA) model, of the electronic structures for the clean surface, for 
first-, second-, and third-period adatoms in fcc site, for the O adatom in hcp and top 
sites, and for a full ($1 \times 1$) monolayer coverage of O.  Also, for comparison, 
O adsorption on the nonpolar TiC($001$) surface is considered.  

According to this model, chemisorption on TiC($111$) is caused by the combination 
of two different types of adatom--substrate couplings:  
(i) a strong, in the NA sense, coupling between the adatom state(s) and the Ti-localized SR 
(TiSR) of TiC($111$) and
(ii) a coupling, whose strength varies significantly between the different adatoms, 
between the adatom state(s) and the TiC($111$) UVB states, in particular the C-localized 
SR's (CSR's) in the low-energy part of the UVB.  

At the same time, our calculations yield characteristic, pyramid-like, trends for 
the atomic fcc adsorption energies $E_{\rm ads}$ within each adatom period, with 
strongest adsorption for the group-VI elements (O and S).  In particular, 
the adsorption energy for fcc O adatom is very high, almost $9$ eV/adatom, reflecting 
the high reactivity of the polar TiC($111$) surface.  

In this section, we discuss how the concerted-coupling model can provide an 
understanding of these adsorption-energy trends and of the extremely large variations 
of adsorption energies between adatoms (from $3.4$ eV/adatom for fcc Al to $8.8$ 
eV/adatom for fcc O) and between surface orientations [the $E_{\rm ads}$ value for O 
on TiC($001$) being $43\%$ smaller than on TiC($111$)].  

In particular, we wish to address the following questions:  
(i) How can we understand the $E_{\rm ads}$ trends within each adatom period?  
(ii) How can we understand the lower $E_{\rm ads}$ values for third-period adatoms?  
(iii) Why is chemisorption strongest in the fcc site?  
(iv) Why is chemisorption stronger on TiC($111$) than on TiC($001$)?  

In the following, the trends in our calculated electronic structures for the 
considered systems are qualitatively correlated to the calculated $E_{\rm ads}$ 
trends.  In this way, we identify features in the concerted-coupling model that 
can be used to understand the calculated $E_{\rm ads}$ trends.  

In order to answer question (i), we start by examining the second-period adatoms.  
As discussed in Sec.\ III.C.3(c), as the adatom nuclear charge $Z$ increases 
within the period, the two components of the concerted coupling vary, in 
different ways.  

The variations in area of the negative $\Delta$DOS peaks around $E_F$ 
indicate variations in the adatom--TiSR coupling strength:  slightly increasing as 
B $\rightarrow$ C and strongly decreasing as 
C $\rightarrow$ N $\rightarrow$ O $\rightarrow$ F.  
The strong decrease from C to F can be understood as a consequence of the 
decreasing energetical overlap between the adlevel and TiSR energies, which decreases 
the strength of the adatom--TiSR coupling.  The slight increase from B to C is based 
on a widening of the negative $\Delta$DOS peak due to the downward shift of the main 
positive peak.  Thus, it may simply be due to a cancellation effect between the positive 
and negative peaks of B, which partially overlap.  

\begin{figure*}
\scalebox{.47}{\includegraphics{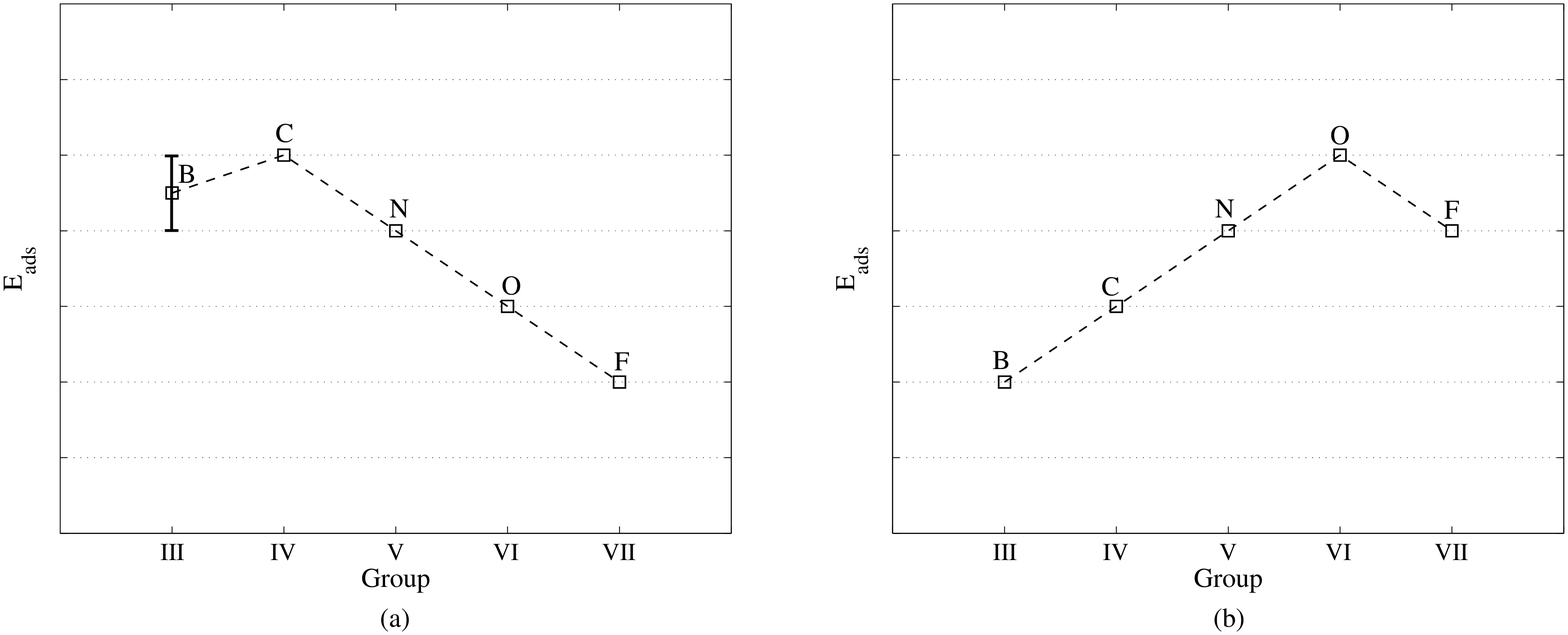}}
\caption{\label{fig:trends}Qualitative graph of the suggested trends in adsorption 
energy $E_{\rm ads}$ for the second-period adatoms caused by 
(a) the adatom--TiSR coupling and (b) the adatom--CSR couplings.  
In (a), the uncertainty in the B $\rightarrow$ C trend (see text for details) 
is marked by an error bar.}
\end{figure*}

Thus, the trend in chemisorption strength that arises from the adatom--TiSR coupling can 
be qualitatively illustrated as in Fig.\ \ref{fig:trends}(a), where we allow an 
uncertainty in the B $\rightarrow$ C trend.

%%%%%%%%% FIGURE %%%%%%%%%%%%%%%%%
%%%%%%%%% FIGURE %%%%%%%%%%%%%%%%%

On the other hand, the total strength of the adatom--CSR couplings increases as the 
adlevel energy approaches the energy of the CSR's, that is, the region between the 
main substrate-UVB peak (CSR4) and the lower edge of the substrate UVB.  
The best energetical overlap between the adlevel and the CSR's is obtained for the 
O adatom, for which all positive $\Delta$DOS peaks of adatom $p$ symmetry below 
$E_F$ lie in the CSR region.  

Thus, the trend in chemisorption strength that arises 
from the adatom-CSR coupling can be qualitatively illustrated as in 
Fig.\ \ref{fig:trends}(b).  

As can ben seen, the adatom-TiSR trend does not agree with the $E_{\rm ads}$ trend, 
while the adatom-CSR does.  We conclude that, in order to reproduce the calculated 
trend in adsorption energy for the second-period adatoms, a description in which 
both the adatom--TiSR and adatom--CSR couplings are present is needed.  
In particular, our analysis points toward the novel result that \emph{the C-localized 
states in the lower part of the TiC UVB, although less important for the bonding nature 
in bulk TiC, play an important role in the description of the nature of chemisorption on 
the TiC(111) surface}.  

We also note that the almost constant value of $E_{\rm ads}$ between C and N can 
be explained by the fact that the two coupling-strength trends go in opposite 
directions as C $\rightarrow$ N.  

For the third-period adatoms, the variations in coupling-strength trends follow 
similar trends to those described above for the second period [see Sec.\ III.C.3(d)].  
The main difference is due to the higher energy of the third-period 
free-adatom level, which causes a better overlap with the CSR's for Cl than for S, in 
contrast to our calculated $E_{\rm ads}$ trend.  As discussed in 
Sec.\ III.C.3(d), however, a weakening of the Cl--CSR coupling is introduced 
by the fact that the Cl adatom is almost fully ionized.  

Finally, we comment on the presence of couplings with the CSR's for the B and Al 
$s$ states [described in Secs.\ III.C.3(c)--(d)].  In the same way as for 
the coupling between the CSR's and the $p$ states of Cl, it should be expected 
that a coupling between CSR's and adatom $s$ states is weak, due to the filled 
nature of the $s$ electron shells for B and Al.  

Question (ii) can now be simply understood.  As discussed in Sec.\ III.C.3(d), 
two main differences, both due to core orthogonalization, can be detected between the 
electronic structures of the second- and third-period adatoms: 
overall smaller negative $\Delta$DOS peaks around $E_F$ for the third-period adatoms 
and overall higher energies for the positive $\Delta$DOS peaks below $E_F$ for 
the third-period adatoms.  The first difference indicates a weaker adatom--TiSR 
coupling for the third-period adatoms.  The second difference causes, within an 
adatom group, a weaker energetical overlap between adlevel and CSR's for the 
third-period adatoms.  Thus, both the coupling to the TiSR and to the CSR's are 
weaker for period-three adatoms, in agreement with our calculated $E_{\rm ads}$ trends.  

Question (iii) can also be answered through the concerted-coupling model.  As discussed in 
Sec.\ III.C.3(e), the preference for fcc site over hcp site is due to the spatial 
extension of the TiSR, which extends toward the fcc site and avoids the hcp site.  
The resulting smaller spatial overlap between hcp adatom state and TiSR, compared 
to the fcc adatom, causes a weaker adatom--TiSR coupling.  This in turn causes a 
smaller downward shift of the bonding adatom--TiSR state and consequently also a 
weaker coupling between the adatom state and the CSR's.  
Thus, both the adatom--TiSR and adatom--CSR couplings are weaker in the hcp site than 
in the fcc site.  

For the top adatom, the lower adatom coordination causes a considerably 
smaller downward shift of the adatom--TiSR bonding state than for both hcp and fcc 
adatoms, again weakening the chemisorption strength.  Also, due to the 
top-site geometry, the spatial overlap between the top-adatom state and the 
C-localized states is negligible, thus strongly weakening also the adatom--CSR 
coupling.  

Finally, question (iv) can be simply understood [Sec.\ III.C.3(f)] from the lack 
of a TiSR on the TiC($001$) surface.  Thus, one of the couplings 
present on the TiC($111$) surface is here absent, causing a considerably weaker 
chemisorption.  

In summary, the concerted-coupling model can be used to successfully explain the 
essential features of the calculated $E_{\rm ads}$ trends.  In particular, the 
coupling of the adatom state to the C-localized states (CSR's) of the TiC($111$) 
surface is expected to play an important role for the chemisorption trends on 
this surface.  This is qualitatively 
different from the case of, {\it e.g.}, adsorption on transition-metal surfaces, 
where the weak coupling to the $sp$ band does not affect the adsorption trends.

%%%%%%%%%%%%%%%%%%%%%%%%%%%%%%%%%%%%%%%%%%%%%%%%%%%%%%%%%%%%
\section{Conclusions}

Chemisorption of atoms in the first three periods in the table of elements 
(H, B, C, N, O, F, Al, Si, P, S, and Cl) on the polar TiC($111$) surface, as 
well as of atomic O on the nonpolar TiC($001$) surface and of full 
($1 \times 1$) monolayer coverage of O on TiC($111$) are studied in a systematic 
way by density-functional theory.  
Adsorption energies and geometries are calculated in fully relaxed configurations.  
Detailed analyses of the electronic structures of the different 
systems, as well as of bulk TiC and of the clean TiC($111$) and ($001$) surfaces 
reveal the nature of chemisorption on this scientifically and 
technologically interesting material.  At the same time, new detailed information 
is obtained on the electronic structures of bulk and surface TiC, showing, 
in particular, the presence of direct C--C interactions and of C-localized 
surface resonances and their importance for the adsorption mechanisms.  

The adsorption on TiC($111$) is found to be more complex and versatile than on 
pure metal surfaces.  The results for the atomic adsorption energies 
$E_{\rm ads}$ of the second- and third-period elements show strong variations, 
characterized by pyramid-shaped trends within each period.  An extraordinarily 
high binding energy is obtained for the O atom, $8.8$ eV/adatom, followed by 
C, N, S, and F.  The lowest energies are obtained for Al, H, and Si.  

To describe the variety in calculated adsorption energies, we propose a 
concerted-coupling model that involves the concerted action of two different 
types of couplings: 
(i) between the adatom state(s) and the Ti-centered Fermi-level 
surface resonance (TiSR) of TiC($111$) and 
(ii) between the adatom state(s) and 
the C-centered surface resonances (CSR's) lying in the lower part of the 
TiC($111$) upper valence band (UVB).  

This model explains successfully the essential features of the calculated 
trends in $E_{\rm ads}$.  For example, the exceptionally strong chemisorption of O 
is understood to arise from this combination of two different adsorption 
mechanisms.  

The strong variations in adsorption strengths show the versatility of TiC 
for technological applications.  The strong O adsorption makes TiC($111$) a good 
substrate for oxide-layer growth, a fact that is used in multilayer CVD 
(chemical-vapor deposition) alumina/Ti(C,N) coatings on wear-resistant cutting 
tools.  At the same time, the high $E_{\rm ads}$ value for S suggests it to be 
a strong candidate in the competition for adsorption sites, which can explain the 
presence of S found in the CVD alumina coatings under certain conditions, 
affecting the alumina phase composition.\cite{Halvarsson}  

Although proper bulk and growth calculations obviously are needed, our calculated 
diffusion-barrier estimates might be indicative for the synthesis and properties 
of the MAX phases.  Their good plasticity relates to a weak bonding between the 
($111$) face of Ti$_6$C layers and, {\it e.g.}, Al or Si.\cite{MAX}  Our 
diffusion-barrier estimates for Al and Si (of the order of $0.1$--$0.3$ eV, 
corresponding to activation temperatures of $\sim 50$--$130$ K)\cite{Bogicevic} 
suggest good lateral mobility between the Al/Si and Ti$_6$C layers.  

The fundamental nature of the concerted-coupling model, based on the 
Newns-Anderson model, should make it apt for general application to 
transition-metal carbides and nitrides.  In separate publications, the 
adsorption features and electronic structures of TiN($111$) are 
compared to those for TiC($111$), showing the applicability of the 
concerted-coupling model also for adsorption on TiN($111$).\cite{SS1,SS2}  
For instance, higher diffusion-barrier estimates are found on TiN($111$), 
which might provide ideas for the fine-tuning of the structural composition 
of the MAX phases.  

Further studies addressing these questions would be highly interesting.  
There are many questions that still remain to be answered.

%%%%%%%%%%%%%%%%%%%%%%%%%%%%%%%%%%%%%%%%%%%%%%%%%%%%%%%%%%%%
\begin{acknowledgments}

The study originates from stimulating discussions with Mats Halvarsson.  
Valuable discussions with Aleksandra Vojvodic and \O yvind Borck are 
gratefully acknowledged, as are financial support from the Swedish Foundation for 
Strategic Research via Materials Consortium \# 9 and ATOMICS and from the Swedish 
Scientific Council, and the allocation of computer time at the UNICC facility 
(Chalmers) and at SNIC (Swedish National Infrastructure for Computing).  

\end{acknowledgments}

%%%%%%%%%%%%%%%%%%%%%%%%%%%%%%%%%%%%%%%%%%%%%%%%%%%%%%%%%%%%

%%%%%%%%%%%%%%%%%%%%%%%%%%%%%%%%%%%%%%%%%%%%%%%%%%%%%%%%%%%%

\begin{thebibliography}
\bibitem{}

\bibitem{Oyama}
        S.T.\ Oyama, editor, {\it The Chemistry of Transition Metal Carbides and 
	Nitrides} (Blackie Academic and Professional, London 1996).

\bibitem{Schwarz}
        K. Schwarz, CRC Crit.\ Rev.\ Sol.\ St.\ Mater.\ Sci.\ {\bf 13}, 
	211 (1987).

\bibitem{Prengel}
	H.G.\ Prengel, W.R.\ Pfouts, and A.T.\ Santhanam, 
	Surf.\ Coat.\ Technol.\ {\bf 102}, 183 (1998).  

\bibitem{MEMS}
	G. Radhakrishnan, P.M.\ Adams, R. Robertson, and R. Cole, 
	Tribol.\ Lett.\ {\bf 8}, 133 (2000).  

\bibitem{Fusion}
	K. Hojou, H. Otsu, S. Furuno, N. Sasajima, and K. Izui, 
	J. Nucl.\ Mater.\ {\bf 239}, 279 (1996);  
	S. Takamura, K. Hayashi, N. Ohno, and K. Morita, 
	{\it ibid.} {\bf 258--263}, 961 (1998).  

\bibitem{Bio}
	M.I.\ Jones, I.R.\ McColl, D.M.\ Grant, K.G.\ Parker, 
	and T.L.\ Parker, J. Biomed.\ Mater.\ Res.\ {\bf 52}, 
	413 (2000).  

\bibitem{Space}
	H.J.\ Boving and H.E.\ Hintermann, 
	Tribol.\ Intern.\ {\bf 23}, 129 (1990).  

\bibitem{Heads}
	X. Li and B. Bhushan, 
	Thin Sol.\ Films {\bf 398--399}, 313 (2001).  

\bibitem{Diamond}
	W.D.\ Fan, K. Jagannadham, and B.C.\ Goral, 
	Surf.\ Coat.\ Technol. {\bf 81}, 172 (1996);  
	M.S.\ Raghuveer, S.N.\ Yoganand, K. Jagannadham, 
	R.L.\ Lemaster, and J. Bailey, Wear {\bf 253}, 1194 (2002).  

\bibitem{Electronics}
	H-S.\ Chen, J.D.\ Parsons, 
	Appl.\ Phys.\ Lett.\ {\bf 65}, 2576 (1994).  

\bibitem{Gunster}
	J. G\"unster, M. Baxendale, S. Otani, and R. Souda, 
	Surf.\ Sci.\ {\bf 494}, L781 (2001).  

\bibitem{Catalysis}
        N.I.\ Ilchenko and Yu.I.\ Pyatnitsky, in 
	{\it The Chemistry of Transition Metal Carbides and 
	Nitrides}, edited by S.T.\ Oyama 
	(Blackie Academic and Professional, London 1996);  
	H.H.\ Hwu and J.G.\ Chen, Chem.\ Rev.\ {\bf 105}, 185 (2005).  

\bibitem{MAX}
	N.I.\ Medvedeva, D.L.\ Novikov, A.L.\ Ivanovski, 
	M.V.\ Kuznetsov, and A.J.\ Freeman, Phys.\ Rev.\ B 
	{\bf 58}, 16042 (1998);  
	Y. Zhou, Z. Sun, X. Wang, and S. Chen, 
	J. Phys.: Condens.\ Matter {\bf 13}, 10001 (2001).  

\bibitem{Jansen}
	S.A.\ Jansen and R. Hoffman, Surf.\ Sci.\ {\bf 197}, 
	474 (1988).  

\bibitem{Tsukada}
	M. Tsukada and T. Hoshino, J. Phys.\ Soc.\ Jpn.\ {\bf 51}, 
	2562 (1982).

\bibitem{Fujimori}
	A. Fujimori, F. Minami, and N. Tsuda, Surf.\ Sci.\ 
	{\bf 121}, 199 (1982).

\bibitem{Aono81}
	M. Aono, C. Oshima, S. Zaima, S. Otani, and Y. Ishizawa, 
	Jpn.\ J. Appl.\ Phys.\ {\bf 20}, L829 (1981).

\bibitem{Oshima81_jlcm}
	C. Oshima, M. Aono, S. Zaima, Y. Shibata, and S. Kawai, 
	J. Less-Common Metals {\bf 82}, 69 (1981).

\bibitem{Zaima}
	S. Zaima, Y. Shibata, H. Adachi, C. Oshima, S. Otani, M. Aono, and 
	Y. Ishizawa, Surf.\ Sci.\ {\bf 157}, 380 (1985).

\bibitem{Tan}
	K.E.\ Tan, M.W.\ Finnis, A.P.\ Horsfield, and A.P.\ Sutton, 
	Surf.\ Sci.\ {\bf 348}, 49 (1996).

\bibitem{Bradshaw}
	A.M.\ Bradshaw, J.F.\ van der Veen, F.J.\ Himpsel, and 
	D.E.\ Eastman, Sol.\ St.\ Comm.\ {\bf 37}, 37 (1980).

\bibitem{Edamoto92_prb}
	K. Edamoto, T. Anazawa, A. Mochida, T. Itakura, E. Miyazaki, 
	H. Kato, and S. Otani, Phys.\ Rev.\ B {\bf 46}, 4192 (1992).

\bibitem{Ahn}
	J. Ahn, H. Kawanowa, and R. Souda, Surf.\ Sci.\ {\bf 429}, 
	338 (1999).

\bibitem{Edamoto90}
	K. Edamoto, Y. Abe, T. Ikeda, N. Ito, E. Miyazaki, H. Kato, 
	and S. Otani, Surf.\ Sci.\ {\bf 237}, 241 (1990).

\bibitem{Souda88}
	R. Souda, C. Oshima, S. Otani, Y. Ishizawa, and M. Aono, 
	Surf.\ Sci.\ {\bf 199}, 154 (1988).

\bibitem{Oshima83}
	C. Oshima, M. Aono, S. Otani, and Y. Ishizawa, 
	Sol.\ St.\ Comm.\ {\bf 48}, 911 (1983).

\bibitem{Edamoto92_ss}
	K. Edamoto, E. Miyazaki, T. Anazawa, A. Mochida, and H. Kato, 
	Surf.\ Sci.\ {\bf 269/270}, 389 (1992).

\bibitem{Edamoto92_prb_O}
	K. Edamoto, A. Mochida, T. Anazawa, T. Itakura, E. Miyazaki, 
	H. Kato, and S. Otani, Phys.\ Rev.\ B {\bf 46}, 7127 (1992).

\bibitem{DFT}
	P. Hohenberg and W. Kohn, Phys.\ Rev.\ {\bf 136}, B864 (1964);  
	W. Kohn and L.J.\ Sham, {\it ibid.} {\bf 140}, A1133 (1965).


%%% SEC. II
\bibitem{dacapo}
	B. Hammer {\it et al.}, 
	Center for Atomic-Scale Materials Physics (CAMP), 
	Danmarks Tekniske Universitet, Lyngby, Denmark.  

\bibitem{PW91}
	K. Burke, J.P.\ Perdew, and Y. Wang, in {\it Electronic Density 
	Functional Theory, Recent Progress and New Directions}, 
	edited by J.F.\ Dobson, G. Vignale and M.P.\ Das (Plenum Press, 
	New York and London 1998); pp.\ 81--111.  

\bibitem{Vanderbilt}
	D. Vanderbilt, Phys.\ Rev.\ B, {\bf 41}, 7892 (1990).

\bibitem{Monkhorst}
	H.J.\ Monkhorst and J.D.\ Pack, Phys.\ Rev.\ B {\bf 13}, 5188 (1976).  

\bibitem{Bengtsson}
	L. Bengtsson, Phys.\ Rev.\ B {\bf 59}, 12301 (1999).  

\bibitem{Bengtsson_PhD}
	L. Bengtsson, {\it Efficient Density-Functional-Based Calculational 
	Methods for Surfaces}, Ph.D.\ Thesis 
	(Chalmers University of Technology 
	and Goteborg University, Goteborg, Sweden 1999); pp.\ 40--3.  

\bibitem{Yourdshahyan}
	Y. Yourdshahyan, C. Ruberto, M. Halvarsson, L. Bengtsson, V. Langer, 
	B.I.\ Lundqvist, S. Ruppi, and U. Rolander, J. Am.\ Ceram.\ Soc.\ 
	{\bf 82}, 1365 (1999).

\bibitem{Ruberto}
	C. Ruberto, Y. Yourdshahayn, and B.I.\ Lundqvist, Phys.\ Rev.\ 
	Lett.\ {\bf 88}, 226101 (2002);  Phys.\ Rev.\ B {\bf 67}, 195412 (2003).

\bibitem{Dudiy}
	S. Dudiy, J. Hartford, and B.I.\ Lundqvist, Phys.\ Rev.\ Lett.\ 
	{\bf 85}, 1898 (2000).  

\bibitem{Rydberg}
	H. Rydberg, M. Dion, N. Jacobson, E. Schr\"oder, P. Hyldgaard, S.I.\ Simak, 
	D.C.\ Langreth, and B.I.\ Lundqvist, Phys.\ Rev.\ Lett.\ {\bf 91}, 
	126402 (2003).  

\bibitem{Bader}
	R.F.W.\ Bader, Chem.\ Rev.\ {\bf 91}, 893 (1991).

\bibitem{Malcolm}
	N.O.J.\ Malcolm and P.L.A.\ Popelier, J. Comput.\ Chem.\ {\bf 24}, 1276 
	(2003).  

\bibitem{Gay}
	J.G.\ Gay, J.R.\ Smith, R. Richter, F.J.\ Arlinghaus, and R.H.\ Wagoner, 
	J.\ Vac.\ Sci.\ Technol.\ A {\bf 2}, 931 (1984).

\bibitem{Boettger_Esurf}
	J.C.\ Boettger, Phys.\ Rev.\ B {\bf 49}, 16 798 (1994).

\bibitem{Fiorentini}
	V. Fiorentini and M. Methfessel, J.\ Phys.:\ Condens.\ Matter {\bf 8}, 
	6525 (1996).  

\bibitem{Boettger_QSE}
	J.C.\ Boettger, Phys.\ Rev.\ B {\bf 53}, 13 133 (1996).

\bibitem{Ruberto_PhD}
	C. Ruberto, {\it Metastable Alumina from Theory:  Bulk, Surface, and 
	Growth of $\kappa$-Al$_2$O$_3$}, Ph.D.\ Thesis 
	(Chalmers University of Technology and G\"oteborg University, 
	G\"oteborg, Sweden 2001).  


%%% SEC. III.A
\bibitem{Dunand}
	A. Dunand, H.D.\ Flack, and K. Yvon, Phys.\ Rev.\ B {\bf 31}, 
	2299 (1985).  

\bibitem{Ihara}
	H. Ihara, Y. Kumashiro, and A. Itoh, Phys.\ Rev.\ B {\bf 12}, 
	5465 (1975).

\bibitem{Johansson}
	L.I.\ Johansson, P.M.\ Stefan, M.L.\ Shek, and A.N.\ Christensen, 
	Phys.\ Rev.\ B {\bf 22}, 1032 (1980).

\bibitem{Blaha83}
        P. Blaha and K. Schwarz, Int.\ J. Quant.\ Chem.\ {\bf 23}, 
	1535 (1983).

\bibitem{Blaha85}
	P. Blaha, J. Redinger, and K. Schwarz, Phys.\ Rev.\ B {\bf 31}, 
	2316 (1985).  

\bibitem{Price}
        D.L.\ Price and B.R.\ Cooper, Phys.\ Rev.\ B {\bf 39}, 
	4945 (1989).

\bibitem{Eberhart}
        M.E.\ Eberhart and J.M.\ MacLaren, in {\it The Chemistry of 
	Transition Metal Carbides and Nitrides}, edited by S.T.\ Oyama 
	(Blackie Academic and Professional, London 1996), Chap.\ 5.

\bibitem{Jhi}
        S.-H.\ Jhi, J. Ihm, S.G.\ Louie, and M.L.\ Cohen, 
	Nature {\bf 399}, 132 (1999).

\bibitem{Matar}
	S.F.\ Matar, Y. Le Petitcorps, and J. Etourneau, 
	J. Mater.\ Chem.\ {\bf 7}, 99 (1997).  




%%% SEC. III.B
\bibitem{Tasker}
	P.W.\ Tasker, J. Phys.\ C {\bf 12}, 4977 (1979).

\bibitem{Price96}
	D.L.\ Price, J.M.\ Wills, and B.R.\ Cooper, 
	Phys.\ Rev.\ Lett.\ {\bf 77}, 3375 (1996).

\bibitem{Tagawa}
	M. Tagawa, T. Kawasaki, C. Oshima, S. Otani, K. Edamoto, and 
	A. Nagashima, Surf.\ Sci.\ {\bf 517}, 59 (2002).

\bibitem{VASP}
	G. Kresse and J. Furthm\"uller, Phys.\ Rev.\ B {\bf 54}, 
	11 169 (1996); 
	Comput.\ Mater.\ Sci.\ {\bf 6}, 15 (1996); 
	G. Kresse and J. Hafner, Phys.\ Rev.\ B {\bf 49}, 14 251 (1994); 
	{\bf 48}, 13 115 (1993); 
	{\bf 47}, 558 (1993).

\bibitem{Dudiy04}
	S.V.\ Dudiy and B.I.\ Lundqvist, Phys.\ Rev.\ B {\bf 69}, 
	125421 (2004).



%%% SEC. III.C.1
\bibitem{Oshima81_ss}
	C. Oshima, M. Aono, T. Tanaka, S. Kawai, S. Zaima, and Y. Shibata, 
	Surf.\ Sci.\ {\bf 102}, 312 (1981).

\bibitem{Souda91}
	R. Souda, T. Aizawa, S. Otani, Y. Ishizawa, and C. Oshima, 
	Surf.\ Sci.\ {\bf 256}, 19 (1991).

\bibitem{Didzilius03}
	S.V.\ Didzilius, K.D.\ Butcher, and S.S.\ Perry, 
	Inorg.\ Chem.\ {\bf 42}, 7766 (2003).

\bibitem{Zhang04}
	Y.-F.\ Zhang, J.-Q.\ Li, and Z.-F.\ Liu, 
	J. Phys.\ Chem.\ B {\bf 108}, 17143 (2004).


%%% SEC. III.C.2

\bibitem{BIL_Tosi}
	See, {\it e.g.}, B.I.\ Lundqvist, in {\it Interaction of Atoms and 
        Molecules with Solid Surfaces}, edited by V. Bortolani, N.H.\ March, 
        and M.P.\ Tosi (Plenum Press, New York and London, 1990), Chap.\ 8.  


%%% SEC. III.C.3
\bibitem{Newns}
	D.M.\ Newns, Phys.\ Rev.\ {\bf 178}, 1123 (1969).

\bibitem{Anderson}
	P.W.\ Anderson, Phys.\ Rev.\ {\bf 124}, 41 (1961).


\bibitem{Desj_Spanj}
	D. Spanjaard and M.C.\ Desjonqu\`eres, in 
	{\it Interaction of Atoms and Molecules with Solid Surfaces}, 
	edited by V. Bortolani, N.H.\ March, and M.P.\ Tosi 
	(Plenum Press, New York and London 1990), Chap.\ 9.
	
\bibitem{Hammer95}
	B. Hammer and J.K.\ N\o rskov, Surf.\ Sci.\ {\bf 343}, 
	211 (1995); {\bf 359}, 306 (1996).

\bibitem{Stokbro97}
	K. Stokbro and S. Baroni, Surf.\ Sci.\ {\bf 370}, 166 (1997).

\bibitem{Hammer00}
	B. Hammer and J.K.\ N\o rskov, in {\it Advances in Catalysis}, 
	edited by B.C.\ Gates and H. Kn\"ozinger (Academic Press, 
	San Diego and London 2000), Vol.\ 45, p. 71.


%%% SEC. V
\bibitem{Halvarsson}
	M. Halvarsson, A. Larsson, and S. Ruppi, Micron {\bf 32}, 
	807 (2001).  

\bibitem{Bogicevic}
	A. Bogicevic, J. Str\"omquist, and B.I.\ Lundqvist, 
	Phys.\ Rev.\ Lett.\ {\bf 81}, 637 (1998).  

\bibitem{SS1}
	A. Vojvodic, C. Ruberto, and B.I.\ Lundqvist, 
	submitted to Surf.\ Sci.\ (2005).  

\bibitem{SS2}
	C. Ruberto, A. Vojvodic, and B.I.\ Lundqvist, 
	submitted to Surf.\ Sci.\ (2005).  


\end{thebibliography}
\end{document}